\documentclass[a4paper,onecolumn,11pt,unpublished]{quantumarticle}
\pdfoutput=1

\usepackage{tikz}
\usepackage{braket}
\usetikzlibrary{shapes,arrows,positioning,shapes.geometric}
\usepackage{graphicx} 
\usepackage{physics}
\usepackage{ulem}
\usepackage[version=4]{mhchem}
\usepackage{multirow}
\usepackage{makecell}
\usepackage{amsmath}
\usepackage{bigstrut}
\usepackage{titlesec}
\usepackage{url}
\usepackage{quantikz}
\usepackage{tabularray}

\usepackage{booktabs}
\usepackage{array}   
\usepackage{ragged2e}
\usepackage{xcolor}  
\usepackage{colortbl} 
\usepackage{siunitx} 
\usepackage{enumitem}
\usepackage{calc} 
\usepackage{booktabs}

\usepackage{subcaption}
\usepackage[export]{adjustbox} 

\definecolor{headcolor}{RGB}{225,225,235}
\newlength{\toolnamewidth}
\usepackage{hyperref} 
\usepackage{cleveref}

\usepackage{comment}
\usepackage{stmaryrd}
\usepackage{amssymb}
\usepackage{amsthm}

\newtheorem{conjecture}{Conjecture}[section]

\DeclareMathOperator*{\argmax}{arg\,max}

\newcommand{\VEC}[1]{\mathbf{#1}}

\newcommand{\RSD}{\text{RSD}}
\newcommand{\setelem}[1]{\ensuremath{\{#1\}}}
\newcommand{\setelemtwo}[2]{\setelem{#1},\setelem{#2}}
\newcommand{\setelemthree}[3]{\setelem{#1},\setelem{#2},\setelem{#3}}

\newcommand{\nBE}{n_{\mathrm{BE}}}

\NewDocumentCommand{\Ptm}{m m}{P^{\mathrm{#1}}_{#2}}

\NewDocumentCommand{\Stm}{m m}{S^{\mathrm{#1}}_{#2}}
\NewDocumentCommand{\Stt}{m m}{S^{\mathrm{#1}}_{\mathrm{#2}}}
\newcommand{\Sin}{\Stm{in}{}}
\newcommand{\Sout}{\Stm{out}{}}
\newcommand{\SJN}{\Stm{JN}{}}
\newcommand{\Saddi}{\Stm{add}{i}}
\newcommand{\Acable}{A_{\mathrm{cable}}}
\newcommand{\Teff}{T_{\mathrm{eff}}}
\newcommand{\Purcell}{\mathrm{Purcell}}
\newcommand{\TPurcell}{T_1^{\Purcell}}

\newcommand{\Tphi}{T_{\phi}}
\newcommand{\Phiext}{\Phi_{\mathrm{ext}}}

\newcommand{\Decay}[3]{\Gamma^{#3}_{{#1}\rightarrow{#2}}}
\newcommand{\DecaySum}[3]{\Gamma^{#3}_{{#1 #2}}}
\newcommand{\DecayBase}[3]{\tilde{\Gamma}^{#3}_{#1 #2}}

\newcommand{\melabs}[3]{\left|\mel{#1}{#2}{#3}\right|}
\newcommand{\Gammaphi}{\Gamma_{\phi}}
\newcommand{\DeltaPthermal}{\Delta P_{\mathrm{th}}}

\newcommand{\xqp}{x_{\text{qp}}}
\newcommand{\defeq}{\overset{\text{def}}{=}}
\newcommand{\nth}{n_{\text{th}}}

\begin{document}

%%
%% The "title" command has an optional parameter,
%% allowing the author to define a "short title" to be used in page headers.
\title{Quantum Design Automation: Foundations, Challenges, and the Road Ahead}

% %%
% %% The "author" command and its associated commands are used to define
% %% the authors and their affiliations.
% %% Of note is the shared affiliation of the first two authors, and the
% %% "authornote" and "authornotemark" commands
% %% used to denote shared contribution to the research.
% \author{XXXX}
% \affiliation{%
%   \institution{XXXX}  
%   \city{XXXX}
%   \state{XXXX}
%   \country{XXXX}
%   \postcode{XXXX}}
% \orcid{}
% \email{}
\author{Feng Wu}
\affiliation{Zhongguancun Laboratory, Beijing, China}
\thanks{These two authors contributed equally.}
\orcid{0000-0003-1652-9243}

\author{Jingzhe Guo}
\affiliation{Department of Computer Science and Technology, Tsinghua University, Beijing 100084, China}
\thanks{These two authors contributed equally.}
\orcid{0000-0002-7921-9771}

\author{Tian Xia}
\affiliation{Independent Researcher}
\orcid{0000-0001-9136-6638}

\author{Linghang Kong}
\affiliation{Zhongguancun Laboratory, Beijing, China}
\orcid{0000-0002-5854-5340}

\author{Fang Zhang}
\affiliation{Zhongguancun Laboratory, Beijing, China}
\orcid{0000-0002-0000-7101}

\author{Ziang Wang}
\affiliation{Zhejiang Institute of Modern Physics and Zhejiang Key Laboratory of Micro-nano Quantum Chips and Quantum Control, Zhejiang University, Hangzhou 310027, China}
\orcid{0009-0002-4931-7781}

\author{Aochu Dai}
\affiliation{Department of Computer Science and Technology, Tsinghua University, Beijing 100084, China}
\orcid{0009-0005-9935-8823}

\author{Ziyuan Wang}
\affiliation{Department of Computer Science and Technology, Tsinghua University, Beijing 100084, China}
\orcid{0009-0009-4476-1382}

\author{Zhaohui Yang}
\affiliation{Department of Electronic and Computer Engineering, The Hong Kong University of Science and Technology, Hong Kong}
\orcid{0000-0003-4698-4378}

\author{Hao Deng}
\affiliation{School of Physical Science and Technology, ShanghaiTech University, Shanghai 201210, China}
\affiliation{School of Information Science and Technology, ShanghaiTech University, Shanghai 201210, China}
\orcid{0009-0006-0647-6327}

\author{Kai Zhang}
\affiliation{Department of Computer Science and Technology, Tsinghua University, Beijing 100084, China}
\orcid{0009-0005-6803-7533}

\author{Zhengfeng Ji}
\affiliation{Department of Computer Science and Technology, Tsinghua University, Beijing 100084, China}
\affiliation{Zhongguancun Laboratory, Beijing, China}
\orcid{0000-0002-7659-3178}

\author{Yuan Feng}
\affiliation{Department of Computer Science and Technology, Tsinghua University, Beijing 100084, China}
\orcid{0000-0002-3097-3896}

\author{Hui-Hai Zhao}
\affiliation{Zhongguancun Laboratory, Beijing, China}
\email{zhaohuihai@iqubit.org}
\orcid{0000-0001-7075-8325}

\author{Jianxin Chen}
\affiliation{Department of Computer Science and Technology, Tsinghua University, Beijing 100084, China}
\email{chenjianxin@tsinghua.edu.cn}
\orcid{0000-0002-9365-776X}

%%
%% By default, the full list of authors will be used in the page
%% headers. Often, this list is too long, and will overlap
%% other information printed in the page headers. This command allows
%% the author to define a more concise list
%% of authors' names for this purpose.

\newcommand\JC[1]{{\color{blue} [JC: #1]}}
\newcommand\XT[1]{{\color{magenta} [XT: #1]}}
\newcommand{\HZ}[1]{{\textcolor{violet} {[HZ: #1]}}}
\newcommand{\jingzhe}[1]{{\textcolor{violet} {[Jingzhe: #1]}}}
\newcommand{\fang}[1]{{\textcolor{violet} {[Fang: #1]}}}
\newcommand{\wufeng}[1]{{\textcolor{brown} {[WF:#1]}}}
\newcommand\YF[1]{{\color{blue} [YF: #1]}}
\newcommand\znote[1]{{\color{red} [ZJ\@: #1]}}
\newcommand\ZY[1]{{\color{purple} [ZY: #1]}}

%%
%% The abstract is a short summary of the work to be presented in the
%% article.
\begin{abstract}
Quantum computing is transitioning from laboratory research to industrial deployment, yet significant challenges persist: system scalability and performance, fabrication yields, and the advancement of algorithms and applications. We emphasize that in building quantum computers---spanning quantum chips, system integration, instruction sets, algorithms, and middleware such as quantum error correction schemes---design is everywhere. 

Historically, hardware and software designs have followed separate development trajectories. In this paper, we advocate for a holistic design perspective in quantum computing, a perspective we argue is pivotal to unlocking innovative co-design opportunities and addressing the aforementioned key challenges. 

To equip readers with sufficient background for exploring co-optimization opportunities, we detail how interconnected computational methods and tools collaborate to enable end-to-end quantum computer design. This coverage encompasses critical stages---such as quantum chip layout automation, high-fidelity system-level simulation, Hamiltonian derivation for quantum system modeling, control pulse simulation, multi-objective design optimization, decoherence mechanism analysis, and physical verification and testing---followed by quantum instruction set design. From this foundation, we then proceed to quantum system and software development, including quantum circuit synthesis, quantum error correction and fault tolerance, and logic verification and testing. 

Through these discussions, we illustrate with concrete examples---including co-optimizing quantum instruction sets with algorithmic considerations, customizing error correction circuits to hardware-specific constraints, and streamlining quantum chip design through tailored code design, among others. It is our hope that the detailed end-to-end design workflow as well as these examples will serve the community by fostering dialogue between the hardware and software communities, ultimately facilitating the translation of meaningful research findings into future quantum hardware implementations.

\end{abstract}

\maketitle

\setcounter{tocdepth}{3}
\tableofcontents

\section{Introduction}

For decades, the exponential growth of classical computing, famously described by Moore's Law~\cite{moore1965moore}, has been driven by the predictable scaling of semiconductor transistors. However, this era has drawn to a close, as the fundamental physical limitations become unavoidable. With the increase in transistor density, the transistor dimensions shrink to the nanometer scale. Therefore, the quantum mechanical effects, such as electron tunneling, lead to significant current leakage, resulting in unmanageable heat generation. This roadblock in classical hardware development has intensified the search for alternative computing paradigms. In this context, quantum computing has emerged as a revolutionary approach to processing information, leveraging principles of superposition and entanglement to solve certain problems that are intractable for any classical computer~\cite{Nielsen2010Quantum}.

The impact of quantum computing is expected to be disruptive, with transformative applications across science and industry. Its most significant uses include the simulation of quantum systems~\cite{Feynman1982}, aiming to revolutionize materials science and drug discovery; the ability to break modern cryptographic standards using Shor's algorithm~\cite{Shor1994}; and the acceleration of optimization and search problems through methods like Grover's algorithm~\cite{Grover1996}, etc.

However, translating these powerful theoretical algorithms into practical solutions on physical quantum hardware presents an immense engineering challenge. The gap between an abstract algorithm and the precise control signals required to execute it on a noisy, resource-constrained quantum processor is huge. Manually navigating this complex translation is not only unscalable but also highly error-prone. In classical computing, this challenge was overcome by Electronic Design Automation (EDA), a critical set of software tools and methodologies that automates the design workflow of integrated circuits, from logic specification down to physical layout. To unlock the quantum advantage for real-world applications, a similar systematic approach is essential to overcome formidable design challenges. This is precisely the role of Quantum Design Automation (QDA), which provides the necessary toolchain for designing, optimizing, and verifying quantum computing systems, transforming the theoretical promise of quantum computing into an engineering reality~\cite{Micheli2022, Almudever2024}.

The QDA workflow can be broadly divided into two interconnected domains:

\begin{enumerate}
\item \textbf{Physical-Level Design.}
This domain focuses on the hardware-centric stack, addressing the multi-scale challenge of designing the quantum processor and the complex physical systems required for its control and operation. 
Research efforts have prioritized customized design methodologies over generalized automation, largely due to the nascent stage of quantum hardware development, where proof-of-concept demonstrations have taken precedence over scaling.
However, with rapid advancements in quantum computing hardware---such as superconducting qubits scaling to hundreds and Rydberg atom systems reaching thousands of qubits---established EDA principles can now be adapted.
These principles include modeling, simulation, and physical verification, mirroring classical computing's design pipeline, which is now essential for scaling quantum hardware.
For superconducting circuits, the focus of this paper, key QDA tasks begin with \textit{chip layout design}, where the geometric patterns and arrangement of qubits, resonators, and control lines are defined and optimized. 
This is followed by \textit{electromagnetic simulation} to model device behavior and extract critical Hamiltonian parameters. The \textit{control pulse simulation} is then performed for high-fidelity gate design.
This level also encompasses \textit{the package and the cryogenic setup design}, which provides the critical electrical I/O and the shielding that isolates the processor within its tightly controlled millikelvin operating environment. 
Because of the fragility of quantum states, it is essential to include comprehensive \textit{decoherence and noise analysis} to identify and mitigate error sources across the entire system, from chip to cryogenic setup in the QDA workflow. Another critical QDA task is yield enhancement, which is achieved by the tools associated with \textit{verification and testing}, as well as the \textit{technology computer-aided design}. 

\item \textbf{Logic-Level Design.}
This domain focuses on the software-centric stack, transforming the high-level description of a quantum algorithm into executable operations tailored for specific quantum hardware. 
The process adapts core principles from classical EDA---abstraction, optimization, and verification---to address the unique challenges posed by the probabilistic and fragile nature of quantum systems. 
Bridged to physical-level design via \textit{quantum instruction set design}, the core tasks of logic-level design encompass four key areas: first, \textit{quantum circuit synthesis}---a process that decomposes complex quantum operations into a set of fundamental, hardware-compatible instructions; second, \textit{circuit optimization}, which aims to minimize resource consumption (e.g., qubit count and gate usage) while alleviating the adverse effects of decoherence; third, \textit{qubit mapping and routing}, which tailors algorithmic requirements to the hardware's constrained physical topology (such as limited qubit connectivity); and fourth, \textit{logic-level verification and testing}, which validates the correctness, reliability, and performance of logic designs before their translation to physical hardware. Additionally, this domain bears responsibility for designing and implementing \textit{quantum error correction (QEC) codes}. These codes are indispensable for shielding fragile quantum states from environmental noise and laying the groundwork for fault-tolerant quantum computation---a critical milestone for advancing practical quantum systems~\cite{Terhal2015QECreview}.

\end{enumerate}

Historically, these research directions have operated in separate silos, with physical-level QDA being dominated by physicists and electronic engineers, and logic-level QDA primarily being driven by computer scientists.
However, bridging this gap by considering the interplay between these two levels is becoming increasingly critical as the field matures.
For instance, the qubit connectivity and placement on the chip layout at the physical level must account for the target quantum error correction scheme and high-level algorithmic constraints, which originate from logic-level abstractions.
Conversely, logic-level QDA tasks---such as circuit optimization---require inputs such as the native gate set or quantum instruction set and noise characteristics of the hardware, which are themselves outputs of physical-level QDA.
This interdependence underscores the need for co-design frameworks that integrate both perspectives to advance scalable, practical quantum computing.

This paper presents an introductory review of the comprehensive QDA workflow, covering its key stages, fundamental challenges, and interdisciplinary connections. 
We systematically examine the QDA pipeline from the physical implementation of quantum hardware to quantum algorithm and quantum error correction scheme development, highlighting recent trends in error correction code design, logic synthesis, circuit optimization, and hardware-aware compilation. 
The review particularly focuses on bridging the traditional divide between physics-oriented physical implementation and computer science-driven logic design approaches. 
While many of the logic-level design concepts and tools discussed are applicable across various hardware platforms, our review of physical-level design will concentrate specifically on the superconducting quantum computing modality, which is currently a leading candidate for building scalable quantum processors~\cite{Kjaergaard2020}.
By focusing on this specific platform, we can thoroughly explore the critical link between the physical implementation of quantum devices and the logical abstractions needed to program them.
By analyzing current co-design methodologies and emerging tools, we identify key bottlenecks in scalability and challenges that must be addressed to enable practical quantum advantage. Furthermore, we discuss how classical EDA principles are being adapted for quantum systems, while noting the unique challenges posed by quantum mechanical constraints.

\section{Comprehensive Overview}

\begin{figure}
\centering
\includegraphics[width=1.0\textwidth]{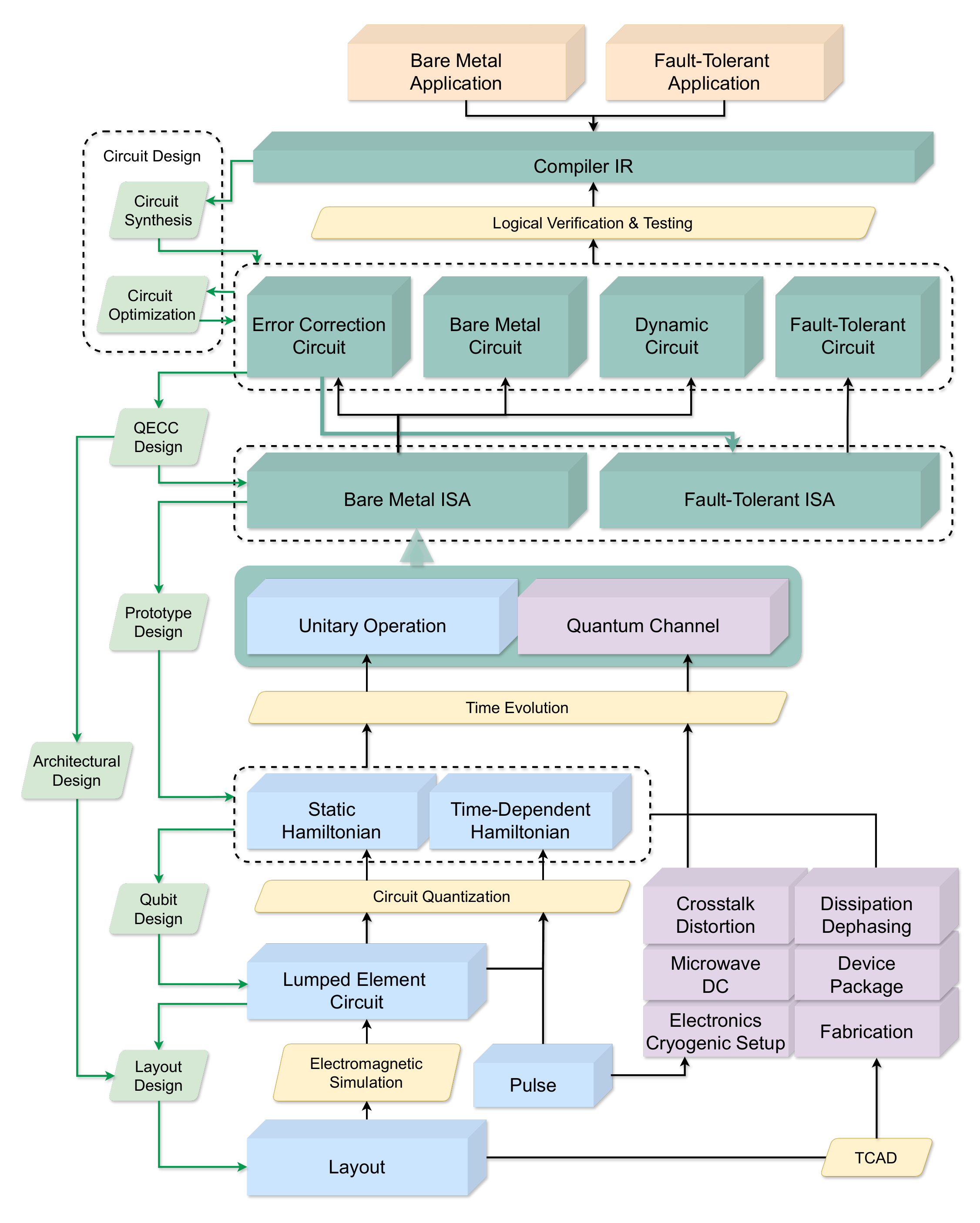}
% \Description{QDA workflow}
\caption{The comprehensive illustration of the QDA workflow, which describe the co-design process connecting the physical-level and the logic-level design stages. The proposed workflow is able to accept a target application and proceed through successive stages of optimization, ultimately producing an optimized circuit compiled for an optimal ISA on an optimized processor. Objects are depicted as cuboids. A black arrow indicates that the performance of the target object can be evaluated based on the source objects, with the corresponding computational process shown in a yellow parallelogram. A green arrow indicates that the target object can be designed by adjusting the source objects, and the associated design process is represented in a green parallelogram. Multiple design processes can be linked together to form a co-design framework.} 
\label{fig:QDA_workflow}
\end{figure}

The QDA involves a series of methods to be implemented as software tools, which work together in a workflow, as shown in~\Cref{fig:QDA_workflow}, to tackle the design issues in quantum computing. 
At the physical level, it includes the design, simulation, optimization, and verification of chips, controls, packages, and cryogenic setups, along with decoherence analysis and technology computer-aided design for yield and performance enhancement.
At the logic level, the design workflow encompasses quantum instruction sets, the synthesis, simulation, and verification of quantum circuits, as well as quantum error correction.
The tools in these two levels provide a complete perspective of the design automation landscape for quantum computing.
As the demand for large-scale quantum computing continues to grow, the QDA will be instrumental in  overcoming the challenges associated with efficient, accurate, and scalable design.
Therefore, in this section, we will examine the spectrum of desired or available QDA tools, clarify their respective purposes, and outline their current state of play. As noted earlier, the scope of this paper centers on the physical realization of superconducting qubits~\cite{Krantz2019A}. That said, certain methodologies---specifically most logic-level approaches and select physical-level techniques---should be adaptable to other leading quantum platforms, including trapped ions~\cite{bruzewicz2019trapped} and neutral atoms~\cite{adams2019rydberg, saffman2010quantum, wu2021concise}.

\subsection{Components of QDA.}

The QDA workflow, as illustrated in~\Cref{fig:QDA_workflow}, is a multi-layer process that integrates the physical level and logic level design stages. The entire process is iterative and interconnected, with feedback loops between different stages to deliver co-optimized design. The workflow can be initiated from either the physical level or the logic level, as the requirements of these two levels are often interdependent. 

At the physical level, the workflow starts with chip layout design, which requires consideration of the qubit connectivity from the architectural design and the qubit type from the qubit design.
The layout design process involves the parametric cell design of reusable components, the floorplan of chip layers, the geometric pattern and placement of qubits, and the routing of control lines. This is followed by electromagnetic simulations based on the geometric layout to extract the parameters for a lumped-element circuit model. The circuit model is then used to derive the static Hamiltonian of the chip through circuit quantization. To predict the performance of the fabricated chip and enhance the yield, the technology computer-aided design (TCAD) is employed to simulate the device behavior and fabrication processes.

With the static Hamiltonian defined, the workflow moves to the design of qubit control. The control pulses, which are transmitted from the electronics to the chip via the cryogenic setup and package, are incorporated into the static Hamiltonian to form a time-dependent (TD) Hamiltonian. The effect of these pulses is analyzed by computing the time evolution of this TD Hamiltonian to obtain the resulting unitary operation. 
The pulse and layout shapes, which determine the parameters of this TD Hamiltonian, are co-optimized to achieve high-fidelity gate execution.

A critical aspect of the QDA workflow is the analysis and mitigation of errors in quantum operations. Various error sources are modeled, and their impact on gate fidelity is assessed. These include dissipation and dephasing from defects in fabrication and environmental noise related to package and cryogenic setup designs, crosstalk between qubits arising from unwanted interactions, and pulse distortion due to control line imperfections. These error sources are collectively integrated into the TD Hamiltonian time evolution simulation in the open quantum system, from which the quantum channel model is extracted to evaluate the gate fidelity under realistic conditions.

The transition from physical to logic level design is bridged by the bare metal instruction set architecture (ISA), which defines the native gate set and qubit connectivity supported by the quantum hardware. This ISA serves as the interface between hardware capabilities and software requirements, providing the prototype design for the quantum processor Hamiltonian and the foundation for quantum circuit synthesis, where high-level quantum algorithms are decomposed into sequences of native gates. The bare metal ISA directly supports noisy intermediate-scale quantum (NISQ) applications through bare metal circuits with the compiler intermediate representation (IR), accompanied by errors from hardware noise. These circuits undergo circuit design optimization to minimize resource usage and error accumulation.

For fault-tolerant quantum computing, the workflow incorporates an additional layer of quantum error correction code design to define how logical qubits are encoded using multiple physical qubits to protect against errors. This leads to the fault-tolerant ISA, which supports operations on error-corrected logical qubits. Through the compiler IR, high-level fault-tolerant applications are synthesized into error correction circuits that implement the syndrome extraction and correction procedures. Dynamic circuits, which modify the circuit based on real-time measurement outcomes, enable adaptive error correction strategies. Eventually, these steps are combined to execute fault-tolerant circuits reliably in the presence of hardware errors.

This entire QDA workflow embodies the co-design principle, where the designs at each stage influence and constrain choices at other stages, as connected by the arrows depicted in ~\Cref{fig:QDA_workflow}. Both the physical-level and logic-level designs are iteratively refined based on feedback from each other. For instance, the choice of error correction code impacts the required qubit connectivity in the layout design and the native gate set in the ISA. Conversely, the capabilities of the hardware may constrain the choice of error correction schemes and circuit optimization strategies at the logic level. The tools and methodologies employed in this workflow should enable the systematic design of quantum computing systems with complex interdependencies, ensuring that physical level and logic level components are co-optimized.

%%%%%%%%%%%%%%%%%%%%%%%%%%%%%%%%%%%%%%%%%%%%%%%%%%%%%%%%%%%%%%%%%%%%%%%%%%%%%%%%%%%%%%%%%
\begin{table}[!ht]

\newenvironment{tightitemize}{
  \begin{itemize}[
    nosep,                
    label=$\bullet$,      
    labelsep=4pt,         
    itemsep=0pt,          
    leftmargin=12pt,      
    topsep=2pt,           
    partopsep=0pt,        
    parsep=0pt            
  ]
  \footnotesize  
}{
  \end{itemize}
}

\newcolumntype{P}[1]{>{\centering\arraybackslash\small}m{#1}}  
\newcolumntype{L}[1]{>{\arraybackslash\small}m{#1}}           

  \centering
  \small  
  \caption{Quantum Design Automation Tools Ecosystem}
  \label{tab:qda_tools}
  
  \begin{tabular}{P{1.55cm}|P{2.6cm}|L{9.6cm}}
    \hline
    \rowcolor{headcolor}
    \textbf{Category} & \textbf{Tool} & \textbf{Key Features} \\  
    \hline

    \multirow{13}{*}{\parbox{1.55cm}{\centering \textbf{Physical}\\ \textbf{Level}}}
    & Qiskit Metal~\cite{Minev2021Qiskit}
    & \begin{tightitemize}
        \item Built-in layout editor with parametric component library
        \item Quantization and Hamiltonian analyses
      \end{tightitemize}
    \\ \cline{2-3}  
    
    & KQCircuits~\cite{Cucurachi2021KQCircuits} 
    & \begin{tightitemize}
        \item Plug-in to KLayout with parametric component library
        \item Fabrication-ready wafer composition and DRCs
      \end{tightitemize}
    \\ \cline{2-3}
    
    & QuTiP~\cite{Johansson2012QuTiP,Johansson2013QuTiP}
    & \begin{tightitemize}
        \item Simulation of dynamics in both open and closed systems
      \end{tightitemize}
    \\ \cline{2-3}
    
    & SuperGrad~\cite{Wang2025SuperGrad}
    & \begin{tightitemize}
        \item Differentiable simulator for superconducting quantum chips
      \end{tightitemize}
    \\ \cline{2-3}
    
    & Scqubits~\cite{Groszkowski2021Scqubits, Chitta2022Computeraided}
    & \begin{tightitemize}
        \item User-friendly analysis for circuits of standard qubit types
      \end{tightitemize}
    \\ \cline{2-3}
    
    & SQcircuit~\cite{Rajabzadeh2023Analysis}
    & \begin{tightitemize}
        \item Superconducting circuit gradient optimization with autodiff
      \end{tightitemize}
    \\ \cline{2-3}
    
    & pyEPR~\cite{Minev2021pyEPR}
    & \begin{tightitemize}
        \item EM to Hamiltonian via energy-participation quantization
      \end{tightitemize}
    \\ \cline{2-3}
    
    & Palace~\cite{Grimberg2023Palace}
    & \begin{tightitemize}
        \item 3D EM solver for superconducting quantum chips
      \end{tightitemize}
    \\ \cline{2-3}
    
    & QuantumPro~\cite{QuantumPro}
    & \begin{tightitemize}
        \item Integrated EM simulation and qubit parameter extraction
      \end{tightitemize}
    \\ \cline{2-3}
    
    & EDA-Q~\cite{Zhao2025EDAQ}
    & \begin{tightitemize}
        \item End-to-end superconducting chip design workflow
      \end{tightitemize}
    \\ \cline{2-3}
    
    & Origin Unit~\cite{OriginUnit}
    & \begin{tightitemize}
        \item Web browser and standalone client layout editors
        \item PDK loading and DRC support
      \end{tightitemize}
    \\ \cline{2-3}
    
    & Tianyi~\cite{Tianyi}
    & \begin{tightitemize}
        \item Parametric component library + auto-routing platform
        \end{tightitemize}
    \\ \hline  
    
    \multirow{15}{*}{\parbox{1.55cm}{\centering \textbf{Logic}\\ \textbf{Level}}}
    & Qiskit~\cite{JavadiAbhari2024Quantum}
    & \begin{tightitemize} 
        \item Full-stack ecosystem
      \end{tightitemize} 
    \\ \cline{2-3}  
    
    & MQT~\cite{Wille2024The}  
    & \begin{tightitemize}
        \item Verified compilation workflows
        \item Design space exploration
        \item Fault-tolerant circuit support
      \end{tightitemize} 
    \\ \cline{2-3}
    
    & Cirq~\cite{Developers2024Cirq}
    & \begin{tightitemize}
        \item Noise-aware circuit manipulation/optimization
        \item Tensorflow-Quantum integration
      \end{tightitemize} 
    \\ \cline{2-3}
    
    & QuTiP-QIP~\cite{Lambert2024QuTiP}
    & \begin{tightitemize}
        \item Quantum circuit simulation extension for QuTiP
        \item Both gate-level and pulse-level simulation
      \end{tightitemize} 
    \\ \cline{2-3}
    
    & Q\# and Azure QDK~\cite{Svore_2018}
    & \begin{tightitemize}
        \item Leverages the LLVM-based QIR~\cite{QIRSpec2021} for interoperability
        \item Resource estimator for FTQC~\cite{damUsingAzureQuantum2024}
      \end{tightitemize} 
    \\ \cline{2-3}
    
    & Tket~\cite{sivarajah2020t}
    & \begin{tightitemize}
        \item Retargetable from multiple frontends to multiple backends
        \item Extensive circuit transformation and optimization passes
      \end{tightitemize} 
    \\ \cline{2-3}

    & CUDA-Q~\cite{CUDA-Q}
    & \begin{tightitemize}
        \item QPU-agnostic hybrid programming model
        \item Seamless integration across GPUs, CPUs, and QPUs
      \end{tightitemize} 
    \\ \cline{2-3}
    
    & Pennylane~\cite{bergholmPennyLaneAutomaticDifferentiation2022}
    & \begin{tightitemize}
        \item Differentiable quantum programming
        \item Integration with machine learning frameworks
      \end{tightitemize}
    \\ \hline  
  \end{tabular}
\end{table}

%%%%%%%%%%%%%%%%%%%%%%%%%%%%%%%%%%%%%%%%%%%%%%%%%%%%%%%%%%%%%%%%%%%%%%%%%%%%%%%%%%%%%%%%%

\subsection{Current status of play.}
With the rapid development of quantum computing hardware, significant progress has also been made in the development of QDA software. To help readers who wish to dive deeper into the QDA, we will briefly walk through several noteworthy QDA software here. Please note that this is not an exhaustive list. Our goal is to provide a broad overview rather than a comprehensive catalog. Further, as this field (and even the listed software themselves) is continuously evolving, the list reflects only our knowledge at the time of this paper's preparation.

Existing physical-level design tools primarily cover the scope from the quantum chip to the Hamiltonian. We start from those developed for chip layout design, among which KQCircuits~\cite{Cucurachi2021KQCircuits} and Qiskit Metal~\cite{Minev2021Qiskit} are representative examples. 
KQCircuits, which employs KLayout as its graphical user interface, specializes in automating the design of lithography-compliant layouts, while Qiskit Metal extends the Qiskit software stack into the domain of chip design, providing a built-in layout editor with interfaces to EM simulation tools. Other notable efforts include: (1) EDA-Q~\cite{Zhao2025EDAQ}, a framework tailored for the workflow from topology definition and device mapping to fabrication process integration; (2) Origin Unit~\cite{OriginUnit} supports automated design and fabrication process specification checks; and (3) Tianyi~\cite{Tianyi} supports layout design with integrated auto-routing algorithms. All these tools provide functionalities ranging from layout editors with parametric component libraries to design rule checks for fabrication manufacturability, thereby simplifying and accelerating the layout design workflow.

To extract quantum chip parameters, EM simulation should be performed based on the layout design. Besides general-purpose EM simulation software, specialized tools for superconducting quantum chip layouts include Palace~\cite{Grimberg2023Palace} and QuantumPro~\cite{QuantumPro}. Palace is a cloud-optimized parallel finite element solver that enables high-performance, large-scale 3D EM simulations of chip components. QuantumPro integrates EM simulation, nonlinear circuit analysis, and automated quantum parameter extraction to streamline the chip design workflow.

Once parameters are extracted via EM simulation, several software tools facilitate the analysis of qubit Hamiltonians in superconducting circuits. Scqubits~\cite{Groszkowski2021Scqubits, Chitta2022Computeraided} and SQcircuit~\cite{Rajabzadeh2023Analysis, Rajabzadeh2024A} are libraries for superconducting circuit analysis, both offering circuit quantization, energy spectrum calculation, coherence time estimation, and interfaces with QuTiP for time-dependent quantum dynamics simulations. The former emphasizes user-friendly simulations of standard qubits, while the latter focuses on gradient optimization for arbitrary superconducting circuits via automatic differentiation. Distinct from them, pyEPR~\cite{Minev2021pyEPR} directly bridges classical EM simulations with quantum Hamiltonian analysis without requiring the intermediate construction of the superconducting circuit. It achieves this via the energy-participation quantization method~\cite{Minev2021Energyparticipation}, which automates the conversion of eigenmode solutions into a parameterized Hamiltonian.

Once the qubit Hamiltonian is derived from the chip layout using the aforementioned tools, QuTiP~\cite{Johansson2012QuTiP,Johansson2013QuTiP} can be employed to simulate the dynamics of both open and closed quantum systems, making it particularly well-suited for quantum control studies. Another tool for qubit time-evolution research is SuperGrad~\cite{Wang2025SuperGrad}: its high-performance differentiable simulator supports multi-qubit time-evolution simulations, gradient optimization of both qubit and control parameters, and fitting of experimental spectroscopy data. 

Other critical physical-level QDA tasks, such as decoherence analysis, package and cryogenic setup design, technology computer-aided design, as well as verification and testing, remain in their infancy, with limited dedicated software available. This gap presents a significant opportunity for future tool development.

For logic-level design, as we noted earlier, IBM's Qiskit~\cite{JavadiAbhari2024Quantum} has established itself as the de facto standard quantum SDK, offering a comprehensive suite of tools and functionalities tailored to quantum algorithm design, optimization, and execution, with error mitigation support. Coupled with its direct integration with IBM's quantum computing platform, it has garnered significant attention from researchers worldwide. Through these collaborative efforts, Qiskit has delivered considerable performance improvements over its earlier versions. Similarly, Google's Cirq~\cite{Developers2024Cirq} and Quantinuum's TKet~\cite{sivarajah2020t} have emerged as prominent alternatives, each with distinct strengths: Cirq excels in low-level control and optimization for near-term quantum hardware, while TKet focuses on advanced circuit compilation and compatibility across diverse quantum architectures---collectively enriching the ecosystem of quantum software. In contrast, Microsoft's Azure Quantum~\cite{Svore_2018} focuses primarily on higher-level capabilities, including resource estimation, quantum programming languages, and intermediate representations, with the goal of enabling more accessible quantum systems via the Azure cloud platform. Recently, NVIDIA has embarked on its quantum computing journey, focusing primarily on enhancing quantum capabilities within its GPU ecosystem. This includes the highly efficient cuQuantum simulator~\cite{cuquantum_sdk_2024} and the quantum expansion of its widely adopted CUDA platform~\cite{CUDA-Q}. 

Beyond the offerings from tech giants, numerous other design-focused projects are available, and we highlight a few here. QuTiP-QIP~\cite{Lambert2024QuTiP}, built on the long-standing QuTiP project, provides quantum information processing capabilities rooted in Hamiltonian simulation. The Munich Quantum Toolkit (MQT)~\cite{Wille2024The} delivers solutions for design tasks through a multi-layer software stack, encompassing classical simulation, compilation, and verification of quantum circuits, along with support for hardware design. PennyLane by Xanadu~\cite{bergholmPennyLaneAutomaticDifferentiation2022} aims to build an open-source framework for integrating quantum computing and machine learning.

These tools, as summarized in~\Cref{tab:qda_tools}, comprise a multi-layer ecosystem that addresses design challenges ranging from physical level operations to logic level implementations.

\section{Quantum Chip Design}\label{sec:chip_design}

\begin{figure}
    \centering
    \includegraphics[width=1.0\textwidth]{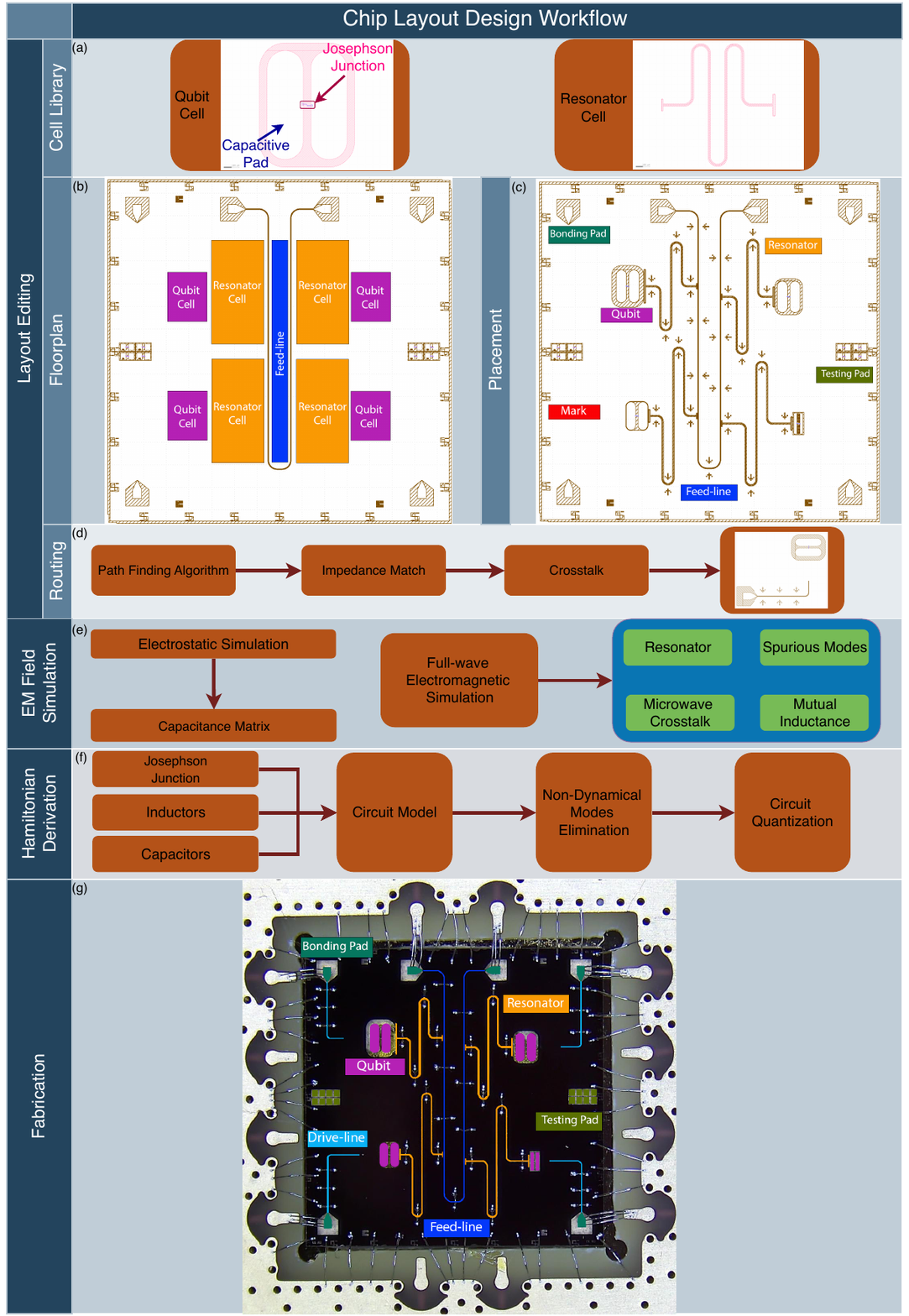}
    \caption{Pictorial illustration of the superconducting quantum chip design workflow, including the key steps of the layout design ((a)-(d)), the EM field simulation (e), and the Hamiltonian derivation (f). An example of a real chip is shown in (g).}
    \label{fig:chip_design}
\end{figure}

The parameters of superconducting qubits are determined by the geometry of superconducting patterned shapes. It is normally necessary to meet the design parameter targets and the design rules in the physical design of a quantum chip, where the capacitive pads, the Josephson junctions, and the inductors form the qubits, the wires serve for driving, and the resonators are used for readout, etc. The placement of the components and the routing between them must be optimized through electromagnetic simulations, Hamiltonian analysis, and design rule checks.
The workflow to design a superconducting quantum chip is depicted in~\Cref{fig:chip_design}(a)-(f). 
First, a detailed layout of the chip, which defines the geometry and placement of capacitive pads, inductors, Josephson junctions, and routed wires, is drafted. 
Next, electromagnetic simulations are performed on this layout to extract the parameters of the circuit model. Then, this circuit model should be further reduced by removing the degrees of freedom that are irrelevant to the dynamics of qubits or resonators. This reduced model is then fed into a circuit quantization procedure to derive the qubit Hamiltonian. 
This entire loop is iterated until the qubit parameter extracted from the chip pattern fall within the target specifications. Finally, the fabrication process is carried out based on the finalized layout design to produce the chip, as exemplified by the one shown in~\Cref{fig:chip_design}(g).

\subsection{Layout design}
The physical chip layout designs of classical digital integrated circuits (ICs) and superconducting quantum chips share some fundamental similarities while exhibiting crucial differences. In classical digital IC design, the process begins with a library of standard cells and a netlist specifying connectivity, followed by the floorplan, cell placement, virtual routing, static timing analysis, clock tree synthesis, and final routing. The primary optimization goals focus on performance, power, and area (PPA), with complexity lying in combinatorial optimization for placement and routing. In contrast, superconducting quantum chip design introduces unique challenges beyond these classical considerations. While following a similar hierarchical approach, quantum chip layouts must account for quantum-specific phenomena, such as qubit energy levels with their transition probabilities, decoherence, and quantum couplings. Unlike classical ICs operating at room temperature, superconducting quantum chips function at cryogenic temperatures, requiring special materials. The performance optimization of the superconducting quantum chip mainly prioritizes coherence time, coupling strength, and crosstalk suppression. Additionally, quantum chip design requires electromagnetic simulation to extract quantum Hamiltonian parameters, an aspect that is unnecessary in the classical design workflow.

Similar to the physical design of classical digital ICs, the quantum chip design follows a structured procedure with several key stages:
\begin{itemize}
    \item \textbf{Cell library creation.}  A library of characterized and optimized cells serves as the building blocks for quantum chips. This includes single-qubit cells, multiple-coupled-qubit cells, resonator cells, and driving coupler cells, etc.~\Cref{fig:chip_design}(a) shows examples of a qubit cell and a resonator cell. Different types of cells are optimized based on different specifications. For instance, the specifications of qubit cells include aspects such as decoherence, qubit parameters, multiple-qubit coupling coefficients, and footprint, while the specifications of resonator cells include aspects such as resonance frequencies, decoherence, and the footprint. Since the couplings between the driving couplers and qubits are usually through local capacitive or inductive couplings, the driving coupler can also be included as part of the qubit cell.
    \item \textbf{Cell characterization.} Most cells in the qubit layout can be designed as the parameterized cells with look-up tables to allow for adjustable parameters. By selecting appropriate geometric parameters for the initial cells, the design parameters of the isolated system can be chosen to be close to the target, leaving the coupling effects to the placement optimization.   
    \begin{itemize}
        \item \textbf{Qubit cell.} Generally speaking, the qubit parameters involve capacitive energy, Josephson junction energy, inductive energy, coupling strength to drivers and resonators, and inter-qubit coupling energy for multiple-qubit cells, etc. Most of these parameters are geometry-dependent. The geometry, in turn, affects the decoherence of the qubit through various types of loss. Therefore, multidimensional look-up tables from pre-characterization are useful for making a good selection of the initial guess.  
        \item \textbf{Resonator cell.} The resonator parameters include the characteristic impedance, resonance frequency, and the coupling strength to the feed-line. These parameters are related to the geometric parameters of a coplanar waveguide (CPW) transmission line in a simple manner that is analytically solvable under reasonable approximations. And these analytical formulae can be used to provide a good initial design for resonators. 
    \end{itemize}
    \item \textbf{Floorplan.} The floorplan highly depends on the packaging scheme, such as the single-layer design with wire bonding, flip-chip with wire bonding, or advanced packaging using through-silicon vias (TSVs).~\Cref{fig:chip_design}(b) shows an example of the floorplan of a single-layer chip. At the floorplan stage, the pin map for the external wiring has to be determined to facilitate routing and leave enough area for cell placement. After the pin assignment, the initial placement of cells can be determined. This stage is usually planned manually for small systems using wire bonding packaging, and the quality of the floorplan strongly affects the feasibility and quality of the final layout. 
    \item \textbf{Feed-line design.} The feed-line is normally a CPW transmission line that is weakly coupled to the resonators for qubit state readout. The through or the reflected signal with a high signal-to-noise ratio is crucial for efficient readout. Therefore, low-curvature routing is usually implemented to avoid unwanted reflections of microwave signals. The feed-line is normally routed after the floorplan and can then be adjusted after the resonator placement. 
    \item \textbf{Placement.} By choosing appropriate cells for qubits, resonators, and couplers, the initial placement from the floorplan is used.~\Cref{fig:chip_design}(c) shows an example of the placement stage. This stage can be regarded as the global placement in the physical design of classical ICs, at which the coupling effects have not been fully considered.
    In contrast to classical IC design, the placement of the superconducting quantum chip prioritizes crosstalk minimization and decoherence suppression~\cite{Zhang2025_QPlacer, Zhang2025_QGDP}. 
    To simulate and optimize the Hamiltonian of the entire circuit, electromagnetic and circuit simulations can be employed to accommodate for the global coupling effects. The resonator parameter can be extracted from eigenmode simulations or scattering parameter extractions to find the resonant frequency and the readout coupling strength (or the coupling quality factor). As for the qubit system, the black-box quantization, energy-participation quantization, or lump-element circuit quantization method can be applied to obtain the full Hamiltonian. To automatically realize the final detailed placement that meets the target Hamiltonian, an optimizer with electromagnetic and quantization engines is used for placement optimization. 
    \item \textbf{Routing.} After the optimized detailed placement, the locations of the qubits, resonators, and coupling drivers are finalized, leaving the interconnects between the drivers and the pins to be routed.~\Cref{fig:chip_design}(d) shows the key steps in the routing stage.
    Some auto-routing algorithms, such as the A* algorithm~\cite{Hart1968A}, can be used to complete the routing. The main factors that have to be considered carefully for routing are the impedance match and crosstalk. The CPWs used to route the interconnect have to match the impedance of package nets, which is usually 50$\Omega$. When air-bridges are applied, both the signal and ground air-bridges should be routed across another net. To limit crosstalk, a minimal lateral distance constraint can be applied for routing.   
    \item \textbf{Package integration.} When integrating the chip with the package, additional components, such as bumps for flip-chip and either air-bridges or wire bonds, need to be applied to the chip after the final routing. These added connections between isolated ground pads can be effective in eliminating the chip mode near the qubit/resonator frequencies.  
    \item \textbf{Design for test (DFT).} Design patterns for room temperature tests to screen the chip with qubits closest to the design target. For example, a pattern with the Josephson junction truncated from the full structure of a qubit, of which the room temperature resistivity can show if the Josephson junction will work as expected. Such patterns are especially useful for pre-screening defective chips before the costly cryogenic measurements. 
    \item \textbf{Design rule checking (DRC).} The DRC ensures that the design can be fabricated reliably and will function correctly once manufactured. These are geometric constraints imposed on the design patterns for acceptable yield and reliability in fabrication, which include specifications for the minimum widths of components and the spacings between them.
\end{itemize}

For the efficient design of large-scale quantum chips, advanced layout design software tools are essential. These tools must address quantum-specific requirements beyond traditional EDA capabilities.
By following the above procedure for designing quantum chips, the layout design tool is preferable to be script-based to improve design efficiency and reduce human error~\cite{Cucurachi2021KQCircuits, Minev2021Qiskit}. To simplify the design process, the tool should include a component library consisting of predefined parameterized qubits, resonators, and driving couplers.
As quantum systems scale up, automated placement optimization and auto-routing features become critical, as the number of control lines grows linearly with the qubit count. Besides incorporating simulation tools to accurately predict quantum Hamiltonian parameters and coherence properties, the software should facilitate design space exploration to optimize quantum-specific metrics and integrate with fabrication process design kits to ensure manufacturability. Furthermore, as quantum processors scale toward error correction requirements, tools must support modular design approaches with standardized interfaces between quantum chips as building blocks.

\subsection{Layout simulation}
\label{sec:layout_simulation}
The superconducting qubit is the quantized electromagnetic (EM) field determined by the pattern of the chip layout. These field patterns arise from the intricate interplay between superconducting materials and dielectric components in the chip's architecture. To predict the quantities of chip parameters prior to costly fabrication processes, it is essential to simulate the EM field by solving Maxwell's equations with the geometries of superconducting objects as the boundary. During the chip design optimization phase, these geometries are iteratively adjusted, so the EM simulation should be performed repeatedly. In this section, we will introduce the method to obtain field distribution via the EM simulation, and subsequently explain how to extract the electromagnetic properties concerned in designing the superconducting chip layout from the obtained EM field distribution, as summarized in~\Cref{fig:chip_design}(e).

\subsubsection{Electrostatic simulation}
The typical working frequency of a superconducting qubit is on the order of gigahertz, so the wavelength of the microwave traveling on the chip is on the order of centimeters.
This wavelength dimension is much larger than the length scale of the qubit footprint, which is about a few hundred micrometers. Therefore, we can construct the lumped-element model of the qubit design pattern using the capacitance matrix extracted from electrostatic simulation. The capacitance matrix is not only used for the derivation of the qubit Hamiltonian, but also for the analysis of unwanted stray couplings that cause charge crosstalk.  

\paragraph{Electrostatic problem.}The electrostatic problem~\cite{BookClassicalelectrodynamics1999, BookIntroductiontoelectrodynamics2024} can be described by Gauss's law 
\begin{equation}
     \nabla \cdot \varepsilon \vec{E}\left(\vec{r}\right) = \rho\left(\vec{r}\right),
    \label{eq:Gauss_law}
\end{equation}
where $\vec{E}\left(\vec{r}\right)$ is the electric field, $\rho$ is the charge density, and $\varepsilon$ is the permittivity. For simplicity, we assume that $\varepsilon$ is homogeneous and isotropic in the following derivation. In classical electrostatics, the electric field $\vec{E}$ is a vector expressed as the gradient of the electrostatic potential $\varphi$
\begin{equation}
    \vec{E}\left(\vec{r}\right)= - \nabla \varphi\left(\vec{r}\right).
    \label{eq:grad_E}
\end{equation}
Substituting~\Cref{eq:grad_E} into~\Cref{eq:Gauss_law}, we obtain
\begin{equation}
    \nabla^2 \varphi\left(\vec{r}\right) = - \rho \left(\vec{r}\right) / \varepsilon,
    \label{eq:Poisson}
\end{equation}
which is Poisson's equation for electrostatics. We consider a chip of $K$ superconducting components, with their surfaces denoted as $S_i \left( i = 1, 2, \ldots, K \right)$. 
The charge is only distributed on the superconducting components, so there is no charge in the solution domain $\tilde{\Omega}$, which is the volume bounded by the surfaces
\begin{equation}
    \tilde{\Gamma} = S_{1}\cup S_{2}\cup\cdots\cup S_{K}. 
\end{equation}
Therefore, Poisson's equation is reduced to Laplace's equation, given by
\begin{equation}
    \nabla^2 \varphi \left(\vec{r}\right) = 0,
    \label{eq:Laplace}
\end{equation}
with the boundary conditions
\begin{equation}
    \varphi \left(\vec{r}\right) = V_i, \; \forall \vec{r} \in S_i  \left( i = 1, 2, \ldots, K \right),
\end{equation}
where $V_i$ denotes the potential of the component $i$. If $\tilde{\Omega}$ is unbounded, we need to add an asymptotic boundary condition, such as $\lim_{\left\Vert \vec{r}\right\Vert \rightarrow\infty}\varphi\left(\vec{r}\right)=0$, to ensure the uniqueness of the solution.

\paragraph{Capacitance matrix.}In the following, we will describe how to compute the entries of the capacitance matrix. Let us denote $\varphi_i\left(\vec{r}\right)$ and $\varphi_j\left(\vec{r}\right)$ as two solutions of~\Cref{eq:Laplace} with the boundary conditions assigned as $V_{k}=\delta_{ik}$ and $V_{k}=\delta_{jk}$, respectively, where $i,j,k=1,2,\ldots,K$. The charge $Q_i$ on the component $i$ induced by the assignment of potentials $V_{k}=\delta_{jk}$ can be computed from the integral form of Gauss's law
\begin{equation}
    Q_{i}=-\varepsilon\int_{S_{i}}\nabla\varphi_{j}\left(\vec{r}\right)\cdot\vec{n}dS,
\end{equation}
where $\vec{n}$ is the outward unit vector normal at the point on $S_i$. According to the definition of the capacitance matrix, the entry $C_{ij}$ is given by $C_{ij} = Q_{i}/V_j = Q_i$. Since $\varphi_{i}\left(\vec{r}\right)=1$ on the surface $S_i$, and $\varphi_{i}\left(\vec{r}\right)=0$ on the other surfaces, we can re-express $C_{ij}$ as the integral over $\tilde{\Gamma}$
\begin{equation}
    C_{ij}=-\varepsilon\int_{S_{i}}\nabla\varphi_{j}\left(\vec{r}\right)\cdot\vec{n}dS
    =\varepsilon\int_{\tilde{\Gamma}}\varphi_{i}\left(\vec{r}\right)\nabla\varphi_{j}\left(\vec{r}\right)\cdot\left(-\vec{n}\right)d\Gamma .
\end{equation}
By applying the divergence theorem, we obtain
\begin{equation}
C_{ij}=\varepsilon\int_{\tilde{\Omega}}\nabla\cdot\left[\varphi_{i}\left(\vec{r}\right)\nabla\varphi_{j}\left(\vec{r}\right)\right]d\Omega
=\varepsilon\int_{\tilde{\Omega}}\nabla\varphi_{i}\left(\vec{r}\right)\cdot\nabla\varphi_{j}\left(\vec{r}\right)
+\varphi_{i}\left(\vec{r}\right)\nabla^{2}\varphi_{j}\left(\vec{r}\right)d\Omega.
\end{equation}
Since $\varphi\left(\vec{r}\right)$ satisfies~\Cref{eq:Laplace} in $\tilde{\Omega}$, we have
\begin{equation}
C_{ij}=\varepsilon\int_{\tilde{\Omega}}\nabla\varphi_{i}\left(\vec{r}\right)\cdot\nabla\varphi_{j}\left(\vec{r}\right)d\Omega.
\label{eq:capacitance_matrix}
\end{equation}
\Cref{eq:capacitance_matrix} tells us that we can compute the capacitance matrix from the potential distributions obtained by solving~\Cref{eq:Laplace} with appropriate potential assignments on the superconducting components.

\paragraph{Finite element method.}We will briefly describe how to apply the finite element method (FEM) to convert the boundary value problem of~\Cref{eq:Laplace} into a set of linear equations~\cite{Jin2010The}.  
In order to perform an efficient numerical simulation, we truncate the infinitely large solution domain $\tilde{\Omega}$ into a finite computational domain $\Omega$. To achieve this purpose, we create a modified boundary $\Gamma$ by adding an artificial boundary $S_0$ to $\tilde{\Gamma}$
\begin{equation}
    \Gamma = \tilde{\Gamma} \cup S_0.
\end{equation}
The widely used boundary conditions for $S_0$ are absorbing boundary conditions and perfectly matched layer boundary conditions~\cite{Jin2010The}. Typically, $\Gamma$ comprises the Dirichlet boundary and the Neumann boundary, denoted as $\Gamma_D$ and  $\Gamma_N$, respectively. 
To seek the solution for $\varphi \left(\vec{r}\right)$, we first expand it using a set of basis functions
\begin{equation}
    \varphi \left(\vec{r}\right) = \sum_{j=1}^{N} a_j h_j \left(\vec{r}\right)
    +\sum_{j=1}^{N_{D}}a_{j}^{D}h_{j}^{D}\left(\vec{r}\right),
    \label{eq:basis_function}
\end{equation}
where $h_j\left( \vec{r} \right)$ denote the basis functions defined in $\Omega$ with $a_j$ as the corresponding expansion coefficients; $h_{j}^{D}$ and $a_{j}^{D}$ denote the corresponding quantities associated with $\Gamma_D$. We assume that $h_j\left( \vec{r} \right)$ vanishes on $\Gamma_D$. 
Using Galerkin's method, we multiply the basis function $h_i\left( \vec{r} \right)$ as the testing function to~\Cref{eq:Laplace}, and integrate over $\Omega$, which yields
\begin{equation}
    \int_{\Omega} d\Omega h_{i}\left(\vec{r}\right)\nabla^{2}\varphi\left(\vec{r}\right) = 0.
\end{equation}
Applying integration by parts, we obtain
\begin{equation}
     \int_{\Omega}\nabla h_{i}\left(\vec{r}\right)\cdot\nabla\varphi\left(\vec{r}\right)d\Omega = \int_{\Gamma}h_{i}\left(\vec{r}\right)\nabla\varphi\left(\vec{r}\right)\cdot\vec{n}d\Gamma.
     \label{eq:integration_by_parts}
\end{equation}
Since $h_{i}\left(\vec{r}\right) = 0$ on $\Gamma_D$, we have
\begin{eqnarray}
    \int_{\Gamma}h_{i}\left(\vec{r}\right)\nabla\varphi\left(\vec{r}\right)\cdot\vec{n}d\Gamma
&=&\int_{\Gamma_{D}}h_{i}\left(\vec{r}\right)\nabla\varphi\left(\vec{r}\right)\cdot\vec{n}d\Gamma+\int_{\Gamma_{N}}h_{i}\left(\vec{r}\right)\nabla\varphi\left(\vec{r}\right)\cdot\vec{n}d\Gamma \nonumber \\
&=&\int_{\Gamma_{N}}h_{i}\left(\vec{r}\right)\nabla\varphi\left(\vec{r}\right)\cdot\vec{n}d\Gamma,
\end{eqnarray}
Substituting~\Cref{eq:basis_function} into~\Cref{eq:integration_by_parts}, we obtain
\begin{eqnarray}
    & &\sum_{j=1}^{N}a_{j}\int_{\Omega}\nabla h_{i}\left(\vec{r}\right)\cdot\nabla h_{j}\left(\vec{r}\right)d\Omega \nonumber \\
    &=&\int_{\Gamma_{N}}h_{i}\left(\vec{r}\right)\nabla\varphi\left(\vec{r}\right)\cdot\vec{n}d\Gamma-\sum_{j=1}^{N}a_{j}^{D}\int_{\Omega}\nabla h_{i}\left(\vec{r}\right)\cdot\nabla h_{j}^{D}\left(\vec{r}\right)d\Omega .
    \label{eq:fem}
\end{eqnarray}
\Cref{eq:fem} can be written more compactly as a representation of a set of linear equations
\begin{equation}
    \sum_{j=1}^{N}W_{ij}a_{j}=b_{i},
    \label{eq:fem_compact}
\end{equation}
where
\begin{equation}
    W_{ij}=\int_{\Omega}\nabla h_{i}\left(\vec{r}\right)\cdot\nabla h_{j}\left(\vec{r}\right)d\Omega,
\end{equation}
and
\begin{equation}
b_{i}=\int_{\Gamma_{N}}h_{i}\left(\vec{r}\right)\nabla\varphi\left(\vec{r}\right)\cdot\vec{n}d\Gamma-\sum_{j=1}^{N}a_{j}^{D}\int_{\Omega}\nabla h_{i}\left(\vec{r}\right)\cdot\nabla h_{j}^{D}\left(\vec{r}\right)d\Omega.
\end{equation}
By solving $a_j$ from~\Cref{eq:fem_compact}, we can obtain the potential distribution $\varphi\left(\vec{r}\right)$, and the capacitance matrix can thereby be computed by~\Cref{eq:capacitance_matrix}.

\subsubsection{Full-wave electromagnetic simulation}

In the design and modeling of superconducting quantum devices, two types of full-wave electromagnetic simulations are particularly useful: the scattering (S) matrix extraction by frequency sweep, and the eigenmode simulation. 

The simulation of the S matrix is one of the widely-used workflows for linear electromagnetic networks in traditional EDA. In a multiple-port system, the S parameter is defined as the ratio of the transmitted voltages at the output port and the input port, while assuming all other ports have no reflections (perfectly matched)~\cite{Pozar2012Microwave}:
\begin{equation}
S_{ij} = \frac{V_i^-}{V_j^+} \bigg|_{
V_k^+ = 0 \text{ for } k \neq j
}
\end{equation}
where $V_i^-$ denotes the voltage propagating out from the network through port $i$, and $V_j^+$ is the voltage propagating into port $j$. The condition $V_k^+ = 0$ indicates that there is no reflection at the port $k$, implying in microwave theory that the load impedance equals the characteristic impedance of the transmission line $Z_0$. 
The S matrix provides a complete description of the network when the system is treated as a linear black-box with input and output ports. By converting it to other types of network matrices, such as impedance (Z) and admittance (Y) matrices, some lumped model parameters can be easily found. 

To obtain all the elements in the S matrix of a network, a full-wave electromagnetic simulation should be performed by assigning a driving field at port $j$. To meet the condition of $V_k^+ = 0$, all other ports are terminated with the matched impedance. After solving Maxwell's equations under the driver $j$, all the transmitted fields can be calculated at all ports to obtain $V_i^-$, thus a column of the S matrix can be found. By repeating the above process using different ports as the drivers, all the elements in the S matrix can be obtained.  

Moreover, since the S matrix encapsulates the electromagnetic response of the black-box system, when a model is assumed for the black-box, all the key parameters in the model can be fitted or extracted from the S parameters. 

An alternative method is the eigenmode simulation. In the configuration of the eigenmode simulation, the chip with the substrate is placed inside an air-box to truncate the finite element simulation domain. 
No drivers exist in the simulation, and Maxwell's equations are converted to a generalized eigenvalue problem, in which the core matrices are the stiffness and mass matrices. The eigenvalues correspond to the resonance frequencies of the entire system, and from each eigenvector, one can find the field distribution at the corresponding resonance frequency. 
The benefit of using the eigenmode simulation is that no frequency sweeps are required to find the resonator modes and the spurious modes.

\paragraph{Resonator extractions.}
When using the S matrix to extract the resonator characteristics, including the resonance frequency $\omega_r$ and the coupling quality factor $Q_c$, the network is modeled as a transmission line coupled to a resonator, with the S parameter given by
\begin{equation}
\label{eq:coupled_resonator_s21}
    S_{21}(\omega) = e^{i \omega \tau} \left [ 1 - \frac{e^{i\phi} Q_l / Q_c}{1+2iQ_l(\omega / \omega_r -1)} \right],
\end{equation}
where $Q_l$, the loaded quality factor, is composed of the internal quality factor $Q_i$ of the resonator and the coupling quality factor $Q_c$, as given by
\begin{equation}
Q_l^{-1} = Q_i^{-1} + Q_c^{-1}.
\end{equation}
By fitting the simulated $S_{21}$ near the resonance frequency $\omega_r$ with the model in~\Cref{eq:coupled_resonator_s21}, one can find the $\omega_r$ and $Q_c$ for each resonator.

When eigenmode simulation is adopted, the resonance frequencies of each resonator, which is weakly coupled to the feed-line, can be found from the solved eigenvalues. 
In a lossless system, only real eigenvalues can be obtained, and it is not possible to directly extract the coupling quality factor $Q_c$. However, one can find $Q_c$ by comparing the frequency shift with the uncoupled resonator, or by loading the two ends of the feed-line with resistors to mimic the signal absorption through the feed-line. 
Then the system becomes lossy, and complex eigenvalues can be used to approximate the coupling quality factor using
\begin{equation}
    Q_c = \frac{\Re[\omega]}{2\Im[\omega]}.
\end{equation}
where the loss of the resonator mode is assumed to be purely due to the electromagnetic coupling to the feed-line, and the energy is perfectly absorbed by the resistors. 

\paragraph{Spurious modes. }
In both methods for finding the oscillation modes with respect to resonator frequencies, other unwanted spurious modes may also exist.
Normally, the resonator modes are the standing-wave modes across the lines of the resonators. Therefore, one should find, from the field distribution in the eigenmode simulation, that the electromagnetic field reaches its peak at the center of the resonator lines. Other unwanted modes, which include on-die electromagnetic modes and the box modes of the package box, will have the electromagnetic field spanning over the whole chip or the whole box. 
The on-die electromagnetic modes can sometimes be simply regarded as LC resonating modes, where the inductance is induced by the big loops that support current flow and the capacitance is due to the conductors in proximity. In such modes, the field distribution will demonstrate itself as the large charge distribution near the adjacently placed conductors with opposite signs. 
These modes can be efficiently eliminated by air-bridging the adjacent conductors, particularly on the ground plane. 
The box mode, due to the symmetry of the box, will show symmetric field patterns on the surfaces of the perfect conductor box. 
However, these are more difficult to suppress because the mode frequencies are related to the size of the packaging box. Still, one can engineer the box by decreasing the overall permittivity inside it, such as by drilling a hole in the package beneath the chip wafer. From the above discussions about the unwanted modes, one can distinguish the resonator modes from the unwanted modes by analyzing the field distributions. 
However, on the other hand, these unwanted modes cannot be easily distinguished from the S parameter because of the black-box nature of the network. 

\paragraph{Other parameters.}
Some other parameters that are important for qubit control modeling and characterization can be directly obtained from the S matrix. For example, in the control line simulation, where the ports are assigned to the pins of the control lines, the off-diagonal terms in the S matrix, $S_{ij}$, are simply the microwave crosstalk. The suppression of crosstalk from electromagnetic simulations will be helpful for control calibrations. 

In the flux line simulations, the mutual inductance $M$ between the flux line and the qubit inductance loop is to be engineered. This can be extracted from the S matrix by assigning ports at the flux line pin and the qubit inductance loop. At the quasi-static limit, the mutual inductance is the dominant coupling effect. By converting the S matrix to the Z matrix, and 
\begin{equation}
    Z_{21} = i \omega M,
\end{equation}
the mutual inductance is the slope of the Z parameter as a function of radial frequency.

\subsection{Hamiltonian derivation}
\label{sec:hamiltonian_derivation}

To construct the Hamiltonian of a superconducting quantum chip, several methods have been developed to extract the parameters of the Hamiltonian using a combination of classical and quantum methods, such as the circuit quantization from the lumped element circuit diagram~\cite{Burkard2004Multilevel, Burkard2005Circuit,Vool2017Introduction}, black-box quantization from impedance simulation~\cite{Nigg2012BlackBox}, energy participation quantization from eigen-mode simulation~\cite{Minev2021Energyparticipation, Yilmaz2024Energy}, etc. 

In this section, we will mainly introduce the method to derive the chip Hamiltonian from the lumped element circuit diagram, which is summarized in~\Cref{fig:chip_design}(f). 
This method allows us to partition the chip layout and extract a subset of circuit parameters from the simulation of the corresponding layout section independently. The layout sections are usually defined to overlap with neighboring sections to ensure the accuracy of circuit parameter extraction. As the simulation cost of each section is almost independent of the chip size, the time cost to construct the circuit diagram of the whole chip roughly linearly depends on the number of qubits. 
In the chip design stage, the chip-wise oscillation of the electromagnetic field should have been eliminated for the sake of decoherence and crosstalk suppression. Therefore, the circuit parameter extraction typically only requires electrostatic simulation, instead of the substantially more expensive full-wave electromagnetic simulation required in the black-box quantization and energy participation quantization methods. 
Therefore, the circuit quantization from the lumped element circuit diagram is a scalable and computationally efficient method for deriving the chip Hamiltonian. 

A superconducting quantum circuit typically includes capacitors, inductors, and Josephson junctions as its components, if we ignore the losses to the external environment, which will be described in~\Cref{sec:decoherence}. 
A Josephson junction is characterized by its Josephson energies, which can be estimated based on the relation between the junction area and the critical current fitted from the measurement data, instead of numerical simulation. The inductor of a qubit is commonly composed of an array of identical Josephson junctions, so the corresponding inductance can also be extracted from the estimated Josephson energy of every array junction. 
On the other hand, the capacitances among superconducting objects of qubits, as well as the effective capacitance and inductance of the readout resonator, should be extracted from the electromagnetic simulation, as described in~\Cref{sec:layout_simulation}. Then, the superconducting circuit diagram of a chip can be constructed with the values of all circuit components determined. We can thereby derive the corresponding Hamiltonian by using techniques from network theory~\cite{Yurke1984Quantum, Burkard2004Multilevel, Burkard2005Circuit}.

A superconducting circuit can be treated as a multi-graph, which consists of a set of branches representing circuit components, as well as a set of nodes representing the points where these components are interconnected. Each of these nodes corresponds to one superconducting object on the chip. 
Subsequently, a spanning tree is formed by connecting every node to each of the other nodes through only one path. 
This arrangement ensures that the fluxes across the branches of the spanning tree constitute a set of independent generalized coordinates, which are employed to describe the circuit dynamics.
The construction of the spanning tree is not unique, but the different choices of the spanning tree are equivalent, as they can be converted from one to another by a coordinate transformation.
By applying Kirchhoff's law, the equations of motion for these fluxes in the spanning tree can be obtained. 
Assume the number of nodes is $K$, thus the number of branches in the spanning tree is $K-1$.
The Lagrangian, along with its Euler-Lagrange equation fulfilling this equation of motion, can be constructed and is given by
\begin{equation}
    \tilde{\mathcal{L}} = \frac{1}{2} \dot{\vec{\Theta}}^T \tilde{\mathcal{C}} \dot{\vec{\Theta}}
    - \frac{1}{2} \vec \Theta^T \tilde{\mathcal{M}} \vec \Theta
    + \sum_{i=1}^{K_J} E_{J, i} \cos\left(\Theta_i / \varphi_0\right),
    \label{eq:circuit_lagrangian}
\end{equation}
where $\vec{\Theta}=\left( \Theta_1, \Theta_2, \ldots, \Theta_{K-1} \right)^T$ is the vector of flux variables corresponding to the branches of the spanning tree with the first $K_J$ entries associated with junction branches, $\tilde{\mathcal{C}}, \tilde{\mathbf{M}}$ are the coefficient matrices of $\dot{\vec{\Theta}}$ and $\vec \Theta$, respectively, $K_J$ is the number of Josephson junctions with $E_{J, i}$ being the Josephson energy of junction $i$, and $\varphi_0 = \frac{\hbar}{2e}$ is the reduced flux quantum. 

There may exist non-dynamical modes in~\Cref{eq:circuit_lagrangian}, which should be eliminated, as they are irrelevant to the dynamics of qubits or resonators. Non-dynamical modes include two types: free modes and frozen modes~\cite{Ding2021Freemode,Chitta2022Computeraided}. A free mode corresponds to a variable $\Theta_c$ that satisfies $\partial \tilde{\mathcal{L}} / \partial \Theta_c = 0$, which means the terms with $\Theta_c$ vanish in $\tilde{\mathcal{L}}$, indicating a sub-circuit coupled with the rest of the circuit via only capacitances. A frozen mode corresponds to a variable $\Theta_f$ that satisfies $\partial \tilde{\mathcal{L}} / \partial \dot{\Theta}_f = 0$, which means the terms with $\dot{\Theta}_f$ vanish in $\tilde{\mathcal{L}}$, indicating a sub-circuit coupled with the rest of the circuit via only inductances. Free or frozen modes cannot be simply eliminated directly, as they may be capacitively or inductively coupled to other modes, respectively. We can construct an orthogonal transformation $W$ on $\vec \Theta$ to obtain an explicit set of free or frozen modes that are completely decoupled from all other modes, allowing us to remove them without affecting the dynamics of qubits and resonators. 
Assuming there are $K_d$ modes left after removing non-dynamical modes, the corresponding reduced Lagrangian describing these $K_d$ modes is given by
\begin{equation}
    \mathcal{L} = \frac{1}{2} \dot{\vec{\Phi}}^T \mathcal{C} \dot{\vec{\Phi}}
    - \frac{1}{2} \vec \Phi^T \mathcal{M} \vec \Phi
    + \sum_{i=1}^{K_J} E_{J, i} \cos\left(\Phi_i / \varphi_0\right),
    \label{eq:circuit_lagrangian_reduced}
\end{equation}
where $\vec \Phi = \left( \Phi_1, \Phi_2, \ldots, \Phi_{K_d} \right)^T$ is a vector formed by removing non-dynamical flux variables from $W \vec \Theta$, and $\mathcal{C}, \mathcal{M}$ are the coefficient matrices formed by removing the corresponding rows and columns from $W \tilde{\mathcal{C}} W^T, W \tilde{\mathcal{M}} W^T$, respectively. It is worth noting that this transformation keeps the flux variables associated with the junction branches unchanged, i.e., $\Theta_i = \Phi_i, \forall i \in \left[1, K_J \right]$.

From the Lagrangian, we can now define the conjugate charge variables to the flux variables by 
\begin{equation}
    Q_i = \frac{\partial \mathcal{L}}{\partial \dot{\Phi}_i} = \sum_{j=1}^{K_d} \mathcal{C}_{ij} \dot{\Phi}_j,
\end{equation}
which can be written in vector form as $\vec Q = \mathcal{C} \dot{\vec{\Phi}}$. After eliminating non-dynamical modes, $\mathcal{C}$ is invertible, so we have $\dot{\vec{\Phi}} = \mathcal{C}^{-1} \vec{Q}$.
By applying the Legendre transformation, the classical Hamiltonian is given by
\begin{equation}
    \mathcal{H} = \dot{\vec{\Phi}}^T \vec{Q} - \mathcal{L}. 
\end{equation}
The quantum Hamiltonian $\hat{\mathcal{H}}$ is obtained through canonical quantization, which can be regarded as replacing all the variables with quantum operators
\begin{equation}
    \begin{aligned}
    \vec{\Phi} & \rightarrow \vec{\hat{\Phi}}, \\
    \vec{Q} & \rightarrow \vec{\hat{Q}}.
    \end{aligned}
\end{equation}
These flux and charge operators obey the canonical commutation relation
\begin{equation}
    \left[ \hat{\Phi}_n, \hat{Q}_m \right] = i \hbar \delta_{nm},
\end{equation}
where $\delta_{nm}$ is the Kronecker delta. Finally, we obtain the quantum Hamiltonian, taking the form
\begin{equation}
    \hat{\mathcal{H}} = \frac{1}{2} \vec{\hat{Q}}^T \mathcal{C}^{-1} \vec{\hat{Q}}
    + \frac{1}{2} \vec {\hat{\Phi}}^T \mathcal{M} \vec{\hat{\Phi}} 
    - \sum_{i=1}^{K_J} E_{J, i} \cos\left(\Phi_i / \varphi_0\right).
    \label{eq:circuit_hamiltonian}
\end{equation}
\Cref{eq:circuit_hamiltonian} can be rewritten as a summation of qubits and resonators with the couplings among them, expressed as
\begin{equation}
    \hat{\mathcal{H}} = \sum_{i=1}^{K_J} \hat{h}_{i}^q 
    + \sum_{i=K_J + 1}^{K_d} \hat{h}_{i}^{r} + \sum_{i=1}^{K_d}\sum_{j=i+1}^{K_d} \hat{h}_{ij},
    \label{eq:circuit_hamiltonian_rewrite}
\end{equation}
where $\hat{h}_{i}^q $ is a qubit term given by
\begin{equation}
    \hat{h}_{i}^q = \frac{1}{2} \left[ \mathcal{C}^{-1} \right]_{ii} \hat{Q}_i^2
    + \frac{1}{2} \left[ \mathcal{M}^{-1} \right]_{ii} \hat{\Phi}_i^2
    - E_{J, i} \cos\left(\Phi_i / \varphi_0\right),
\end{equation}
$\hat{h}_{i}^r $ is a resonator term given by
\begin{equation}
    \hat{h}_{i}^r = \frac{1}{2} \left[ \mathcal{C}^{-1} \right]_{ii} \hat{Q}_i^2
    + \frac{1}{2} \left[ \mathcal{M}^{-1} \right]_{ii} \hat{\Phi}_i^2,
\end{equation}
and $\hat{h}_{ij}$ is a coupling term given by
\begin{equation}
    \hat{h}_{ij} = \left[ \mathcal{C}^{-1} \right]_{ij} \hat{Q}_i \hat{Q}_j 
    + \left[ \mathcal{M}^{-1} \right]_{ij} \hat{\Phi}_i \hat{\Phi}_j,
\end{equation}
which can be the interaction between two qubits, two resonators, or one qubit with one resonator.

The procedures described above in this section demonstrate that the chip Hamiltonian can be derived from a general superconducting circuit in a systematic workflow. This chip Hamiltonian is essential for qubit control and decoherence analysis, which will be described in subsequent sections.

\subsection{What's next}
Superconducting qubit properties are influenced by both the EM field distribution over the entire millimeter-scale chip and at nanometer-scale interfaces of superconducting thin films. Simulating the electromagnetic behavior of this multi-scale system, which spans several orders of magnitude, demands extensive computational resources. Therefore, the time-consuming electromagnetic simulation is a critical roadblock in the layout design workflow, severely limiting the ability to explore the vast parameter space required to optimize performance and yield. A large-scale superconducting quantum processor typically consists of a multi-layer planar chip where out-of-plane charge and current are confined to elements such as flip-chip bump bonds, TSVs, and air-bridges. These elements should be located far from the qubit patterns in order to have a negligible impact on qubit properties. To improve simulation efficiency, a dedicated simulator, serving as an alternative to general-purpose EM simulation software, can be developed and specifically optimized for qubit property extraction from the chip with an ideal layered structure. 

A paradigm-shifting approach is to create surrogate models for EM simulation through the use of machine learning. By training neural networks on data generated from analytical solutions and accurate EM simulations, it is possible to develop a surrogate model that allows for the fast prediction of the qubit properties of the layout. The AI-driven model can serve as a rapid preliminary screening, enabling designers to quickly converge on qualitatively promising designs before performing costly iterations with EM simulations for precise validation of parameters.

\section{Hamiltonian Design}\label{sec:control}

\begin{figure}
  \centering
  \includegraphics[width=3.75in]{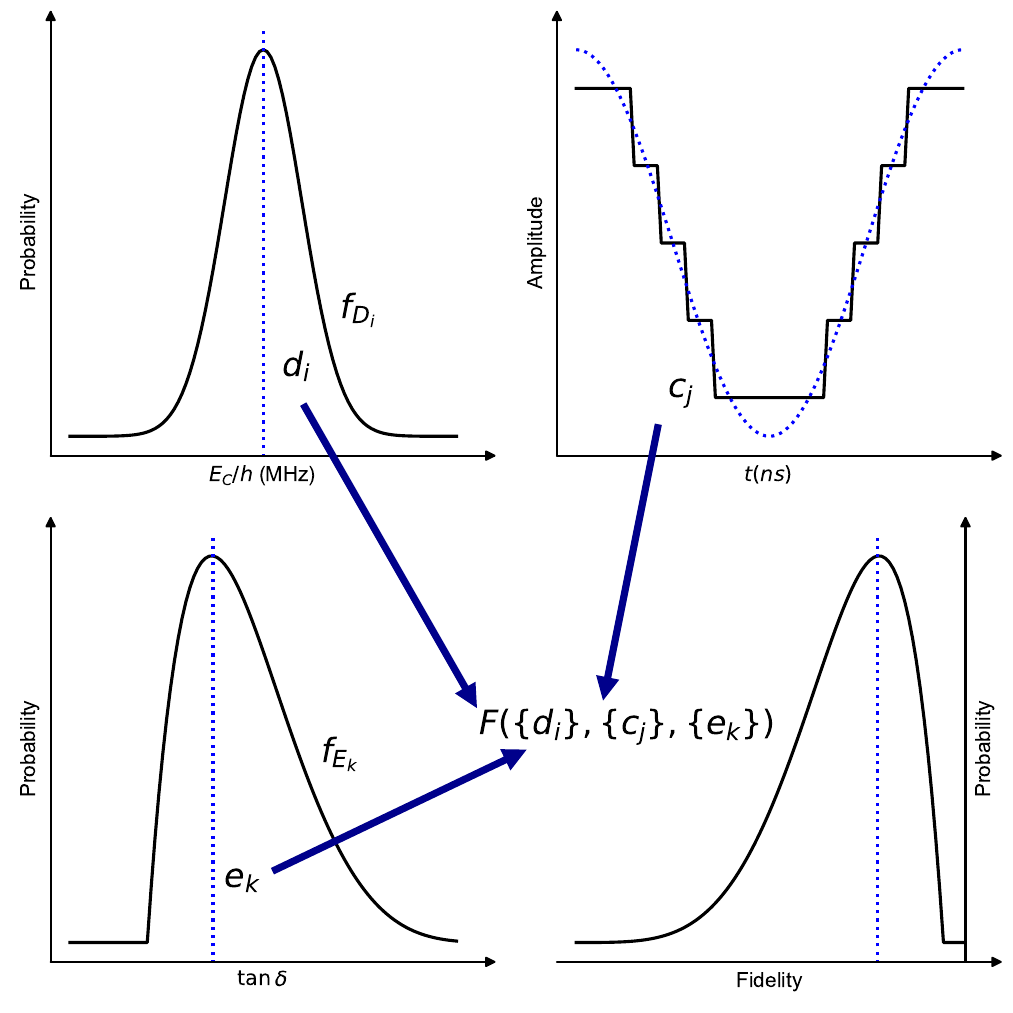}
  \caption{The diagram of the design parameters on the Hamiltonian level in the quantum processor. The fidelity of the operation $F(\setelem{d_i},\setelem{c_j},\setelem{e_k})$ is related to three sets of parameters: device parameters, control parameters and decoherence parameters. The device and decoherence parameters may fluctuate due to instability in the fabrication. The resolution of the control parameters is limited by the instrument.}
  \label{fig:system_design_param}
\end{figure}

\subsection{Quantum model}
The practical design of quantum computing systems must account for the non-ideal nature of their physical realization. Quantum mechanics provides a sufficiently accurate framework for modeling the physical systems relevant to quantum computation. This framework supplies the formalism to map the abstract concepts of quantum information processing onto measurable physical quantities.

Specifically, the Hamiltonian and the Schr\"odinger equation are the primary tools for describing the coherent evolution of the quantum system, detailing both the device and its corresponding controls. Furthermore, decoherence and noise processes are incorporated through extensions to this model, notably through master equations. This section introduces methods for analyzing quantum processors at the level of their governing Hamiltonian.

For a physical quantum computing system operating with a defined instruction set (a finite set of operations), the system's behavior can be represented by the model illustrated in~\Cref{fig:system_design_param}. This model is characterized by three primary sets of parameters, whose values evolve through successive stages of design, fabrication, and operation:
\begin{enumerate}
  \item A quantum processor is defined by device parameters $\setelem{d_i}$, control parameters $\setelem{c_j}$, and decoherence parameters $\setelem{e_k}$. The fluctuation of a parameter can be described by a probability density function (PDF). The system Hamiltonian can be written as $H(\setelemtwo{d_i}{c_j},t)$, while the decoherence parameters can be included in different ways.
  \item Device parameters $\setelem{d_i}$ provide the basic physical properties of the device. The actual device parameters $\setelem{d^f_i}$ of the fabricated device follow the PDF $\setelem{D^f_i[d^d_i](d^f_i)}$ determined by the design and the fabrication together. After calibration, in the long time execution $d^f_i$ may still drift to $d^r_i$, following another PDF $D^r_i[d_i^f](d_i^r)$.
  \item Control parameters $\setelem{c_j}$ give the shapes, frequencies, and phases of control pulses. $\setelem{c_j}$ is optimized both in the design stage as $\setelem{c^d_j}$ and in the calibration process as $\setelem{c^f_j}$. Due to the resolution limit and noise of the instruments, the actual input $\setelem{c_j^r}$ to the device follows PDF $\setelem{C^r_j[c_j^f](c^r_j)}$.
  \item Each decoherence parameter $e_k$ describes one type of noise source coming from the fabrication process or the instruments. These parameters also fluctuate in the fabrication process follows $\setelem{E_k^f[e_k^d](e_k^f)}$. $e_k$ here can refer to a single parameter, such as the dielectric loss tangent $\tan\delta_C$, or a function like the power spectrum density $S(\omega)$.
  \item In summary, at the design stage, a set of parameters $\setelem{d_i^d},\setelem{c_j^d},\setelem{e_k^d}$ are determined. For a fabricated device, actual $\setelem{d_i^f},\setelem{e_k^f}$ are measured, and $\setelem{c_j^f}$ are optimized in the calibration. After calibration processes, the device and control parameters may change to $\setelem{d_i^r},\setelem{c_j^r}$ due to fluctuations. Fluctuations can be written as PDFs depend on previous stages: $\setelem{D^f_i[d_i^d](d_i^f)},\setelem{D^r_i[d_i^f](d_i^r)},\setelem{C_j^r[c_j^f](c_j^r)},\setelem{E_k^f[e_k^d](e_k^f)}$.
\end{enumerate}

There are plenty of packages to handle Hamiltonians with physical parameters like CircuitQ~\cite{Aumann2022CircuitQ}, SQcircuit~\cite{Rajabzadeh2023Analysis}, Scqubits~\cite{Groszkowski2021Scqubits} and SuperGrad~\cite{Wang2025SuperGrad}. We list some widely used Hamiltonians of qubits with constant microwave driving for reference:
\begin{itemize}
  \item Transmon~\cite{Koch2007Chargeinsensitive}
        \begin{align}
          \hat{H}_0 & = 4E_C(\hat{n}-n_g)^2-E_J\cos\hat{\phi}, \label{eq:ham_transmon}  \\
          \hat{H}_d & = A\cos(\omega t + \varphi)\hat{n}. \label{eq:ham_transmon_drive}
        \end{align}
        Here $\setelem{d_i}=\setelem{E_C,E_J,n_g}$, $\setelem{c_j}=\setelem{A, \omega, \varphi}$.
  \item Fluxonium~\cite{Manucharyan2009Fluxonium,Nguyen2019HighCoherence}
        \begin{align}
          \hat{H}_0 & = 4E_C\hat{n}^2 + \frac{1}{2}E_L\hat{\phi}^2-E_J\cos\left(\hat{\phi}-\frac{\pi\Phiext}{\Phi_0}\right), \label{eq:ham_fluxonium} \\
          \hat{H}_d & = A\cos(\omega t + \varphi)\hat{n}. \label{eq:ham_fluxonium_drive}
        \end{align}
        Here $\setelem{d_i}=\setelem{E_C,E_J,E_L,\Phiext}$, $\setelem{c_j}=\setelem{A, \omega, \varphi}$.
  \item 0-$\pi$ qubit~\cite{Gyenis2021Experimental,Gyenis2021Moving}
        \begin{align}
          \hat{H}_0 & = 4E_C^{\theta}(\hat{n}_{\theta}-n_g^{\theta})^2 + 4E_C^{\phi}\hat{n}_{\phi}^2 - 2E_J\cos\theta\cos\left(\hat{\phi}-\frac{\pi\Phiext}{\Phi_0}\right)+E_L\hat{\phi}^2, \label{eq:ham_0pi} \\
          \hat{H}_d & = A(t)\cos(\omega t + \varphi)\hat{n}_{\theta}. \label{eq:ham_0pi_drive}
        \end{align}
        Here $\setelem{d_i}=\setelem{E_C^{\theta},E_C^{\phi},E_J,E_L,n_g,\Phiext}$, $\setelem{c_j}=\setelem{A, \omega, \varphi}$.
\end{itemize}
Alternatively, device modeling is not limited to lumped-element descriptions~\cite{Ganjam2023Improving}, like black-box quantization~\cite{Nigg2012BlackBox}, energy-participation ratio~\cite{Minev2021Energyparticipation},inductive-energy participation ratio~\cite{Yu2024Using}, and machine-learning based models~\cite{Genois2021QuantumTailored,Hangleiter2024Robustly,An2025DualCapability}. The selection of a specific formalism is contingent upon the problem under investigation, as each offers distinct advantages in the derivation.

Based on this model, we can generalize the design criterion, that with design parameters $\setelemthree{d_i}{c_j}{e_k}$, the device must be able to execute a target quantum circuit with enough fidelity and a high enough yield.
\begin{align}
  P(F >F_t) = & \int dd^f_i \cdots \int de^f_k \cdots \int dc^r_j \cdots \int dd^r_{i'} \cdots \nonumber                                                                                                                                              \\
              & \prod_i D^f_i(d^f_i)\prod_k E_k^f(e^f_k) \prod_j C^r_j[\setelem{d^f_i},\setelem{e^f_k}](c^r_j) \prod_{i'} D^r_{i'}[\setelem{d^f_i}](d^r_{i'})  \nonumber \\
             & \times \prod_l \Theta(F(\setelem{d^r_{i'}},\setelem{c^r_j},\setelem{e^f_k}) - F_t) \nonumber \\
  >           & Y_t, \label{eq:design_criteria_yield_instruction}
\end{align}
where $P$ is the probability, $F_t$ is the target fidelity of the circuit, $Y_t$ is the target yield, and $\Theta$ is the step function that is 1 for positive inputs and 0 for negative inputs. Here we use the fidelity~\cite{Nielsen2002A} $F$ to represent the performance of the controls, but other error metrics are also valid. Various metrics are employed for quantifying errors under diverse conditions, including trace distance, state fidelity, average gate fidelity, entanglement fidelity, diamond distance, cross-entropy, and so on\cite{Hashim2025Practical}. For practical applicability, the selected error metric must be both fast for simulation within the QDA workflow and compatible with experimental calibration protocols.

In its general form, this probabilistic criterion is computationally intractable to solve directly. It is therefore practical to decompose it into a set of more manageable separated criteria. We note that there are some assumptions often valid for the simplification:
\begin{enumerate}
  \item A general quantum processor is defined with a quantum instruction set with multiple quantum operations $\setelem{O_l}$. $F_l(\setelem{d_i},\setelem{c_j},\setelem{e_k})$ denotes the fidelity $F$ of a quantum operation $O_l$. In practice, optimization efforts focus on a few critical operations. For example, in many superconducting quantum architectures, two-qubit gates are the primary performance bottleneck; consequently, optimizing two-qubit gate fidelity is often the main driver for improving overall processor performance.
  \item As the processor must be calibrated after fabrication, we must ensure the control scheme can always reach a good enough fidelity, which means finding $\setelem{c_j}$ for a fixed set of $\setelem{d_j}$. These means we only need to consider one point, $\setelem{d_i}$, in the possible distribution, for these controls if we can prove the control can be calibrated to similar fidelity for most $\setelem{d^f_i}$. Also, if the control is robust, then the fidelity should be high enough for most points in $\setelem{d^r_i}$.
  \item The electronic digital to analog and analog to digital conversions often behave as step functions and the noise is much smaller than the resolution limit. In some control schemes like single qubit rotations, the fidelity is related to the integration of the pulse amplitudes which is not limited by the resolution per time point. In these cases, it is OK to treat $c_j$ as exact, while in other cases the resolution limit can be modeled by a uniform distribution within $c_j\pm\delta c_j$.
  \item Under the same fabrication condition and with the same instrument setup, we can assume $e_k$ is a fixed value to skip the study of $E_k$ at the start of the design, and only optimize $\setelemtwo{d_i}{c_j}$ with this assumption. In some other cases, $E_k$ follows a long-tail distribution and require different treatment. For example, in the superconducting quantum processor, strongly coupled TLSs plague the system. When a fast decay strongly coupled TLSs and a qubit are very close in the frequency, the qubit will have a very short $T_1$. This means the qubit coherence is fully deteriorated and cannot recover. For this long-tail distribution, a more practical choice is to find fail-safe method to run applications with defects on the processor. For example, using frequency tunable qubits helps to remedy this issue in the architecture level as it can be selected to operate at frequency away from TLSs. Quantum error correction schemes can also be designed to be able to tolerate a few defects in a large system.
\end{enumerate}

Under these assumptions,~\Cref{eq:design_criteria_yield_instruction} can be decomposed. First we define that the optimization of the operation $O_l$ can reach maximum fidelity $F^B_l$ with control parameters $\setelem{c_j^B}$:
\begin{align}
  \setelem{c_j^B}(\setelemtwo{d_i}{e_k}) & = \argmax_{c_j} F_l(\setelemthree{d_i}{c_j}{e_k}),                                         \\
  F^B_l(\setelemtwo{d_i}{e_k})           & = \max_{c_j} F_l(\setelemthree{d_i}{c_j}{e_k}). \label{eq:design_criteria_optimal_control}
\end{align}
\begin{enumerate}
  \item Fidelity: The control can achieve high enough fidelity with optimal control:
        \begin{align}
          F^B_l(\setelemtwo{d_i}{e_k}) & \ge F_{t,l}.\label{eq:design_criteria_fidelity_one_point}
        \end{align}
  \item Reachability: The control can always reach similar fidelity with fluctuated parameters after the fabrication with calibration:
        \begin{align}
          F^E_l(\setelemtwo{d^d_i}{e_k}) & = \min_{d^f_i\in D^f_i(d^f_i) > 0} F^B_l(\setelemtwo{d^f_i}{e_k}) \ge F_{t,l}. \label{eq:design_criteria_reachable}
        \end{align}
  \item Robustness: The fixed control can keep high enough fidelity if the parameters fluctuate after optimization:
        \begin{align}
          F^R_l(\setelemtwo{d^f_i}{e_k}) & = \min_{d^r_i\in D^r_i(d^r_i) > 0, c^f_j\in C^f_j(c^f_j)>0} F_l(\setelemthree{d^r_i}{c^f_j}{e_k}) \ge F_{t,l}. \label{eq:design_criteria_robust}
        \end{align}
  \item Yield: The yield of the device that can achieve high fidelity is sufficiently high: 
        \begin{align}
          P(F^B_l > F_{t,l} \forall l)  = & \int d d^f_i \cdots \int d d^r_m \prod_i D^f_i(d^f_i) D^r_m(d^r_m) 
          \nonumber\\ 
          & \times \Theta(F_l(\setelemthree{d^r_m}{e_k}{c_j^F(\setelem{d^f_i},\setelem{e_k})})-F_{t,l}) > Y_t. \label{eq:design_criteria_yield_opt_fidelity}
        \end{align}
\end{enumerate}

This framework provides a foundational workflow for the design and control of the system Hamiltonian. This workflow involves two central tasks: first, computing the objective function (fidelity or other error metrics) from the system's time evolution, and second, optimizing the resulting metric by tuning the parameters $\setelemthree{d_i}{c_j}{e_k}$.

\subsection{Time evolution}

In this part, we will discuss how to compute the time evolution of a quantum system in brief. In the coherence limit where we can neglect the decoherence $\setelem{e_k}$, we can write the system in the time-dependent Schr\"odinger equation:
\begin{align}
  \frac{\partial }{\partial t}\ket{\psi(t)} & = H(t)\ket{\psi(t)}, \label{eq:psi_time_evo}                                     \\
  U(t_0,t_g)                                & = T \exp \left\{-i \int_{t_0}^{t_g} H(t) dt \right\}, \label{eq:u_from_time_evo}
\end{align}
where $\ket{\psi(t)}$ is the state at time $t$, $H(t)$ is the time-dependent Hamiltonian of the system at time $t$, $t_0$ and $t_g$ are the initial time and the final time of the evolution, $T$ is the time-ordering operator, $U(t_0, t_g)$ is the unitary operation in this period. In practice, these equations are often reformulated in the interaction picture. This approach is particularly advantageous when the Hamiltonian can be separated into a dominant, time-independent component and a smaller, time-dependent perturbation. When $H$ and $\ket{\psi}$ are expanded in the matrix form,~\Cref{eq:psi_time_evo} becomes a set of linear ordinary differential equations (ODEs). The unitary operator $U$ can then be constructed by solving this ODE system~\Cref{eq:psi_time_evo} for a complete basis of initial states at $t=0$. 

To study model systems, we often take $H(t)$ as a constant $H$ to make the equation solvable and easy to understand. In most cases, for time varying $H(t)$, the equations are not analytically solvable. The numerical solution of ODE systems is a long-lasting problem that has been extensively studied~\cite{Hairer1993Solving,Hairer1996Solving,Hairer2006Geometric,DAmbrosio2023Numerical}. Consequently, various established techniques have been developed to solve these equations approximately.
\begin{enumerate}
  \item Runge-Kutta (RK) family methods are well known methods to solve ODEs, like the Dormand-Prince method~\cite{Dormand1980A}. Given the Taylor expansion of the value $y(x_{n+1})$ of the next time step $y(x_{n+1})=y(x_n)+hy'(x_n)+\frac{1}{2!}h^2y''(x_n)\cdots$, when the local truncation error is $\mathcal{O}(h^{k+1})$, the global error will be $\mathcal{O}(h^k)$.
  \item Störmer-Verlet (SV) method is a symplectic version of the Runge-Kutta method. The symplecticity is a characteristic property of a twice continuously differentiable Hamiltonian systems. In another word, the volume of states of $H(t)$ in the variable phase space is conserved. Symplectic methods can keep the symplecticity, and they could reach better numerical precision with similar computational cost than non-symplectic methods.
  \item Suzuki-Trotter decomposition (STD) is a well known technique to compute the time evolution of Hamiltonians~\cite{Trotter1959On,Suzuki1976Generalized,Suzuki1976Relationship}. The evolution can be approximated as time slices as discussed above, and with $n$-th order STD,~\Cref{eq:u_from_time_evo} can be decomposed to
        \begin{align}
          U(t_0,t_g)       & = U(t_g - \delta t, t_g) U(t_g-2\delta t, t_g-\delta t)\cdots U(t_0, t_0+\delta t), \label{eq:u_time_split}              \\
          H(t)             & = \sum_{j=1}^J H_j(t), \label{eq:local_hamiltonian}                                                                      \\
          U(t, t+\delta t) & \approx S_n(t, \delta t) = \prod_i \exp(-i H_{j(i)}(t) p_{n,i} \delta t), \label{eq:std_terms}                           \\
          S_1(t, \delta t) & = \prod_{j=1}^J \exp(-i H_j(t)\delta t), \label{eq:trotter_1}                                                            \\
          S_2(t, \delta t) & = \prod_{j=1}^J \exp(-i H_j(t)\frac{1}{2}\delta t)\prod_{j=N}^1 \exp(-i H_j(t)\frac{1}{2}\delta t). \label{eq:trotter_2}
        \end{align}
        where $H_j$ are $j$-th local Hamiltonian terms, which are all 1-body or 2-body terms in superconducting quantum computing, $p_{n,j}$ are the coefficients of the $n$-th order STD. Note in~\Cref{eq:std_terms} there can be more than $J$ terms in higher order STD and each $H_j$ may appear more than one times, of which the index is indicated as $H_{j(i)}$. The construction of higher order terms has been studied in relevant papers~\cite{Yoshida1990Construction,Morales2022Greatly,Ostmeyer2023Optimised}. The error of Trotterization is in the order of $\mathcal{O}(\delta t^{n+1})$.  Note this method is also symplectic.
\end{enumerate}

There are already plenty of packages developed to perform the time evolution with methods mentioned above, like QuTip~\cite{Johansson2013QuTiP,Johansson2012QuTiP,Lambert2024QuTiP}, Qiskit-Dynamics~\cite{Puzzuoli2023Algorithms}, rbqoc~\cite{Propson2022Robust}, JuQbox.jl~\cite{Petersson2022Optimal,Petersson2020Discrete}, Qibo~\cite{Efthymiou2021Qibo}, Supergrad~\cite{Wang2025SuperGrad}, pygrape~\cite{Reinhold2019Controlling}, DYNAMO~\cite{Machnes2011Comparing}, Spinach~\cite{Hogben2011textitSpinach}, Krotov~\cite{Goerz2019Krotov}, RedCRAB~\cite{Heck2018Remote,Zoller2018Optimal}, Q-CTRL~\cite{Ball2021Software} and so on. To speed up calculations, GPU acceleration techniques are widely used in recent packages.

To include all important effects during the time evolution, the control simulation must also include the effect from the decoherence. The Schr\"odinger equation can be generalized to the Liouville von Neumann equation:
\begin{align}
  \frac{d}{dt}\rho & = -i[H, \rho], \label{eq:liouville_von_neumann}        \\
  \mathcal{L}\rho  & = -i[H,\rho], \label{eq:liouvillian}                   \\
  \rho(t)          & = \exp(\mathcal{L} t) \rho(0), \label{eq:rho_time_evo}
\end{align}
where $\mathcal{L}$ is the Liouvillian of the system, which is a super operator. The time evolution of $\rho(t)$ can be formally written in $\exp(\mathcal{L}t)$ similar to the form of $\psi(t)$ in the Hamiltonian $\exp(-iHt)$.

If the decoherence effect is Markovian, with appropriate approximations, the control can be simulated by the master equation in the Lindblad form
\begin{align}
  \mathcal{L}\rho & = -i[H,\rho] + \sum_i \gamma_i(c_i\rho c_i^{\dagger} - \frac{1}{2}c_i^{\dagger}c_i\rho - \frac{1}{2}\rho c_i^{\dagger}c_i). \label{eq:lindblad}
\end{align}
This $\mathcal{L}$ can be written in other forms, like Choi matrices or Kraus matrices~\cite{Nielsen2010Quantum}, and $c$ is a collapse operator which is an operator on the system which reflects the interaction to the environment.

The time evolution methods described previously can be directly extended to solve the master equation when formulated in the super operator representation. The primary challenge in this extension is computational: a system with a Hilbert space dimension of $d$ requires a $d^2 \times d^2$ density matrix representation, leading to a significant increase in computational cost. Given this cost, and assuming that decoherence is a perturbative effect in recent devices, approximations are frequently employed. The most common simplification is to simulate the ideal, coherent evolution $U(t)$ first, and then apply a post-processing correction to the final fidelity (e.g., subtracting first-order error terms proportional to $t_g/T_1$, and the coefficients can be deduced from model Hamiltonian). Alternatively, more integrated approaches exist. The OpenGRAPE framework~\cite{Chen2025Robust}, for instance, applies correction terms sequentially within the time evolution (e.g., after each step $U(t)$) to model the combined effects of decoherence and parameter fluctuations.

Some types of noises are non-Markovian and they require other treatment~\cite{Keeling2025processTensorNonMarkovian}. $1/f$ type noise diverges in the low-frequency, it can be partially treated by quasi-static noise with repeated simulations~\cite{OBrien2017Densitymatrix,Rol2019Fast}. In general, a variety of methods have been developed to model open quantum systems subject to non-Markovian noise. The stochastic Schr\"odinger equation (SSE) provides a powerful framework for this purpose and has found broad application across quantum mechanics\cite{Diosi1997The,Diosi1998NonMarkovian,Strunz1999Open}. One can directly analyze the property of the evolution under noise by geometric correspondence method\cite{Hai2025Geometric} or solve the equation by numerical approaches, which are often rooted in the Feynman-Vernon path integral formalism\cite{Feynman1963The,Caldeira1983Path}. Here we list several numerical techniques for treating non-Markovian dynamics, Hierarchical Equations of Motion (HEOM) and the Hierarchy of Pure States (HOPS). 

HEOM expands the system-bath correlation using a series of auxiliary density operators. Derived from the path integral approach, it is particularly effective for reservoirs with specific spectral densities\cite{Tanimura1989Time,Tanimura2020Numerically,Xu2022Taming,Nakamura2025Entanglement,Chen2025Simulation}. HOPS represents the system state as an ensemble of stochastic wavefunctions, rather than a single density matrix. It employs auxiliary pure states to decompose the bath correlation function into a system of coupled differential equations, which offers faster convergence and is well-suited for parallelization\cite{Suess2014Hierarchy,Suess2015Hierarchical}. These methods can also utilize a matrix product state representation for futher speed up\cite{Gao2022NonMarkovian}.

The optimal choice of method depends on the characteristics of the noise source and the required balance between computational cost and accuracy. An appropriate technique can then be incorporated into the QDA workflow for the simulation and optimization of device and control parameters.

Certain phenomena can be directly simulated using a static Hamiltonian for the sake of simplicity and clarity. For instance, frequency crowding is a critical issue in large multi-qubit chips that must be addressed to minimize interference between qubits (see the following subsection). Another significant phenomenon is the phase transition between many-body localization and quantum chaos, which inherently exists in quantum many-body systems like large multi-qubit chips. Quantum computation is assumed to work in the many-body localized phase, where distinct energy levels can be clearly resolved. Therefore, the probability of the phase transition must be sufficiently low to ensure chip stability~\cite{Varvelis2024Perturbative,Berke2022Transmon}.

\subsection{Optimization \label{sec:control_optimization}}
The simulation of quantum system time evolution is a computationally intensive task, demanding significant time and memory, and often represents the primary bottleneck in the QDA process.  Optimization at the Hamiltonian level, which is crucial for improving processor performance, requires iterative evaluations of the time evolution, causing the total computational cost to scale rapidly with the number of device and control parameters. Consequently, designing a quantum processor for large-scale, error-corrected quantum computing necessitates handling both large parameter spaces and complex quantum systems, which in turn demands substantial computation resources.

The control optimization problem,~\Cref{eq:design_criteria_fidelity_one_point} has been extensively studied in the quantum optimal control (QOC)~\cite{Werschnik2007Quantum,Skinner2012Optimal,Koch2022Quantum,Rembold2020Introduction,Wilhelm2020An}. QOC methods can be categorized into two approaches. Open-loop QOC is based on a pre-built model and does not require additional experimental feedback, while closed-loop QOC requires that the fidelity is measured and the control pulses are optimized based on the experimental measurements.

Open-loop QOC encompasses two main strategies. The first strategy focuses on a specific model and develops analytical or semi-analytical schemes tailored to study and correct specific errors. The derivative removal with adiabatic gate (DRAG) method~\cite{Gambetta2011Analytic} is one of the most successful methods in eliminating leakage to the non-computational space, which is the standard protocol in microwave driven single qubit gates and also applicable in other gates. The CPMG sequences ~\cite{Carr1954Effects,Meiboom1958Modified,Uhrig2007Keeping,Uhrig2011Exact} and dynamic decoupling (DD)~\cite{Viola1999Dynamical,Zanardi1999Symmetrizing,Vitali1999Using,Vitali2002Mirror,Khodjasteh2005FaultTolerant,Witzel2007Multiplepulse,Lidar2014Review} are important techniques in suppressing dephasing errors, and DD is the standard protocol to preserve the coherence of idle qubits in long-time execution of circuit. Another key field is the development of fast adiabatic operation~\cite{Chen2010Transient,delCampo2013Shortcuts,Demirplak2003Adiabatic,Martinis2014Fast}, and a Fourier basis expansion of the near-optimal solution is widely adopted recently~\cite{Martinis2014Fast}. More recently, mitigating crosstalk in scaling multi-qubit systems has garnered increasing attention, though it remains an active area of research with many open challenges~\cite{Hortensius2015Pulse,Gambetta2012Characterization,Barends2014Superconducting,Huang2020Alibaba,Takita2017Experimental,Mundada2019Suppression,Dai2021Calibration,Xie2022Suppressing,Tripathi2022Suppression,Niu2021Analyzing,Zhao2022Quantum,Ni2024Superconducting}.

The second strategy employs general-purpose numerical optimization algorithms that are largely model-agnostic. Gradient-free methods, like Nelder-Mead~\cite{Nelder1965A} for local minimum and covariance matrix adaptation evolution strategy (CMA-ES) for global minimum~\cite{Beyer2002Evolution}, are always applicable, but often exhibit slow convergence. Gradient-based local optimization methods like BFGS and L-BFGS-B~\cite{Zhu1997Algorithm,Morales2011Remark} generally offer better efficiency. Considering the representation of the pulse, gradient ascent pulse engineering (GRAPE) is one of the most well known QOC method which is originally designed for NMR experiments~\cite{Khaneja2005Optimal} and widely applied in other quantum systems. In short, GRAPE can be treated as taking gradient of a pulse written in piecewise constant by adjoint state method (see discussion of gradient below). Krotov's method~\cite{Reich2012Monotonically,Sklarz2002Loading} works similarly, but sequentially optimizes control parameters one by one to achieve better convergence. Chopped random basis algorithm (CRAB) method~\cite{Caneva2011Chopped,Doria2011Optimal,Rach2015Dressing,Muller2022One} can work on a chopped random basis without gradients. Gradient optimization of analytic controls (GOAT)~\cite{Machnes2018Tunable} further extends this concept to more general analytical basis functions.

A significant limitation of traditional QOC methods is the requirement for manual derivation of analytical gradients, a process that restricts flexibility in the QDA process. Recently, automatic differentiation (AD)\cite{Baydin2018Automatic}, a core technique in modern machine learning, has been applied to quantum control problems. AD automates gradient computations with respect to system parameters, offering superior flexibility, computational speed, and precision compared to numerical approaches like the finite difference method.

The integration of AD into quantum control has been demonstrated in frameworks like TensorFlow\cite{Abadi2016Tensorflow} for both closed and open systems\cite{Abdelhafez2019Quantum,Abdelhafez2019Gradientbased,Abdelhafez2020Universal}.  Several software packages now incorporate AD capabilities to support automatic differentiation of device or control parameters in the time evolution: Qiskit-Dynamics~\cite{Puzzuoli2023Algorithms}, QuTip~\cite{Johansson2013QuTiP,Johansson2012QuTiP,Lambert2024QuTiP}, Scqubits\cite{Groszkowski2021Scqubits, Chitta2022Computeraided} and QuOCS\cite{Rossignolo2023QuOCS} leverage JAX~\cite{jax2018github}, while SQcircuit\cite{Rajabzadeh2024A} uses PyTorch\cite{Paszke2019PyTorch}.  Supergrad~\cite{Wang2025SuperGrad} further supports the differentiation with respect to both control and device parameters. This integration also enables combining qubit property calculations with neural networks for AI-assisted qubit design\cite{Rajabzadeh2024A}. For better effeciency, a semi-AD approach by manually separating the time propagation and the evaluation of the functional when calculating the gradient is developed\cite{Goerz2022Quantum}. 

Despite its advantages, AD via backpropagation presents two primary drawbacks: high memory consumption compared to the finite difference method, and greater computational overhead than manually derived analytical gradients. Unlike the finite difference method, where function evaluations are independent, backpropagation requires caching all intermediate results of the forward computation pass. Consequently, memory usage scales with the simulation duration. This memory cost can be reduced by adjoint state method~\cite{Blondel2024The}. The idea is that the partial derivative terms contains also similar $U(t, t+\delta t)$ terms which can be partially replaced by a reverse time evolution, instead of storing all intermediate results. In the implementation, this reverse time evolution can also be done similar to Suzuki-Trotter decomposition with only local operators, which further improves the performance~\cite{Wang2025SuperGrad}. To improve computational efficiency, one can apply a semi-AD approach, which manually separates the time propagation from the evaluation of the objective functional during gradient calculation~\cite{Goerz2022Quantum}.

In parallel with developments in gradient-based optimization, reinforcement learning (RL) has emerged as another powerful paradigm, like AlphaZero~\cite{Silver2018A} and Proximal Policy Optimization (PPO)~\cite{Schulman2017Proximal}. QOC with reinforcement learning methods~\cite{Niu2019Universal,Zhang2019When,Dalgaard2019Global,Bukov2018Reinforcement,An2019Deep,Wang2020Deep,Sivak2022ModelFree} supports global optimization under various constraints and can find unexpected hidden structures beyond human experience. For instance, RL-based optimization achieved a two-qubit gate fidelity of 99.92\% on a fluxonium-based architecture, demonstrating its practical efficacy~\cite{Ding2023HighFidelity}.

For achieving ultimate performance in physical experiments, closed-loop QOC is indispensable~\cite{Feng2018Gradientbased,Li2017Hybrid,Wittler2021Integrated}. Adaptation by hybrid optimal control (Ad-HOC)~\cite{Egger2014Adaptive} combines the model-based gradient search and a model-free Nelder-Mead search. Adaptive control via randomized optimization nearly yielding maximization (ACRONYM)~\cite{Ferrie2015Robust} introduces stochastic optimization algorithm to efficiently deal with the randomness in the experiments. Optimized randomized benchmarking for immediate tune up (ORBIT)~\cite{Kelly2014Optimal} uses a subset of randomized benchmarking as the error indicator to optimize operations with Nelder-Mead. The Snake Optimizer, a closed-loop QOC framework, has been successfully deployed on a 53-qubit processor, where it demonstrated performance exceeding that of a human expert~\cite{Kelly2018Physical,Klimov2020The}.

\Cref{eq:design_criteria_reachable} and \Cref{eq:design_criteria_robust} belong to the robust control, which also attract lots of studies in recent years~\cite{Zhang2014Robust,Chen2014Samplingbased,Liu2024Robust,Kosut2013Robust,Allen2020Robust,Harutyunyan2023Digital,Yi2024Robust,Cao2024Robust,Chen2025Robust,Carvalho2021ErrorRobust,Weidner2025Robust,Poggi2024Universally,Chen2013ClosedLoop}. A single qubit gate with robust control is demonstrated to show $~10\times$ coherent error reduction on real quantum hardware~\cite{Carvalho2021ErrorRobust}.Techniques include sampling-based approaches~\cite{Khaneja2005Optimal}, trajectory optimization~\cite{Propson2022Robust}, and policy optimization~\cite{Howell2023Direct}. 
Lyapunov-based control, a staple of classical control theory, has also been adapted for quantum systems to ensure the stability of the desired evolution of the Hamiltonian. With an appropriately selected Lyapunov function $V$, a designed control laws can be obtained when the first order derivative of $V$ is kept non-positive to avoid the divergence of the system~\cite{Grivopoulos2003Lyapunovbased,Mirrahimi2005Lyapunov,Cong2007Quantum,Grivopoulos2008Optimal,Kuang2008Lyapunov,Hou2012Optimal,Hou2014Realization,Kuang2017Rapid,Li2022Lyapunovbased,Cong2020Lyapunovbased,Silva2024On}.

Directly optimizing the manufacturing yield objective in\Cref{eq:design_criteria_yield_opt_fidelity} is often intractable due to the difficulty in obtaining the function $\tilde{F}_l$. A common strategy is to translate this problem into constraints directly on physical device parameters. The most well-known case is the frequency crowding (spectral crowding) problem~\cite{Schutjens2013Singlequbit,Egger2014Optimal,Theis2016Simultaneous}, where the frequencies of $0-1$ and $1-2$ transitions of multiple qubits are so close that these qubits cannot be manipulated individually. Here, the gate fidelity problem is mapped to a problem in the frequency domain. In a transmon, we have the 0-1 frequency $hf_{01}\approx \sqrt{8E_CE_J}-E_C$, so the relationship between the relative standard deviation error of the resistance and that of the frequency is $\RSD_f=\RSD_i/2$. The best standard deviation of the Josephson junction resistance $\sigma_R$ and that of the Josephson energy, $\sigma_{J}$, which is proportional to the resistance, is 2\% at the time of the fabrication, and the corresponding standard deviation of the frequency is 1\%~\cite{Kreikebaum2020Improving}. For transmon near 5\text{GHz}, this translates to $\sigma_f\approx50\text{MHz}$. The Monte-Carlo sampling based on the standard deviation shows that, for the processor structure with a square lattice large enough to support $d=7$ surface code, to achieve the yield of a collision-free process above 1\%, the maximum tolerable frequency spread is $\sigma_f=7\text{MHz}$~\cite{Hertzberg2021Laserannealing}. This means that with current fabrication, one can barely make any useful chip. Post-processing technique, laser annealing, can reduce the frequency RSD to 0.15\% with an appropriate design of the frequency distribution. This shows that the laser annealing is mandatory on a large-scale processor with fixed frequency transmon qubits~\cite{Hertzberg2021Laserannealing,Kim2022Effects,Kreikebaum2020Improving,Zhang2022Highperformance}. Besides the fabrication, the collision chance can be reduced by the design of processor topology and frequency distribution~\cite{Yang2023A}.

The optimization of the processor parameters for given instruction sets, can be further combined with the overall performance of the quantum processor on quantum applications. More studies on the whole workflow have been proposed in recent years~\cite{Zhao2025EDAQ,LevensonFalk2025A,Li2023Quantum,Li2023Towards,Yang2024A}.

\subsection{What's next}

The theoretical foundation of quantum computing posits that perfect device control would yield a flawless quantum computer. In any physical implementation, of course, this ideal is unattainable. Therefore, the primary research effort is dedicated to improving the fidelity of control operations to approach the level based on that a fault-tolerant quantum computer can be implemented.  A comprehensive theoretical framework already exists for the forward design problem: evaluating the performance of a given qubit and control Hamiltonian against simple metrics like average gate fidelity or more complex ones like robustness and hardware compatibility. In contrast, the inverse design problem, the creation of novel qubit and control architectures with desired properties, remains a significant challenge. This process has long relied on the intuition of scientists to explore the gigantic design space. However, two promising paths are advancing the field. First, improvements in the accuracy and speed of forward simulations, driven by a deeper understanding of the underlying physics and the use of automatic differentiation, are enabling optimization-based approaches to inverse design. Second, direct inverse design methods, including those based on artificial intelligence, are being actively developed to help scientists discover novel architectures that can overcome the limitations of current superconducting devices.

\section{Decoherence \label{sec:decoherence}}

\begin{figure}
\centering
\includegraphics[width=1.0\textwidth]{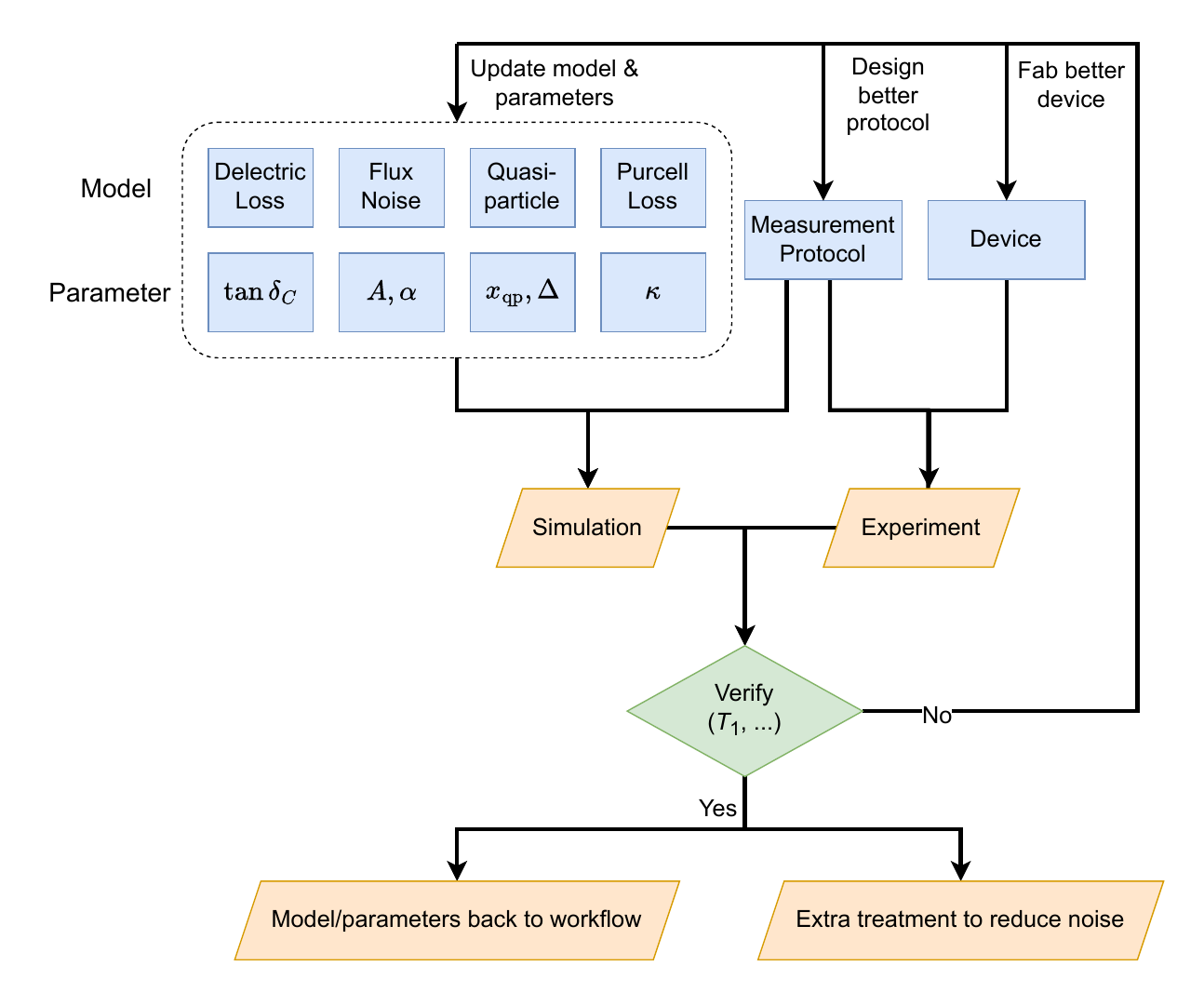}
\caption{Illustration of the decoherence related workflow, including the representation of decoherence, steps to determine the decoherence models and parameters and how they are used in QDA. For detailed description about how to verify the models and parameters, see~\Cref{subsec:noise_extraction}.}
\label{fig:decoherence_workflow}
\end{figure}

While Hamiltonian parameters, such as $\setelemtwo{d_i}{c_j}$, are primarily determined by the chip design, decoherence parameters $\setelem{e_k}$ typically arise from less controllable processes, such as device fabrication. Therefore, integrating decoherence effects into the QDA workflow requires an experimental calibration procedure. This iterative process, illustrated in~\Cref{fig:decoherence_workflow}, begins by postulating a decoherence model with an initial set of parameters. For a given measurement protocol, the outcomes are both simulated using this model and measured experimentally. Discrepancies between the simulated and experimental results guide the refinement of the model.  The refinement may call for updating the decoherence model or its parameters to achieve better agreement. In some cases, redesigning the measurement protocol is required if it proves to be insensitive to the parameters of interest.  On the other hand, it is possible that device anomalies are identified, which may require more experiments on a new device. Once the model and its parameters are calibrated to sufficient precision, they can be incorporated into the QDA workflow to improve designs. Moreover, identifying the physical sources of decoherence can inform specific mitigation experimental techniques for the device. This section focuses on the physics of the decoherence models, whereas the methodology for their experimental validation and parameter extraction is detailed in~\Cref{subsec:noise_extraction}.

\subsection{Noise model}
Most studies of quantum computing are based on the Bloch-Redfield model~\cite{Wangsness1953The,Redfield1957On,Bloch1957Generalized,Geva1995On}, which separates the decoherence of quantum systems into two parts, dissipation (depolarization or longitudinal relaxation) and dephasing (transverse relaxation). This model works when the noise is weak and Markovian (short-correlated), which is true in most cases in quantum computing.
The evolution of a system that follows the Bloch-Redfield model is~\cite{Ithier2005Manipulation,Ithier2005Decoherence}
\begin{align}
  \rho(t) & = \begin{pmatrix}
                1 + (|\alpha|^2-1)\exp(-t\Gamma_1)                          & \alpha^*\beta\exp(-t\Gamma_1/2)f_{z,R}(t)\exp(-i\Delta\nu t) \\
                \alpha\beta^*\exp(-t\Gamma_1/2)f_{z,R}(t)\exp(i\Delta\nu t) & |\beta|^2\exp(-t\Gamma_1)
              \end{pmatrix}  \label{eq:bloch_redfield_general} \\
          & \approx   \begin{pmatrix}
                        1 + (|\alpha|^2-1)\exp(-t\Gamma_1)               & \alpha^*\beta\exp(-t\Gamma_2)\exp(-i\Delta \nu t) \\
                        \alpha\beta^*\exp(-t\Gamma_2)\exp(i\Delta \nu t) & |\beta|^2\exp(-t\Gamma_1)
                      \end{pmatrix}  , \label{eq:bloch_redfield}
\end{align}
where the initial state is a pure state $\rho(t=0)=\alpha\ket{0}+\beta\ket{1}$, and $\Delta\nu$ is the detuning of the driving. \Cref{eq:bloch_redfield} is the most used equation with the decoherence time $T_1$ and $T_2$. \Cref{eq:bloch_redfield_general} is the more general equation that supports non-exponential decay functions of pure dephasing, especially when the noise is singular near $\omega\approx 0$. The overall coherence time $T_2$, the dissipation coherence time $T_1$ and the pure dephasing time  $\Tphi$, along with the corresponding rate $\Gamma_2,\Gamma_1,\Gammaphi$ follows:
\begin{align}
  T_1      & = 1/\Gamma_1,  \label{eq:t1_gamma1}                 \\
  \Tphi    & = 1/\Gammaphi, \label{eq:tphi_gammaphi}             \\
  T_2      & = 1/\Gamma_2,  \label{eq:t2_gamma2}                 \\
  \Gamma_2 & = \Gamma_1 / 2 + \Gammaphi, \label{eq:gamma2_1_phi} \\
  1/T_2    & = 1/2T_1 + 1/\Tphi. \label{eq:t2_1_tphi}
\end{align}

This can be written in the collapse operator in the Lindblad master equation for any two levels $i,j$:
\begin{align}
  \hat{c}_{1,ij}    & = \sqrt{\Gamma_{1,ij}}\sigma_{x,ij}, \label{eq:collapse_t1}                \\
  \hat{c}_{\phi,ij} & = \sqrt{\frac{\Gamma_{\phi,ij}}{2}}\sigma_{z,ij}. \label{eq:collapse_tphi}
\end{align}
Note~\Cref{eq:collapse_tphi} is only correct when the pure dephasing is also exponential, as~\Cref{eq:bloch_redfield}. These collapse operators from channels should be considered in the time evolution of the Lindblad master equation~\Cref{eq:lindblad} to include the decoherence effect in the optimization of device and control parameters.
It should be noted that in a multi-level qudit, if the pure dephasing comes from the flux fluctuation and the frequency sensitivity to flux, $\Gamma_{\phi,ij}$s between different $i,j$ are not independent. In this case, $\Gamma_{\phi,ij}$
must be carefully constructed to meet the relationship~\cite{Rol2019Fast}.

A standard protocol to measure $T_1$ is to populate the qubit to any state and to see how it decays back to the thermal state as
\begin{align}
  \Delta P & = (\Delta P(t=0) - \DeltaPthermal)e^{t/T_1} + \DeltaPthermal, \label{eq:t1_exp_decay}
\end{align}
where $\Delta P=P_1-P_0$ is the difference between the population of $\ket{0}$ and that of $\ket{1}$. Note $T_1$ includes both relaxations $0\rightarrow 1$ and $1\rightarrow 0$.
Similar experiments can be used to demonstrate $T_2$ by putting the qubit at a superposition state at the start. Due to historical reason, in lots of studies the $T_2$ in the free-induction experiment, or the Ramsey experiment, is assigned with the symbol $T_2^*$, while the Hahn echo experiment that contains a $\pi$-gate at the half of the decay time, is assigned with the symbol $T_2$.

A major source of the dissipation is the noise on the charge or the phase quantum operators, which connects the two levels where unwanted transition happens. As the noise is generally weak, this type of noise can be computed by the time-dependent perturbation theory and Fermi's Golden rule. Assuming the interaction between the system and the environment can be written as
\begin{align}
  \hat{H}_{se} = \hat{O} \otimes \hat{\lambda}, \label{eq:ham_system_env}
\end{align}
where $\hat{O}$ is in the system subspace and $\lambda$ is in the environment.
$\Gamma_1$ between two levels $i,j$ follows
\begin{align}
  \Gamma_{1,ij} & = \frac{1}{\hbar^2}\left|\mel{i}{\frac{\partial \hat{H}}{\partial \lambda}}{j}\right|^2S_{\lambda}(\omega_{ij}). \label{eq:t1_fermi_golden_rule_1} \\
\end{align}
The relaxation processes are in two directions $i\rightarrow j$ and $j\rightarrow i$. The noise can be assumed to be from a shunting admittance $Y$. With the relationship of quantum noise $S(\omega)=S(-\omega)\exp(\hbar\omega/k_B T)$, the PSD at finite temperature can be written as~\cite{Nguyen2019HighCoherence,Pop2014Coherent}
\begin{align}
  S_{\lambda}(+\omega_{ij})                             & = \hbar\omega_{ij}\Re Y(\omega)\frac{1}{2}\left(\coth\left(\frac{\hbar\omega}{2k_BT}\right)+1\right), \label{eq:psd_plus_y}                                               \\
  S_{\lambda}(-\omega_{ij})                             & = -\hbar\omega_{ij}\Re Y(-\omega)\frac{1}{2}\left(\coth\left(\frac{\hbar\omega}{2k_BT}\right)-1\right), \label{eq:psd_minus_y}                                            \\
  S_{\lambda}(+\omega_{ij})+S_{\lambda}(-\omega_{ij})   & = \hbar\omega_{ij}\Re(\omega)\coth\left(\frac{\hbar\omega}{2k_BT}\right), \label{eq:psd_sum_sign}                                                                         \\
  \Gamma_{1,i\rightarrow j} + \Gamma_{1,j\rightarrow i} & = \frac{1}{\hbar^2}\melabs{i}{\frac{\partial \hat{H}}{\partial \lambda}}{j}^2 \hbar\omega_{ij}\Re(\omega)\coth\left(\frac{\hbar\omega}{2k_BT}\right). \label{eq:t1_y_sum}
\end{align}
Therefore, for noise sources that adhere to the aforementioned assumptions, the parameters describing the noise can be obtained by computing $\melabs{i}{\frac{\partial \hat{H}}{\partial \lambda}}{j}$ and $\Re Y(\omega)$ of the corresponding model.

The pure dephasing term, $\Gamma_{\phi}$, primarily originates from the fluctuation of a quantity upon which the qubit frequency depends. The measured value of $\Gamma_{\phi}$ is strongly dependent on the specific measurement protocol employed. With a given window function $W_T(\omega)$ determined by the measurement protocol, the random phase $\Delta\phi$ accumulated and the phase decay $f_z(t)$ is~\cite{Ithier2005Decoherence,Nguyen2020Toward}
\begin{align}
  \Delta \phi & = \frac{\partial \omega_{01}}{\partial \lambda}\int_0^t dt' \delta \lambda (t'), \label{eq:random_phase_accumulated}                                                                                           \\
  f_z(t)      & = \expval{\exp(i\Delta\phi)} = \exp(-\frac{1}{2}\expval{\Delta\phi^2}), \label{eq:phase_decay_random}                                                                                                           \\
  \expval{\Delta\phi^2} &= \left(\frac{\partial \omega_{01}}{\partial \lambda}\right)^2\int_{-\infty}^{+\infty}\frac{d\omega}{2\pi}S_{\lambda}(\omega)W_T(\omega), \label{eq:phase_window_integral}
\end{align}
As an example, the spin-echo experiment corresponds to the decay function
\begin{align}
  f_{z,E}(t)  & = \exp \left( -\frac{t^2}{2} \left(\frac{\partial \omega_{01}}{\partial \lambda}\right)^2\int_{-\infty}^{+\infty}d\omega S_{\lambda}(\omega)\sin^2\frac{\omega t}{4}\mathrm{sinc}^2\frac{\omega t}{4}  \right), \label{eq:tphi_echo}
\end{align}

The preceding discussion has implicitly assumed that decoherence occurs within a single qubit. In a multi-qubit system, decoherence can also occur between multi-qubit energy levels. However, in most cases, such inter-qubit decoherence processes are considered negligible, as the inter-qubit coupling strength is typically much smaller than intra-qubit energy scales. For high-precision control schemes, incorporating these multi-qubit processes into noise simulations may be necessary.

From a design and optimization perspective, the decoherence parameters $\setelem{e_k}$ in~\Cref{fig:system_design_param} can be classified into two categories.
\begin{enumerate}
  \item Nominal-Is-Best (NIB): Decoherence sources arise because qubits must be coupled to the environment to enable manipulation. For example, the readout resonator is a necessary component coupled to the qubit, and it introduces both dissipation (Purcell loss) and dephasing (photon-shot noise). Drive lines also contribute to Purcell loss. Consequently, a trade-off between coupling strength and induced decoherence must be managed during the design stage. The resonator decay rate, $\kappa$, is therefore a critical parameter in this context.
  \item Lower-Is-Better (LIB): Decoherence sources that are not utilized in the control should be minimized. In this context, $\setelem{e_k}$ are constrained by experimental techniques, and these practical values must be factored into the design. Dielectric loss and inductive loss are two well-known phenomenological models used to describe losses from the charge and the phase operators. The corresponding dielectric ($\tan\delta_C$) and inductive ($\tan\delta_L$) loss tangents, as well as the effective temperature $T$, must be minimized. Similarly, quasiparticle density $\xqp$, thermal photon $\nth$, and other such parameters fall into this category. Given the ubiquity of these channels, protected qubit designs have been developed to reduce the coupling between the loss channel and the qubit, thereby strongly suppressing these decoherence effects.
\end{enumerate}

We note that some parameters, like $\tan\delta_C$, are not directly related to the parameters in the Hamiltonian. But it is influenced by the distribution of the electric field, which is determined by the layout design. Therefore, achieving optimal performance necessitates a joint optimization process. Several important loss channels and their corresponding $\setelem{e_k}$ are discussed below for reference.

\subsection{Purcell loss and photon shot noise}
A dispersively coupled resonator is often placed near the qubit to perform measurements. The Purcell effect describes the relaxation of a qubit coupled to a lossy resonator~\cite{Houck2008Controlling,Johnson2011Controlling,Sete2014Purcell}, while photon shot noise is the dephasing due to the fluctuation of the coupled resonator photon number~\cite{Gambetta2006Qubitphoton,Clerk2007Using,Rigetti2012Superconducting,Zhang2017Suppression,Kou2017FluxoniumBased}. The key decoherence parameter is the decay rate of the resonator $\setelem{e_k}=\setelem{\kappa}$, which is shared between two channels.

The Purcell relaxation rate can be derived from the Lindblad master equation of the qubit-resonator systems, though this calculation can be computationally complex. In typical superconducting computing systems, the decay rate and the coupling are similar, and the detuning between the qubit and the resonator $\Delta$ is much larger, $\Delta \gg \kappa \approx g$. In this parameter regime, the Purcell relaxation can be written as~\cite{Sete2014Purcell}
\begin{align}
  \Gamma_{1}^{\text{Purcell}} & = \frac{\kappa g^2}{\Delta^2}. \label{eq:t1_purcell}
\end{align}

The dephasing rate from the photon fluctuation and its low photon number limit is
\begin{align}
  \Gamma_{\phi}^{ps}                          & = \frac{\kappa}{2}\Re\left(\sqrt{\left(1+\frac{i\chi}{\kappa}+\frac{4i\chi \nth}{\kappa}\right)} - 1\right), \label{eq:tphi_photon_shot} \\
  \lim_{\nth\rightarrow 0} \Gamma_{\phi}^{ps} & \rightarrow \frac{\nth \kappa \chi^2}{\kappa^2 + \chi^2}, \label{eq:tphi_photon_shot_low}
\end{align}
where $\chi$ is the dispersive shift difference between $\ket{0}$ and $\ket{1}$. Minimizing the thermal photon number $\nth$ is ideal for reducing the photon shot noise. However, it is not always possible because the readout resonator must be pumped for readout routinely, which inevitably brings residual photons.

The decay rate $\kappa$ is a value that needs to be determined in the design stage. A minimal $\kappa$ is generally preferred to achieve long coherence times, as it suppresses both dissipation and dephasing. On the other hand, small $\kappa$ inherently slows the readout process, thereby degrading measurement performance. Consequently, a joint optimization of readout fidelity and qubit coherence time $T_1$, $T_2$ is necessary to determine the optimal $\kappa$.

\subsection{Dielectric and inductive loss}
Dielectric loss is often the major loss channel in a charge-type qubit, such as a transmon, and in a flux-type qubit at the flux frustration point.  It is found that the dielectric loss mostly comes from materials, or more specifically, from the defects (TLSs) in the materials~\cite{Martinis2005Decoherence}. Phenomenologically, we need one decoherence parameter, the dielectric loss tangent, $\setelem{e_k}=\setelem{\tan\delta_c}$, to describe the dielectric loss:
\begin{align}
  \Gamma_1^{\text{diel}} & = \frac{1}{4E_C}\left|\mel{0}{\hat{\phi}}{1}\right|^2\hbar \omega^2\tan\delta_C \coth\left(\frac{\hbar\omega}{2k_BT}\right). \label{eq:t1_dielectric}
\end{align}
From quantum harmonic oscillators at $T=0$, the above equation can be simplified:
\begin{align}
  \mel{0}{\hat{\phi}}{1} & = \left(\frac{2E_C}{E_L}\right)^{1/4}, \label{eq:phi01_lc}           \\
  \hbar\omega            & = \sqrt{8E_C E_L}, \label{eq:omega01_lc}                             \\
  \Gamma_1^{\text{diel}} & = \omega \tan\delta_C \defeq \omega / Q, \label{eq:t1_dielectric_lc} \\
  Q                      & = 1 / \tan\delta_C. \label{eq:q_dielectric_lc}
\end{align}
Here we can see that an oscillator can be described by a quality factor $Q$, which is the inverse of $\tan\delta_C$. For qubits with weak anharmonicity, such as transmon, $Q$ is also a good indicator of the decoherence and is widely used in the literature.

Dielectric loss originates primarily from the substrate and the surface oxidation layers of metals. Consequently, significant research has focused on optimizing both substrate and metal materials to mitigate this loss. Nowadays, Si and sapphire are the most common substrates due to their low dielectric loss tangent~\cite{Martinis2014UCSB}. The superconducting material itself has also become a focus of recent investigation. Materials exhibiting superior surface properties, like tantalum~\cite{Place2021New}, have been demonstrated to produce 2D transmon qubits with state-of-the-art coherence times\cite{Bland2025Millisecond}. Conversely, Nb is a good metal for superconductivity due to its very high $T_C$, but its oxides on the surfaces are known to be particularly detrimental. The fabrication process can be improved to reduce the TLSs on surfaces. For example, capping the metal layer with more stable materials to prevent the formation of unwanted metal oxides has been shown to significantly improve the qubit lifetime~\cite{Bal2024Systematic}.

From a design perspective, mitigating dielectric loss involves minimizing the electric field participation ratio of lossy surfaces~\cite{Wang2015Surface,Gambetta2017Investigating,Woods2019Determining,Martinis2022Surface,Deng2023Titanium}. The dielectric loss mostly comes from surfaces near the component vertices due to the larger electric fields. However, these boundaries are necessary for building the capacitance between two components of the quantum processors. Therefore, a careful optimization is necessary to balance these competing design constraints.

Inductive loss is described by a model similar to that of dielectric loss. It comes with one parameter, the inductive loss tangent, $\setelem{e_k}=\setelem{\tan\delta_L}$, and similar effective admittance
\begin{align}
  \Re Y(\omega) & = \tan\delta_L / \omega L. \label{eq:t1_inductive}
\end{align}

It should be noted that both dielectric and inductive loss channels are based on a generalized phenomenological model. This framework can incorporate more specific physical loss mechanisms by utilizing frequency-dependent loss tangents, $\tan\delta_C(\omega)$ and $\tan\delta_L(\omega)$.

\subsection{Quasiparticle}
In the context of superconducting quantum computing, the term "quasiparticle" refers to an unpaired electron in the superconductor~\cite{Catelani2011Relaxation,Glazman2021Bogoliubov,Gustavsson2016Suppressing,Catelani2011Quasiparticle,Catelani2012Decoherence}. The interaction between qubits and quasiparticles constitutes a fundamental source of decoherence.

The quasiparticle density $\xqp$ (the number of quasiparticles divided by the number of Cooper pairs) in equilibrium is related to the working temperature $T$ and superconducting gap $\Delta$\cite{Catelani2012Decoherence}. For typical values $T=20\text{mK}$ and $\Delta=50\text{GHz}$, the average number of quasiparticles in one device should be negligibly low~\cite{Catelani2019Nonequilibrium}. It is known that the non-equilibrium quasiparticle is a primary source of the decoherence~\cite{Martinis2009Energy,Shaw2008Kinetics,Vool2014NonPoissonian,Wang2014Measurement}. However, the generation mechanism of this non-equilibrium phenomenon is still not fully understood. Potential sources, such as cosmic rays, can cause a cluster of correlated error events that cannot be fixed by error correction~\cite{McEwen2022Resolving}, which may become the ultimate limitation for this field.

Given the low density of quasiparticles, the number of quasiparticles is often modeled as following a Poisson distribution. The $T_1$ can be written as~\cite{Pop2014Coherent}:
\begin{align}
  p_{\lambda}(n) & = \lambda^n e^{-\lambda} / n! , \label{eq:qp_poisson}                            \\
  \expval{P(t)}  & = e^{\lambda \exp(-t / T_{1\text{qp}})-1}e^{-t/T_{1\text{R}}}. \label{eq:t1_pt_qp}
\end{align}
where $\lambda$ is the average number of quasiparticles, $T_{1\text{R}}$ is the $T_1$ from other sources, and $T_{1\text{qp}}$ is the $T_1$ from one quasiparticle. $P(t)$ in~\Cref{eq:t1_pt_qp} is double-exponential instead of the common single exponential, which means $P(t)$ deviating from exponential behavior in $T_1$ measurements may come from quasiparticles. We note that quasiparticles may have different energies, and the energy distribution of quasiparticles can affect the above conclusion~\cite{Catelani2019Nonequilibrium,Vool2014NonPoissonian}.

$T_{1\text{qp}}$ can be computed in the model where the quasiparticle tunnels across the junction and induces the change of the qubit states~\cite{Catelani2011Quasiparticle,Catelani2011Relaxation}:
\begin{align}
  \Gamma_{1\text{qp}} & = \left|\mel{0}{\sin\left(\frac{\hat{\phi}}{2}\right)}{1}\right|^2 \xqp \frac{8E_J}{\pi\hbar}\sqrt{\frac{2\Delta}{\hbar \omega_{01}}}. \label{eq:t1_qp_junction}
\end{align}
For designs with junction arrays, the dissipation when quasiparticles tunnel across a junction in the array is~\cite{Catelani2011Quasiparticle}
\begin{align}
  \Gamma_{1\text{qp,array}} & = \left|\mel{0}{\frac{\hat{\phi}}{2}}{1}\right|^2 \xqp \frac{8E_L}{\pi\hbar}\sqrt{\frac{2\Delta}{\hbar \omega_{01}}}. \label{eq:t1_qp_array}
\end{align}
Overall, to describe the quasiparticle effect, decoherence parameters necessary in the design are $\setelem{e_k}=\setelem{\xqp, \Delta}$.

To reduce the quasiparticle effect from the design, several strategies can be deployed. An effective shielding is necessary to prevent the generation of quasiparticles, and the details will be discussed in the shielding~\Cref{subsec:shielding}. Another technique is "quasiparticle trapping", which aims to move the quasiparticle away from the qubit. Quasiparticles can be trapped in normal metal, which also helps to speed up the depletion of quasiparticles, though this introduces loss from normal metal part~\cite{Riwar2016Normalmetal}. Superconductors with different gap energies can also act as traps. For example, the vortices in the superconductor generated by the residual magnetic field can be treated as regions with diminished gap that trap quasiparticles, but the residual magnetic field will also affect the decoherence~\cite{Wang2014Measurement}. Gap engineering by adding other kinds of superconducting metals and/or varying the thickness of superconducting metal films can also create a region with different gap energy to trap quasiparticles to improve $T_1$~\cite{Riwar2019Efficient,Pan2022Engineering}. Phonon trap is another type of technique to reduce quasiparticles by dissipating the phonon energy that can break the Cooper pairs. It has the advantage that the trap is not necessarily directly connected to qubit, and no significant additional loss is introduced~\cite{Patel2017Phononmediated,Bargerbos2023Mitigation,Martinis2021Saving,Henriques2019Phonon,deVisser2021PhononTrappingEnhanced}.

\subsection{Flux noise}
Flux noise is an inherent source of decoherence in solid-state devices that mostly originates from surface defects~\cite{Yoshihara2006Decoherence,Koch2007Model,Galperin2007NonGaussian,Kakuyanagi2007Dephasing,Quintana2017Observation,Kumar2016Origin,Wang2018Hydrogen}. It can lead to both relaxation and dephasing. It is especially important in flux-tunable qubits, where the qubit frequency depends on the external flux.  

Relaxation due to the flux noise:~\cite{Nguyen2019HighCoherence,Sun2023Characterization}
\begin{align}
  \Gamma_{ij}^{\Phi} &= \frac{1}{(2e)^2L^2}\melabs{i}{\hat{\phi}}{j}^2S_{\Phi}(\omega_{01}).\label{eq:t1_flux_noise_psd} 
\end{align}

Dephasing due to the white flux noise:
\begin{align}
  \Gamma_{\phi} & = \frac{1}{2}\frac{\partial \omega_{01}}{\partial \Phi}C_{\Phi}. \label{eq:tphi_flux_white_noise_psd}
\end{align}

Dephasing due to the $1/f$ type flux noise~\cite{Nguyen2020Toward}:
\begin{align}
  S(\omega)                     & = 2\pi \frac{A^2}{\omega}, \label{eq:psd_1f} \\
  \Gamma_{\phi}^{\text{echo}} & = \frac{\partial \omega_{01}}{\partial \Phi} A \sqrt{\ln 2}. \label{eq:tphi_flux_1f_noise_psd}
\end{align}
Note $\Gamma_{\phi}^{\text{echo}}$ corresponds to a Gaussian type decay. However, experimental evidence indicates that the exponent $\alpha$ in $1/f^{\alpha}$ noise can deviate significantly from 1 under certain conditions, such as elevated temperatures\cite{Anton2012Pure}. This deviation results in a decay profile that is neither purely exponential nor Gaussian~\cite{NavaAquino2022Flux,Gao2025The}.

In contrast to dielectric loss, flux noise is typically characterized by its PSD instead of a single value. Consequently, the selection of the decoherence parameter $\setelem{e_k}$ is contingent upon the assumed form of the PSD. For instance, if the noise is assumed to be white, the parameter is $\setelem{e_k}={C_{\Phi}}$. If it is assumed to be $1/f$, $\setelem{e_k}={A}$. For $1/f^{\alpha}$ noise, the parameters are $\setelem{e_k}={A,\alpha}$.

Numerous studies attribute flux noise to surface spins~\cite{Koch2007Model,Braumuller2020Characterizing}. This attribution is supported by first-principle calculations showing that spin lattices on the surface, such as those from adsorbed hydrogen and oxygen, can generate $1/f$-like noise~\cite{Wang2018Hydrogen} in magnitude similar to that observed in experiments. This understanding informs two primary mitigation strategies. First, for a fixed density of spin defects, the device layout can be optimized to minimize flux noise. Second, the spin defect density itself can be reduced; for example, appropriate surface treatments to remove adsorbed molecules have been shown to suppress static spin susceptibility by an order of magnitude~\cite{Kumar2016Origin}. It should also be noted that external sources, such as the input lines controlling the external magnetic field, can contribute additional flux noise (see \Cref{subsec:input_lines}).

\subsection{What's next}
Advancing the study of decoherence requires a more detailed identification of its underlying physical origins. This understanding is crucial for developing comprehensive models and effective approximations that can be incorporated into the design of quantum hardware. However, the characterization of quantum noise is constrained by fundamental physical principles. The amount of information that can be extracted from a quantum system is inherently limited, and simulating such systems exactly on classical computers is often computationally intractable. Consequently, the pursuit of more accurate and complex decoherence models must be balanced with the need for simplicity and elegance. A successful model must not only be descriptive but also computationally tractable for simulation and verifiable through experimentation.

\section{Package and Cryogenic Setup Design}
\subsection{Package design}
The package is a carrier that shields the superconducting chip from environmental electromagnetic noise while providing impedance-matched and low-crosstalk interconnects, as well as removing the heat to anchor the chip at around $10\mathrm{mK}$. 
In both single-layer and flip-chip quantum chip designs, the chip is typically placed at the center of a printed circuit board (PCB), where the pins of the chip are wire bonded to corresponding pads located on the PCB.
The number of connections is linear to the dimension of the chip because the pins are located around the perimeter of the chip. In the TSV package~\cite{Rosenberg2020TSV, Tamate2022ScalablePackaging}, however, the control lines are wired to the bottom bumps of the silicon wafer. Since the whole area is used for interconnects, the number of connections increases quadratically as chip dimension increases. After the connection, the chip is then covered by a metal box for shielding.
A few aspects need to be considered in the design of the package and the interconnects~\cite{Huang2021Package}: 
\begin{itemize}
    \item \textbf{Impedance match.} Impedance match is one of the signal integrity considerations, which should be optimized and well addressed in the package design. The net impedance in the package can be standardized, with the impedance matched to 50$\Omega$. The major mismatch occurs at the interconnection between the chip and package through wire bonds. The bonding-wires are naturally inductive, so some capacitive compensation needs to be added to reduce the reflections. These capacitive effects can be designed at the pin pads of the chip. Therefore, the pad size and shape should be optimized based on the bond wire configurations through electromagnetic simulation to reduce overall reflections. 
    \item \textbf{Crosstalk.} Crosstalk is another key factor. The techniques used in package design to reduce the crosstalk can be directly applied. At the interconnects between the chip and package, the crosstalk can be optimized within the simulation for impedance match, where the ground bonding wires are crucial for reducing the crosstalk. 
    \item \textbf{Spurious box-modes.} The box mode is the electromagnetic mode within the box that covers the chip. These box modes are of low quality factors due to the lossy dielectrics in the PCB, and the conductive losses on the box surfaces, thus providing loss channels for qubit modes through the Purcell effect. One way to suppress these loss channels is to increase the detuning: the difference between the qubit frequency and the box mode frequency $f_c \sim \frac{1}{L \sqrt{\epsilon_r}}$, where $L$ is the size of the box and $\epsilon_r$ is the permittivity of the environment. The box size should be just slightly larger than the chip size, so as to minimize $L$. The more effective way to increase $f_c$ is to add a hole underneath the chip to decrease the effective $\epsilon_r$. 
\end{itemize}

\subsection{Cryogenic setup design}

Superconducting quantum processors, much like their classical counterparts, require physical wiring to connect with external control and readout electronics, as well as an appropriate environment to control the temperature. However, a critical challenge arises in quantum systems because the quantum chip operates at cryogenic temperature, and these wires bridge the substantial thermal gradient between the room-temperature control hardware and the cryogenic quantum chip. Consequently, the wiring acts as a direct conduit for thermal and electromagnetic noise from the high-temperature environment to the sensitive processor.

The mechanisms of noise in cryogenic systems are a subject of extensive research, with direct applications in quantum computing and sensing~\cite{Viennot2014Charge,Gabelli2006Evidence, Corcoles2011Protecting, Goetz2017Secondorder,Krinner2019Engineering,Yan2018Distinguishing,Baur2012Realizing,Boehme2016Characterisation,Krantz2016The,Wang2018The,Nguyen2020Toward,Zhang2021Universal,Arute2019Quantum,Wu2021Strong,Bao2022Fluxonium}. To utilize the quantum properties of a processor, environmental noise must be suppressed well below the quantum limit. Any noise with an energy comparable to or greater than the qubit's energy spacing will induce decoherence and thermal excitations. This is equivalent to raising the system's effective temperature into the classical regime, where the thermal occupation of excited states becomes significant ($e^{-\Delta E / kT}\gg 1$). For a typical superconducting qubit operating at a frequency of $f \approx 4$ GHz, the photon energy is $E=hf \approx 2.6\times10^{-24}$ J. To ensure the qubit is initialized and remains in its ground state, the environmental temperature must be significantly lower than the characteristic temperature $T = E/k_B \approx 0.2$ K. Modern dilution refrigerators meet this stringent requirement, providing a base temperature below 10 mK, which is sufficiently cold to suppress thermal noise and maintain the quantum state of the qubits.

While a dilution refrigerator provides the necessary low-temperature environment, achieving high-fidelity quantum control requires that the entire electronic system, from room-temperature instruments to the cryogenic stage, transmits signals with minimal added noise. Therefore, the design of the cryogenic control lines and associated shielding within the cryostat is of critical importance. 

This section outlines the fundamental principles for designing the control lines and shielding to mitigate noise that can degrade device performance. A comprehensive treatment of general cryogenic engineering, including topics such as materials and heat transfer, is beyond the scope of this work. For such details, readers are referred to specialized publications in the field~\cite{BookCryogenics2022,BookTheArtOfCryogenics2008,BookAnIntroductionToMillikelvinTechnology1989,BookCryogenicEngineering2004}.

\subsubsection{Input lines\label{subsec:input_lines}}
Superconducting qubits are controlled using microwave pulses and DC currents. These signals are delivered from room-temperature electronics to the quantum device via transmission lines. However, these lines also act as thermal conduits, propagating thermal photons from the room temperature electronics down to the chip, regardless of the precision of the electronics. As discussed previously, these thermal photons will destroy any quantum information stored in the quantum device.

To mitigate this, the signal lines must be able to effectively filter out high-energy thermal photons. In an ideal scenario, a single perfect filter at the base temperature stage would remove all thermal noise, except the base temperature thermal photon, without affecting the control signal. In practice, however, real-world components are imperfect and introduce several critical design challenges:

\begin{enumerate}
\item \textbf{Signal Attenuation:} The filtering required to reduce thermal noise inherently attenuates the desired control signal power.
\item \textbf{Component Self Heating:} Every component (e.g., attenuators, filters) will emit its own thermal photons, adding noise to the line.
\item \textbf{Heat Load:} The energy from the attenuated signal photons and absorbed thermal photons is dissipated as heat, which imposes a thermal load on the refrigerator's cooling power at each stage.
\item \textbf{Signal Distortion:} Components can introduce frequency-dependent variations in attenuation and phase, distorting the temporal shape of control pulses.
\end{enumerate}

The first three challenges must be addressed through careful cryogenic setup design, while the final issue of signal distortion can be partially controlled in the cryogenic setup, and partially compensated during the calibration of control pulses. Consequently, the standard engineering approach is to implement a cascaded chain of attenuators and filters distributed across the different temperature stages of the dilution refrigerator.

The attenuation can be described either by the power or the photon occupation number, and it follows the relationship~\cite{Clerk2010Introduction,Viennot2014Charge,Gabelli2006Evidence,Nguyen2020Toward,Krinner2019Engineering,Sears2012Photon,Yan2018Distinguishing,Pechal2016Microwave}:
\begin{align}
  P            & = \int_{-\infty}^{+\infty}S(f)df ,\label{eq:def_psd}                                                     \\
  \SJN(f,T)    & = hf \nBE(f,T) = \frac{hf}{e^{hf/k_B T} - 1}, \label{eq:bose_distribution}                               \\
  \Stm{out}{i} & = D^2 \Stm{in}{i} + (1-D^2)\SJN(f,T_i) = A \Stm{in}{i} + (1-A)\SJN(f,T_i), \label{eq:power_input_output} \\
  n_{i+1}(f)   & = A n_i(f) + (1-A) \nBE(f,T_i), \label{eq:photon_occ_chain}
\end{align}

where $f$ is the frequency, $S$ is the power spectral density (PSD) of power $P$, $n$ is the photon occupation number, and $\nBE$ is the Bose-Einstein distribution function, which can also be interpreted as the photon occupation number. $i,j$ are indices of stages in a cryogenic system, where larger $i$ means lower temperature. $A$ is the attenuation factor of the energy, $D$ is the transmission of the signal or voltage, and we have the relationship $A=D^2$. Note $A$ and $D$ are generally frequency dependent.
In practice, the properties of attenuators are often labeled in the unit of decibel $d$, where $D=10^{-d/20}$, and $A=10^{-d/10}$. All equations of the PSD above can be extended to the PSD of the voltage $S_{VV}$ or the PSD of the current $S_{II}$. For a more detailed description of the properties of the thermal noise, or Johnson-Nyquist noise, one can refer to relevant reviews\cite{Clerk2010Introduction}.

We note that the thermal noise can be approximated by white noise at low frequencies, where $hf/(e^{hf/k_B T}-1)\approx k_B T$, indicating that the system is in the classical regime. However, this approximation is not always valid in the cryogenic environment. As the corresponding frequencies of the temperatures with $k_B T = hf$ shown in~\Cref{table:temperature_frequency_relationship}, we can see that when $T<1{\text{K}}$, the thermal photon energy is comparable to the typical control system (MHz to 10 \text{GHz}), thus the white noise assumption is no longer valid.

\begin{table}[h!]
  \centering
  \begin{tabular}{ccccccc}
    \hline\hline
    Stage                     & RT   & 50K   & 4K   & Still & CP       & MXC     \\
    \hline
    Temperature (K)           & 298  & 35    & 2.85 & 0.882 & 0.082    & 0.006   \\
    $k_BT$ ($10^{-24}$ Joule) & 4.1k & 0.48k & 39   & 12    & 1.1      & 0.083   \\
    Frequency (Hz)            & 6.2T & 0.73T & 59G  & 18G   & 1.7G     & 0.13G   \\
    Cooling Power (W)         & -    & 30    & 1.5  & 40m   & 200$\mu$ & 19$\mu$ \\
    Cable length (mm)         & -    & 200   & 290  & 250   & 170      & 140     \\
    Cable attenuation (dB/m)  & -    & 9.7   & 8.7  & 8.2   & 8.2      & 8.2     \\
    \hline\hline
  \end{tabular}
  \caption{The selected parameters of the temperature, the corresponding photon frequency, the typical cooling power, the cable length and attenuation of a dilute fridge. The temperature, cooling power and cable length is from the parameters of BlueFors XLD400~\cite{Krinner2019Engineering}. The cable attenuation constants are estimated from those of stainless steel cable at 6GHz~\cite{Kurpiers2017Characterizing}.}

  \label{table:temperature_frequency_relationship}
\end{table}

We can list the following conditions that must be satisfied in a cryogenic system with $M$ stages. Here $\SJN$, $\Sin$ and $\Sout$ indicate the Johnson-Nyquist type noise PSD from the components, the input PSD and the output PSD, while $\Stt{}{S}$ and $\Stt{}{N}$ indicate the signal and noise correspondingly.
\begin{enumerate}
  \item Large enough signal at all required frequencies $f$ at the last stage:
        \begin{equation}
          \Stt{min}{S}(f) \le \Stt{in}{S}(f) \prod_{i=1}^{M}A_i . \label{eq:cryo_cond_1_signal}
        \end{equation}
        To control the qubit, the input signal $\Stt{in}{S}$ must be larger than the threshold $\Stt{min}{S}$ at the required frequency to drive the qubit at the required speed.
  \item Small enough noise at all relevant frequencies $f$:
        \begin{align}
          \Stt{max}{N}(f) & \ge \Stt{in}{N}(f)\prod_{i=1}^{M}A_i  + \sum_{i=1}^{M}\left(  \Saddi \prod_{j=i+1}^{M}A_j\right), \label{eq:cryo_cond_2_noise_chain} \\
          \Saddi          & = \SJN(f,T_i)(1-A_i). \label{eq:cryo_sadd}
        \end{align}
        In this chained relationship, we must count both the noise generated from the room-temperature devices, including the instrument noise and the room temperature thermal noise as the input noise $\Stt{in}{N}$, and the noise  $\Saddi$ added by components at each stage.

        Cables connecting these components also contribute to the noise. Unlike discrete components, which can be modeled as isothermal, cables are NOT thermalized and span large temperature gradients. It must be modeled by a continuous temperature distribution over the line. Given that the temperature is linearly distributed on the cable, the overall effect of a cable can be modeled as a series of attenuators which can be written as an integral in the continuous limit:
        \begin{align}
          \Stt{cable}{}(f,T_0,T_1,\Acable) = & (1-\Acable^{1/L})\sum_{i=1}^{L} \SJN(f, \frac{i}{L} (T_1 - T_0) + T_0) \Acable^{1-i/L}            \\
          \xrightarrow{L\rightarrow\infty}   & -\ln \Acable \int_0^1 dl \SJN(f, l(T_1 - T_0) + T_0) A^{1-l}. \label{eq:cable_linear_attenuation}
        \end{align}
        Here $T_0$ and $T_1$ are the temperatures of two stages connected by the cable, $A$ is the total attenuation of the cable, and $l$ is a dimensionless variable to indicate the relative position on the cable. In this way, the cable noise $\Stt{cable}{}$ from~\Cref{eq:cable_linear_attenuation} and its attenuation $\Acable$ can be used as one stage in the chain relationship~\Cref{eq:cryo_cond_2_noise_chain} $\Saddi$ and $A_i$.

  \item Large enough cooling power to dissipate heat at each stage $k$:
        \begin{equation}
          \Ptm{cool}{k} \ge \int df (1-A_k)\Stt{in}{S}\prod_{i=1}^{k-1}A_i. \label{eq:cryo_cond_3_cooling_power}
        \end{equation}
        For high-fidelity operations, the input signal power $\Stt{in}{S}$ should be much larger than any noise at any level, so all noise terms can be neglected in~\Cref{eq:cryo_cond_3_cooling_power}. Note~\Cref{eq:cryo_cond_3_cooling_power} may overestimate the upper bound by times, because the heat transfer process must be considered to determine whether the cooling power is enough, and heat sources other than the dissipated signal exist. The cables are connected on both sides, and their cooling should also be on both sides. As an estimation of the upper bound, we can assign the dissipation of the cables to the low temperature stage.
        A naive design approach might place all attenuators at the base stage to achieve the lowest possible thermal noise. This strategy is unfeasible, as the heat dissipated from attenuating the strong input signal would overwhelm the cooling power available at the base stage, which is the smallest among all stages. Therefore, a distributed attenuation scheme is essential. Attenuators must be strategically placed at various temperature stages to leverage the greater cooling capacities available at warmer stages.

\end{enumerate}

To illustrate the above design principles, this section analyzes a typical cryogenic setup, calculating the input power, dissipated power, and thermal noise at each stage in~\Cref{table:temperature_frequency_relationship}. There are 2 typical distinct lines often used in superconducting quantum computing: microwave lines which are used to transfer photons with a specific frequency to drive the qubit, and flux lines which are used to transfer DC current that is often used to tune the qubit frequency. These two types of lines require different bandwidths, power and noise limitations, as well as the cryogenic setup.

The power of a microwave line to drive a qubit can be computed from the Rabi frequency $\Omega_0$, the drive frequency of the qubit 0-1 transition $\omega_{01}$, and the coupling between the qubit and the line. The voltage on the microwave line $V$ applied on the qubit islands $V_q$ can be described by a factor $\beta$ as $V_q=\beta V$, which relies on the coupling capacitance and the qubit capacitance~\cite{Rasmussen2021Superconducting}:
\begin{align}
  \beta & = \frac{C_c}{C_c+C_q}. \label{eq:ratio_voltage_line_qubit}
\end{align}
Due to the coupling, the voltage noise $S_{VV}$ on the line will also affect the qubit via the charge operator $2e\hat{n}$, leading to Purcell loss. Based on Fermi's Golden rule, the Purcell loss due to the microwave line can be written as~\cite{Pechal2016Microwave}
\begin{align}
  \Decay{1}{0}{\Purcell}     & = \frac{1}{\hbar^2}4e^2 \melabs{0}{\hat{n}}{1}^2 \beta^2 S_{VV}(-\omega_{01}) = \DecayBase{0}{1}{\Purcell} (1+\nBE(\omega_{01})) , \label{eq:purcell_drive_10}                                                         \\
  \Decay{0}{1}{\Purcell}     & = \frac{1}{\hbar^2}4e^2 \melabs{0}{\hat{n}}{1}^2 \beta^2 S_{VV}(\omega_{01}) = \DecayBase{0}{1}{\Purcell} \nBE(\omega_{01}), \label{eq:purcell_drive_01}                                                               \\
  \DecaySum{0}{1}{\Purcell}  & = \Decay{1}{0}{\Purcell} + \Decay{0}{1}{\Purcell} = \DecayBase{0}{1}{\Purcell}\coth{\frac{\hbar\omega_{01}}{2k_BT}} \xrightarrow{\hbar\omega_{01}\gg k_BT} \DecayBase{0}{1}{\Purcell}, \label{eq:purcell_drive_01_sum} \\
  \DecayBase{0}{1}{\Purcell} & = \frac{8e^2}{\hbar} n_{01}^2 \beta^2 \Re Z(\omega_{01})\omega_{01}, \label{eq:purcell_drive_base}
\end{align}
where $\DecayBase{0}{1}{\Purcell}$ is a common factor in the $0\rightarrow 1$ and $1\rightarrow 0$ rates.
The relationship between the Rabi rate $\Omega_0$ in the drive Hamiltonian $\hat{H}_d=\Omega_0\cos(\omega t)\hat{n}$ and the voltage on the microwave line $V$ when the drive frequency is $\omega_{01}$ is
\begin{align}
  \hbar\Omega_0 & = \beta V2e\melabs{0}{\hat{n}}{1}= 2e\beta V n_{01}. \label{eq:rabi_voltage}
\end{align}
Given that the qubit transition energy $\hbar\omega_{01}$ is much higher than $k_B T$, from~\Cref{eq:purcell_drive_01_sum} and~\Cref{eq:rabi_voltage}, we have the power dissipated on the line with an external voltage source:
\begin{align}
  P & = \left(\frac{V}{2\Re Z(\omega_{01})}\right)^2 \Re Z(\omega_{01}) = \hbar \omega_{01} \Omega_0^2 / \DecaySum{0}{1}{\Purcell} = \frac{1}{2}\hbar \omega_{01} \Omega_0^2 \TPurcell. \label{eq:microwave_line_rabi_purcell}
\end{align}

To illustrate a practical design, we analyze the configuration of a microwave line here. Given that we need to drive a qubit with a 20 ns long $\pi$-pulse with a Gaussian envelope of $\sigma=3.3$ns, the peak Rabi frequency is $\Omega_0\approx 2\pi \times 60$ MHz. The average power at the base stage that determines the heat load is estimated to be -78 dBM,  because the pulse shape and the idle time of the $\pi$-pulse must be considered. \Cref{table:cryo_input_line_detail} shows the detailed results of the design of a microwave line with 20dB(50K) + 20dB(CP) + 20dB(MXC) configuration for this target power~\cite{Krinner2019Engineering}. The dissipated powers of 25 lines at each stage are all smaller than 5\% of the target heat loads, and the effective temperature of the installed qubits at MXC stage is reasonable (40 mK). This configuration represents an optimal balance: shifting more attenuation to low temperature stages would increase the heat loads, while shifting more attenuation to high temperature stages will increase the noise of the qubit.

\begin{table}
  \centering
  \begin{tabular}{cccccc}
    \hline\hline
    Name  & Power(dBm) & Ratio                & Dissipate (W)        & $n_{\mathrm{Noise}}$ & $\Teff$  (K) \\
    \hline
    RT    & -9         & 0.0\textperthousand  & $0                $  & $1.0\times 10^{3}$   & 298          \\
    Cable & -11        & 0.0\textperthousand  & $4.6\times 10^{-5}$  & $8.6\times 10^{2}$   & 247          \\
    50K   & -11        & 0.0\textperthousand  & $0                $  & $8.6\times 10^{2}$   & 247          \\
    Cable & -13        & 0.0\textperthousand  & $3.6\times 10^{-5}$  & $5.1\times 10^{2}$   & 146          \\
    4K    & -33        & 1.5\textperthousand  & $4.5\times 10^{-5}$  & $14              $   & 4.3          \\
    Cable & -35        & 0.0\textperthousand  & $1.7\times 10^{-7}$  & $11              $   & 3.3          \\
    Still & -35        & 0.0\textperthousand  & $0                $  & $11              $   & 3.3          \\
    Cable & -37        & 0.0\textperthousand  & $7.8\times 10^{-8}$  & $8.4             $   & 2.6          \\
    CP    & -57        & 51.1\textperthousand & $2.0\times 10^{-7}$  & $0.11            $   & 0.13         \\
    Cable & -58        & 0.0\textperthousand  & $4.8\times 10^{-10}$ & $0.089          $    & 0.12         \\
    MXC   & -78        & 4.1\textperthousand  & $1.6\times 10^{-9}$  & $8.9\times 10^{-4}$  & 0.041        \\
    \hline
  \end{tabular}
  \caption{Computed input power per line, dissipated power per line, ratio between the dissipated power and the cooling power assuming there are 25 lines, the thermal photon number, the effective temperature from the thermal photon number based on the cryogenic parameters in~\Cref{table:temperature_frequency_relationship} for microwave input lines at 6GHz. We assume there are 20dB attenuators in the 4K, CP and MXC stages. The target power -78 dBm follows~\cite{Krinner2019Engineering}. The effective temperature is estimated from thermal noise photon based on~\Cref{eq:bose_distribution}.}
  \label{table:cryo_input_line_detail}
\end{table}

The flux line is designed to provide magnetic flux to tune the qubit, which enables the ability to tune the $\omega_{01}$ of tunable transmon qubits and fluxonium qubits. A tuning range of $0.5\Phi_0$ is often sufficient, as the important superconducting qubit Hamiltonian depends on a $\cos(2\pi \Phi/\Phi_0)$ style term, which is periodic. To achieve the magnetic flux in the qubit SQUID loop, the required current is
\begin{equation}
  IM = \Phi_0 / 2, \label{eq:current_flux_relationship}
\end{equation}
where $M$ is the mutual inductance between the qubit SQUID loop and the flux line. From this we can get the PSD of the flux
\begin{align}
  S_{\Phi} & = M^2 S_{II}.  \label{eq:psd_phi_i}
\end{align}
This implies
\begin{align}
  S_{\Phi} & = \frac{\Phi_0^2}{4I^2} S_{II}.
\end{align}
 To minimize the impact of current noise from the control electronics, a large mutual inductance is desirable, as it allows the target flux to be generated with a smaller current. However, a large $M$ can also increase magnetic crosstalk between adjacent qubits. Conversely, a small $M$ reduces crosstalk at the cost of requiring a larger bias current, which is constrained by the output range of the current source and can lead to higher dissipated power. Therefore, $M$ must be carefully optimized during the chip design phase.

In a typical design, $M$ is about $1\mathrm{pH}$, although this value can reach up to 12 pH~\cite{Moskalenko2021Tunable,Moskalenko2022High}. From~\Cref{eq:current_flux_relationship}, the typical bias current $I\approx 1\mathrm{mA}$. This corresponds to a 1.6 V input, which falls within the typical range of electronic instruments~\cite{ZurichAWGParameter}, with 30 dB attenuation from room temperature to the base stage.

In contrast to microwave drive noise, which primarily contributes to energy relaxation $T_1$, flux noise is a dominant source of both relaxation (for flux type qubits) and dephasing $T_\phi$. Follows \Cref{eq:t1_flux_noise_psd} and~\Cref{eq:tphi_flux_white_noise_psd}, values related to qubit decoherence due to the flux line are estimated and presented in~\Cref{table:decoherence_flux_line}. Notably, the $T_\phi$ values are significantly lower than the state-of-the-art values for superconducting qubits. This indicates the challenge in enhancing $T_2$ values when the qubit operates off its sweet spot, particularly given the presence of flux noise. A particularly notable observation in the table is the low $T_1$ value ($< 10 \mu\mathrm{s}$) for the fluxonium qubit at its sweet spot, which is attributed to the flux line. This is a consequence of the strong coupling between the flux line and the fluxonium qubit. To enable effective operation of a fluxonium qubit, additional low-pass filters are required to attenuate noise above the qubit frequency by at least 20 dB relative to the values presented in the table, while maintaining a sufficiently large static current for system tuning.

\begin{table}
  \centering
  \begin{tabular}{ccc}
    \hline\hline
    Property                                           & Fluxonium                                              & Tunable Transmon        \\
    \hline
    Mutual inductance                                  & \multicolumn{2}{c}{1 pH}                                                         \\
    Total attenuation                                  & \multicolumn{2}{c}{30 dB}                                                        \\
    RT Voltage for $\Phi_0/2$                          & \multicolumn{2}{c}{1.6V}                                                         \\
    RT Instrument white noise~\cite{ZurichAWGParameter} & \multicolumn{2}{c}{$35\mathrm{nV}/\sqrt{\mathrm{Hz}}$}                           \\
    $\sqrt{S_{\Phi}}$                                  & \multicolumn{2}{c}{10.7n$\Phi_0/\sqrt{\mathrm{Hz}}$}                             \\
    $|\mel{0}{\hat{\phi}}{1}|$ (sweet spot)            & 2.7                                                    & -                       \\
    $T_1$ (sweet spot)                                 & 3$\mu$s                                                & $\infty$                \\
    $df_{01} / d\Phi$ (biased)                         & $6 \mathrm{GHz}/\Phi_0$                                & $2 \mathrm{GHz}/\Phi_0$ \\
    $\Tphi$ (biased)                                   & 12$\mu$s                                               & 111$\mu$s               \\
    \hline
    \hline
  \end{tabular}
  \caption{Qubit decoherence from a flux line with a typical parameter sets. $df_{01}/d\Phi$ in both fluxonium qubits or tunable transmon qubits varies with $\Phiext$ in the corresponding Hamiltonian, here we pick a value that is $1/4$ of the maximum.}
  \label{table:decoherence_flux_line}
\end{table}

The preceding calculations assume that ideal attenuators provide uniform attenuation across all frequencies. Unfortunately, practical attenuators have a limited working bandwidth. Especially, typical commercial attenuators have a high frequency cutoff at 10-40 GHz~\cite{Nguyen2020Toward}. This limitation is critical in superconducting quantum computing, as high frequency noise close to or above the superconducting gap will excite quasiparticles in the system, which is a significant source of decoherence~\cite{Catelani2011Quasiparticle,Barends2011Minimizing,Corcoles2011Protecting,Pop2014Coherent,McEwen2022Resolving,Vepsalainen2020Impact,Martinis2021Saving,Serniak2018Hot,Kreikebaum2016Optimization}. To mitigate the quasiparticle poisoning, specialized cryogenic low-pass filters are necessary. Eccosorb filter is widely adapted in superconducting computing~\cite{Peterer2012Investigating,Nguyen2020Toward,Krinner2019Engineering,Bao2022Fluxonium} to suppress the noise $>100 \text{GHz}$ thus preventing the generation of the quasiparticles. Metal powder filter, which attenuates the radio-frequency signals by skin-effect damping of the large surface area of powder, is another commonly used radio-frequency low-pass filter~\cite{Martinis1987Experimental,Fukushima1997Attenuation,Lukashenko2008Improved,Baselmans2009Long,Barends2011Minimizing,Masluk2012Reducing,Geerlings2013Improving}. This filter can be made by different materials including copper, bronze, brass, manganin, stainless steel and so on. Both the Eccosorb filter and powder filter can be home-made to adjust the attenuation by tuning the lengths, and the choice of the filter depends on the appropriate length and the impedance match of the system~\cite{Geerlings2013Improving}.

It is worth noting an alternative approach to apply DC signals on multiple qubits. Instead of employing one flux line per qubit, a multi-loop metal coil can be considered to generate uniform magnetic field across the entire chip. This architecture significantly reduces the the overall heating and control complexity. However, it presents a considerable engineering challenge in achieving uniform magnetic field across all qubits.

\subsubsection{Output lines}
On the other hand, we must be able to detect extremely weak output signals from qubits, requiring an output line optimized for high-fidelity readout~\cite{Walter2017Rapid,Eichler2012Characterizing}. Consequently, amplification is essential before the signals reach the detector. As superconducting qubit outputs are at the single-photon level, amplifiers operating near the quantum limit are required~\cite{Blais2021Circuit,Clerk2010Introduction,Caves1982Quantum}. Examples include the Josephson parametric amplifier (JPA)~\cite{Castellanos-Beltran2007Widely,Bergeal2010Phasepreserving,Hatridge2011Dispersive,Yamamoto2008Fluxdriven,Roch2012Widely,Eichler2014Controlling} and its variants, such as the traveling-wave parametric amplifier (TWPA)~\cite{Macklin2015A,Heinsoo2018Rapid}. A JPA typically provides 20-30 dB of gain within a sub-GHz bandwidth, whereas a TWPA offers a significantly larger bandwidth of several GHz. The second stage amplifier is a high electron mobility transistor (HEMT) low-noise amplifier (LNA) at 4 K, and it provides around 40 dB gain with effective noise temperature around 1.5 K~\cite{Zeng2023LowPower,Zeng2023100W}. The final stage is a room temperature amplifier or a chain of amplifiers, which gives 40-80 dB gain in total. This output line setup is common in microwave-based cryogenic systems, including both superconducting and spin qubit quantum computing platforms~\cite{Prabowo2024Modeling}. The primary design objective for the output line is to maximize the signal-to-noise ratio (SNR), thereby achieving optimal readout fidelity.

Based on the noise of a classical cascaded system without impedance mismatch,~\cite{Pozar2012Microwave}
\begin{align}
  N_{i+1}            & = G_{i+1}(N_i +N_{\text{th},i+1})  \label{eq:noise_output_cascade}                                                  \\
                     & = \prod_{k=1}^i G_k N_0 + \sum_{j=1}^i \prod_{k=1}^j G_k N_{\text{th},i}, \label{eq:noise_output_cascade_per_level} \\
  T_{i+1,\text{avg}} & = T_{i+1} + T_{i} / G_{i} + T_{i-1} / G_i G_{i-1} + \cdots T_1 / \prod_{k=1}^i G_k
\end{align}
where $B$ is the bandwidth considered, $N_0$ is the input power, $N_i$,$N_{\text{th},i}$ and $G_i$ are the output power, the thermal noise, and the gain of the $i$-th stage. It can be easily seen that the noise characteristic are dominated by the noise characteristics of earlier stages that are amplified more. This can also be reflected in the average temperature $T_{i, \text{avg}}$ at a given stage.

The components in the quantum regime require different treatment. The parametric amplifiers, in the ideal condition with large gain $G\gg 1$, can be described by an effective temperature
\begin{align}
  \frac{k_B \Teff}{G_0} & = \frac{1}{2}\hbar \omega_S + \frac{1}{2}\hbar \omega_S , \label{eq:teff_jpa}
\end{align}
where $\omega_S$ is the frequency of the signal. Here, the first half photon is the vacuum noise of the quantum signal source, which is generally a resonator, and the second half photon is the added noise of the parametric amplifier~\cite{Clerk2010Introduction}.

Combined~\Cref{eq:noise_output_cascade_per_level} and~\Cref{eq:teff_jpa}, we can estimate the noise and signal power of an output line as~\Cref{table:cryo_output_line_detail}. Here We analyze two configurations for comparison: one with the JPA and one without. We can see that in this typical cryogenic setup, the SNR with the inclusion of the JPA is much better than that without. The JPA reduces the required gain from subsequent room-temperature amplifiers, which would otherwise amplify the noise introduced by the HEMT. Without a JPA, the HEMT noise dominates the system's total noise and degrades the SNR, even when using a low-noise HEMT with $T_N\approx 2\mathrm{K}$. This highlights the critical importance of the parametric amplifier with high gain, large bandwidth and low noise for superconducting quantum computing.

\begin{table}
  \centering
  \footnotesize
  \setlength{\tabcolsep}{4.5pt}
  \begin{tabular}{cc|ccccc|ccccc}
    \hline \hline
    Type                   & Component       & \multicolumn{5}{c|}{with JPA} & \multicolumn{5}{c}{no JPA}                                                         \\
    \hline
                           &                 & \makecell{Stage                                                                                                    \\ Power} & \makecell{Stage \\ Gain} & \makecell{Total \\ Gain} & \makecell{Output \\ Power} & \makecell{Stage \\ Ratio}
                           & \makecell{Stage                                                                                                                      \\ Power} & \makecell{Stage \\ Gain} & \makecell{Total \\ Gain} & \makecell{Output \\ Power} & \makecell{Stage \\ Ratio}         \\
    \hline
    \multirow{7}{*}{Noise} & Qubit           & -206                          & 0                          & 111  & -34.6 & 38\% & -206 & 0   & 111 & -34.6 & 5\%  \\
                           & JPA             & -206                          & 15                         & 111  & -34.6 & 38\% &      &     &     &       &      \\
                           & Cable           &                               & -3                         &      &       &      &      & -3  &     &       &      \\
                           & HEMT (2K)       & -196                          & 40                         & 99   & -36.6 & 24\% & -196 & 40  & 114 & -21.6 & 93\% \\
                           & Cable           &                               & -3                         &      &       &      &      & -3  &     &       &      \\
                           & RTA (300K)      & -174                          & 65                         & 62   & -51.8 & 1\%  & -174 & 80  & 77  & -36.8 & 3\%  \\
                           & Cable           &                               & -3                         &      &       &      &      & -3  &     &       &      \\
                           & Total           &                               &                            &      & -30.4 &      &      &     &     & -21.3 &      \\
    Signal                 &                 & -130                          &                            & 111  & -19.0 &      & -130 &     & 111 & -19.0 &      \\
    SNR                    &                 &                               &                            & 11.4 &       &      &      & 2.3 &                    \\
    \hline \hline
  \end{tabular}
  \caption{Computed signal and noise per output line with/without JPA. The readout frequency is 8.26GHz, and the effective temperature is 80mK based on a thermal photon measurement~\cite{Yan2018Distinguishing}. The cable loss is estimated to be 3dB from base to the HEMT, 3dB from HEMT to the room temperature amplifier, and 3dB from the room temperature amplifier to the receiver. Effects of other components, like circulators, are combined into the cables. Noise from cables in the final output is neglected. The bandwidth is assumed to be 1MHz. The signal power is -130dBm, which corresponds to around 20 photons in one measurement. The stage power is the noise / signal power before the gain, and the stage gain is the gain at this stage. The total gain is the summation of gain from this stage to the receiver. The output power is the power received considering the total gain. The stage ratio represents the proportion of the output noise from this stage in the total output noise. }
  \label{table:cryo_output_line_detail}
\end{table}

\subsubsection{Shielding\label{subsec:shielding}}
As discussed above, the radiation from the line can severely degrade the performance of the superconducting quantum processor. The photon can propagate directly in the space besides via the lines, so the system must be protected by shields to block such radiation~\cite{Corcoles2011Protecting}. Standard dilution fridges are equipped by nested metal cans to block photons from outsides, while additional enclosures may be installed around the chip for further protection. To prevent radiation leakage, it is crucial to seal any gaps or seams in these metallic structures~\cite{Geerlings2013Improving,Nguyen2020Toward}.

External magnetic fields pose an additional environmental challenge. Macroscopic fields, originating from the geomagnetic field or nearby laboratory instrumentation, can introduce undesired magnetic flux into the superconducting qubit loops. $\mu$-metal, a material characterized by very high magnetic permeability, is widely employed to shield these fields and is often formed into customized enclosures surrounding the sample~\cite{Nguyen2020Toward,Bao2022Fluxonium,Zhang2021Universal}.

Cosmic rays and environmental radioactivity are ultimate sources of non-equilibrium quasiparticles in superconducting materials. High-energy particles, with energies exceeding the superconducting gap, can penetrate standard shielding, break Cooper pairs, and generate quasiparticles. Experiments show that ionization radiation from \ce{^64Cu} can deteriorate the performance of the qubits, and 10-cm-thick lead bricks outside the dilute fridge can reduce the external radiation and slightly increase $T_1$~\cite{Vepsalainen2020Impact}. The shielding can be further exaggerated by putting the whole instruments in a deep-underground facility under 1.4 km of granite to reduce the rate of the correlated QP burst by up to a factor of 30~\cite{Cardani2021Reducing}.  The decoherence limit imposed by cosmic rays in a typical lab environment (approx. 3 ms~\cite{Vepsalainen2020Impact}) is already approaching the current state-of-the-art $T_1$ times (ca. 1 ms~\cite{Somoroff2023Millisecond}). Even in the current large-scale chip with shorter coherence time (e.g. $T_1$ is lower than 100$\mu$s), this error could be significant as it is chip-wide that breaks the assumption of uncorrelated errors and thus halts the quantum error correction~\cite{McEwen2022Resolving}. Therefore, protecting systems from high-energy radiation is a critical task for the future of superconducting quantum computing.

\subsection{Multiplexing}
Unlike classical processors that operate on multi-bit words (e.g., bytes), quantum computing systems require individual qubit addressability~\cite{Krantz2019A}. In current physical implementations, this typically means each qubit has dedicated control cables connecting to ports on room-temperature electronic instruments. This one-to-one architecture presents a significant scalability bottleneck. The cables and electronics consume limited space in the dilute fridges and introduce a substantial thermal load to strain the cooling power, which is a fundamental limitation that hinders the expansion of the quantum system. Multiplexing is therefore an essential strategy for large-scale quantum computing. However, implementing multiplexing while preserving individual qubit addressability requires specialized component design.

All readout resonators multiplexed on one line must be separated in the frequency domain. A typical processor architecture with multiplexing readout is that each qubit is dispersively coupled to a unique readout resonator, and these resonators are attached to one transmission line, sharing the same cryogenic setup. This configuration requires that the frequencies of readout resonators are approximately uniformly distributed (or staggered) in a specific frequency range. A typical case involves 4 transmon qubits whose readout resonator frequencies lie within a 100 MHz range, with all resonators are coupled to a common filter resonator and a JPA~\cite{Jeffrey2014Fast}.  Similarly, a single 1.2 GHz wide readout channel for 5 transmon qubits has been demonstrated using a TWPA with high selectivity~\cite{Heinsoo2018Rapid}. Another architecture is using on-chip Josephson bifurcation amplifiers (JBA)~\cite{Vijay2009Invited,Manucharyan2007Microwave,Mallet2009Singleshot,Hao2010High} to replace resonators in conventional architectures, which provides better intrinsic readout fidelity and eliminates the need of an extra quantum-limit amplifier to improve the fidelity, thereby removing associated bandwidth limit. On the other hand, the on-chip JBAs significantly increase the complexity of the design and the fabrication. Various experiments of multiplexing readout are listed in~\Cref{table:multiplexing_examples}.

\begin{table}[!ht]
  \centering
  \begin{tabular}{ccc}
    \hline\hline
    Parallel Readout                               & Frequency Range & Setup                   \\
    \hline
    6 flux qubits~\cite{Jerger2012Frequency}        & 780MHz          & Resonators              \\
    4 transmon qubits~\cite{Jeffrey2014Fast}        & 100MHz          & Resonators, JPA 600MHz  \\
    4 transmon qubits~\cite{Schmitt2014Multiplexed} & 226MHz          & JBA                     \\
    5 transmon qubits~\cite{Heinsoo2018Rapid}       & 1.2GHz          & Resonators, TWPA 1.3GHz \\
    \hline
    \hline
  \end{tabular}
  \caption{Experiments of the multiplexing readout in superconducting processors. In these experiments, the readout frequency range is designed to avoid readout crosstalk and to meet the limitation of parametric amplifiers. We note in these experiments the maximum number of qubits that can be simultaneously readout is not that the number of qubits connected on the same line, as the used room temperature electronics bandwidth is also limited, which in principle can be improved by better equipment.}
  \label{table:multiplexing_examples}
\end{table}

\section{Technology Computer-Aided Design}
TCAD is a branch of the EDA that utilizes numerical simulation to analyze the device and fabrication process using the detailed descriptions of the materials and physical models that govern the behavior of devices and fabrication flows, enabling the prediction of the device properties before they have been manufactured. The goal of the TCAD tools is to reduce the costly and time-consuming experimental iterations in optimizing the performance of devices and the yield of fabrication processes.  It is worth noting that the TCAD tools in the current semiconductor industry are widely applicable to quantum computing. Here we focus on a unique component in superconducting quantum computing, the Josephson junction.

\subsection{Device TCAD}
The Josephson junction is a key component of superconducting quantum devices, which is usually made of an Al/\ce{AlO_{x}}/Al trilayer structure, since the aluminum oxide tunnel junction between aluminum leads has proven to be among the highest quality for superconducting qubits. To study the Josephson junction numerically, the microscopic simulation that goes beyond the lumped-element approximation is required. Ref.~\cite{Mizel2024Theory} develops a microscopic theory using self-consistent Bogoliubov-de Gennes equations and gauge transformations, maintaining consistency with lumped element theory while enabling new physical insights. In addition, the full quantum approach based on Bardeen–Cooper–Schrieffer(BCS) theory can not only simulate the Josephson junction, but also derive the effective superconducting circuit Hamiltonian from a microscopic fermionic model for interacting electrons~\cite{Liao2024Circuit}, establishing the connection between the microscopic and the macroscopic description of the qubit. 

To simulate a realistic Josephson junction with complex 3D geometry, Ref.~\cite{Pham2023Fluxbased} develops a computational approach to solve Maxwell's equations coupled with the nonlinear Schr\"odinger equation by using gauge-invariant flux fields and discrete exterior calculus (DEC), which is capable of capturing nonlinear responses and induced currents of Josephson junctions. 
The simulation results qualitatively agree with the experimental observation of the higher harmonics of the current-phase relation of Josephson junctions~\cite{Willsch2024Observation}.
This method synthesizes accurate low-energy quantum Hamiltonians and extends circuit quantum electrodynamics (cQED) to complex 3D systems, offering insights for designing devices while taking fabrication imperfections into account.

The junction barrier is a thin \ce{AlO_{x}} insulating layer which is only a few nanometers thick. The morphology of the \ce{AlO_{x}} layer and the Al/\ce{AlO_{x}} interfaces has a significant impact on the parameters of qubits and the coherence time.
Ref.~\cite{Kim2020A} utilizes density functional theory (DFT) to model the atomic structure of both amorphous and crystalline junctions, enabling the prediction of the critical current as a design parameter derived from ab initio simulations.
By employing molecular mechanics and DFT simulations, Ref.~\cite{DuBois2013Delocalized} shows that delocalized oxygen atoms in amorphous aluminum oxide create two-level systems (TLS) responsible for the decoherence. Simulation by DFT in a reasonably large atomic model can show the different electronic properties for different structures. For example, different kinds of defects like \ce{Al}/\ce{O} vacancy in the metal cap or the oxide, trapped atoms including \ce{H}, \ce{OH} and \ce{O2}, and other substitutions are studied in~\cite{Reshef2021Method,Qiu2024Manipulation}, while the thickness of \ce{Al2O3} layers and terminations are studied in~\cite{Shan2022Oterminated}. 

To model large amorphous \ce{AlOx} structures which are beyond the scale that DFT simulation is able to handle, the follow-up studies~\cite{DuBois2015Constructing, Zeng2016Atomic} develop atomistic models of junctions using molecular mechanics, DFT, and melt-quench simulations, providing a framework for optimizing barrier properties to minimize decoherence in superconducting qubits. Similarly, the simulated annealing method~\cite{Hertzberg2021Laserannealing} or the molecular dynamics~\cite{Bayros2024Influence} can be used to generate large random \ce{AlOx} atomic structures, of which the electronic structure and the current flow can be computed by atomistic semi-empirical methods, including density functional tight binding (DFTB) and non-equilibrium Green's function. 
This shows the variability of the atomic structure with stoichiometry and its effect on the current-voltage characteristics, resistance, and critical current, as well as the metallic conduction pathways, pinholes and weak links that dominate the charge transport~\cite{Lapham2022Computational,Bayros2024Influence}. To directly match the experiments including current-voltage curves and scanning transmission electron microscopy energy dispersive X-ray spectroscopy (STEM-EDS), the inhomogeneity of the junction can be investigated by the phenomenological Simmons model~\cite{Simmons1963Generalized}, which gives current-voltage properties based on the potential barrier height and its thickness, with Monte Carlo simulation of thickness, and it is found that a skewed distribution of the barrier thickness with standard deviation 15\%-20\% is compatible with experimental data~\cite{Kennedy2025Analysis}.

\subsection{Process TCAD}
A Josephson junction can be fabricated by the oxidation of a deposited Al thin film to form the insulating \ce{Al2O3} layer between two superconducting metal layers like \ce{Al} and \ce{Nb}. This oxidation process is critical to the thickness and  morphology of \ce{Al2O3} layers and \ce{Al2O3}-metal interfaces. This interface movement can be modeled by the level set method with FEMs based on logarithmic growth kinetics~\cite{Weingartner2021Modeling}. The movement speed depends on various physical quantities, including temperature, oxygen pressure, and \ce{Al}/\ce{Al2O3} properties, where the oxidation becomes harder to happen when the oxides thicken. Based on the process simulation further considering the metal layers, we can get the whole JJ structure and apply the device simulation techniques above to obtain the critical current, conductance, and capacitance of the JJ. This interface can also be investigated by molecular dynamics to understand the atomic level of structures, including voids and arrangements. The relationship among the aluminum deposition rate, the crystal orientations of the substrate, the oxidation depth, the stoichiometric \ce{Al}:\ce{O} ratio, the coordination number of atoms, and the number of atomic holes can be modeled by simulations of molecular dynamics of the process of oxidation and upper aluminum layer deposition on the aluminum substrate~\cite{Cyster2021Simulating, Han2023Computational}. Localized conduction channels can be observed in a junction of uniform thickness under specific conditions~\cite{Cyster2020Effect}.   Ab-initio grand canonical Monte Carlo can also be used to explore the formed interface structures and demonstrate the form of aluminum vacancies in the \ce{Al2O3/Al} interface, unlike \ce{ZrO2}/\ce{Zr} that tends to form interfacial oxide~\cite{Somjit2022Atomic}. The process simulation provides valuable guidance for fabrication experiments.

\subsection{What's next}
The development of TCAD tools is essential for enhancing the yield of fabrication, as well as the performance and reliability of devices. However, it presents significant challenges, primarily in correlating the microscopic simulations with actual device properties. To precisely predict device behaviors and optimize fabrication processes, accurate models for Josephson junctions, dielectrics, and other materials used in superconducting qubits should be developed, with empirical parameters extracted from characterization data of devices from extensive experiments in addition to numerical simulations.

\section{Physical-level Verification and Test}

In electronic design automation, besides the design of the functions, there are numerous issues that must be addressed, such as power efficiency, signal integrity, and so on. Similar issues also exist in the quantum design automation and should be treated with both classical physics and quantum physics. In this section, we will discuss several topics about handling these issues by verification in the design process and tests in the experiments.

\subsection{Crosstalk}
The crosstalk, representing unwanted interference from one component to another, is a pervasive issue in diverse systems. In recent error correction experiments on superconducting qubits, the crosstalk takes 11\% of the total error~\cite{Acharya2024Quantum}, ranking as the third largest among eight components. Crosstalk in superconducting quantum devices can be categorized into three types. 
\begin{itemize}
  \item \textbf{Layout crosstalk}. This refers to the unintentional influence of electromagnetic fields from a control line on non-target qubits. It arises from the diffuse nature of electromagnetic fields that typically decay as $E\propto r^{-2}$. In typical superconducting devices, the qubit-qubit separation often exceeds the qubit dimensions by only one order of magnitude or less. This distance translates to only several percent of the wavelength of the electromagnetic field around GHz. It means an external pulse applied on a drive line can significantly affect components other than the connected qubit. 
  \item \textbf{Circuit crosstalk}. When the qubit is floating, any external source applied on it can be reflected to other coupled qubits. This can be shown in the circuit analysis in~\Cref{sec:hamiltonian_derivation}, as the inversion from $\mathcal{C}$ to $\mathcal{C}^{-1}$ in the Hamiltonian and the construction of qubit modes by linear combination will hybridize the qubit capacitance and coupling capacitance from the layout together. Here we demonstrate the analysis on two floating transmon qubits $Q_1$ and $Q_2$, coupled capacitively, and a voltage source is attached on the $Q_1$ by capacitance $C_V$ where $C_V$ is much smaller than any other capacitance values. The circuit Hamiltonian includes
        \begin{align}
          \VEC{C_V}   & = \begin{pmatrix} C_V \\ 0 \\ 0 \\ -C_V\end{pmatrix}, \label{eq:2q_capacitive_circuit_cv_raw} \\
          \mathcal{C} & = \begin{pmatrix}
                            C_{g2}+C_{Q1} & -C_{g2}       & -C_{g2}       & -C_{g2}       \\
                            -C_{g2}       & C_{g2}+C_{Q2} & C_{g2}        & C_{g2}        \\
                            -C_{g2}       & C_{g2}        & C_{c,}+C_{g2} & C_{g2}        \\
                            -C_{g2}       & C_{g2}        & C_{g2}        & C_{g1}+C_{g2}
                          \end{pmatrix}. \label{eq:2q_capacitive_circuit_c_raw}
        \end{align}
        After the free model removal~\cite{Ding2021Freemode}, $\mathcal{C}^{-1}$, and $\VEC{C_V}$ are transformed to keep only two qubit modes by the transformation matrix $\VEC{W}$:
        \begin{align}
          (\VEC{W}\mathcal{C}\VEC{W}^T)^{-1} & = \frac{
          \begin{pmatrix}
            C_{g2}C_{Q2} + C_c C_{g2} + 2C_c C_{Q_2} & C_c C_{g2}                               \\
            C_c C_{g2}                               & C_{g2}C_{Q1} + C_c C_{g2} + 2C_c C_{Q_1}
          \end{pmatrix}
          }{2C_c C_{Q1} C_{Q2}+ C_{g2}C_{Q1}C_{Q2} + C_c C_{g2}C_{Q1} + C_c C_{g2}C_{Q2}} , \label{eq:2q_capacitive_circuit_cinv_free_removed} \\
          \VEC{W}\VEC{C_V}                   & = C_V\begin{pmatrix}
                                                      \frac{C_{g1}}{C_{g1}+C_2} \\
                                                      \frac{C_2}{C_{g1}+C_2}
                                                    \end{pmatrix}, \label{eq:2q_capacitive_circuit_cv_free_removed}
        \end{align}
        where $C_2=C_cC_{g2}/(C_c+C_{g2})=(C_c^{-1}+C_{g2}^{-1})^{-1}/2$. This shows that the distribution of the voltage source from $Q_1$ to $Q_2$ is determined by the ratio of $C_{g1}$ to half of the harmonic mean of $C_c$ and $C_{g2}$. Note that this does not happen if the external voltage source is connected to a grounded qubit, which can be treated as an individual block. However, grounding has its own disadvantage, precluding its universal application as a crosstalk mitigation strategy.

  \item \textbf{Hamiltonian crosstalk}. Coupling is inherent in multi-qubit systems. Due to state hybridization, a drive applied to one qubit inevitably affects the states of coupled qubits while also modifying the drive's effect on the target qubit itself. This manifests, for example, as a tilt in the rotation axis of single-qubit gates. Tunable coupler architectures can mitigate this effect by reducing unwanted qubit-qubit coupling to near-zero values. However, it is important to note that tunable couplers do not eliminate the previously discussed circuit crosstalk. Furthermore, residual crosstalk may persist even with tunable couplers, potentially originating from parasitic coupling involving higher energy levels within the transmon qubits.~\cite{Zhao2022Quantum}.
\end{itemize}

To deal with the crosstalk, we must characterize the crosstalk at each level and minimize their influence on the control of the qubits. The layout crosstalk and the circuit crosstalk are classical, while the Hamiltonian crosstalk is fundamentally quantum-mechanical. In practice, gate-level measurements often capture a conflation of these classical and quantum contributions. Therefore, specialized protocols are necessary to measure these two types of crosstalk independently. When it is not feasible, some types of crosstalk can be estimated from simulation, while others can be extracted from the measured overall crosstalk by subtracting the simulated ones.

Discrepancies between measured and simulated crosstalk values can be attributed to three main factors. First, crosstalk originates primarily from weak, unavoidable mutual capacitances or inductances, and accurately extracting these values requires computationally expensive electromagnetic simulations. Second, predicting the current distribution on the superconducting film is challenging, largely due to difficulties in accurately modeling an imperfect ground plane within the simulation. Third, a spurious mode that mediates coupling may be absent from the simulation due to an inappropriate or incomplete model setup. Consequently, reliable crosstalk estimation for chip design should integrate simulation data with empirical corrections derived from the analysis of measurement data.

\subsection{Defect inspection}
Following device fabrication, optical inspection is essential for identifying fabrication flaws. Chips that are free of detectable defects near functional patterns, such as qubits, resonators, and control lines, are selected for packaging and subsequent cryogenic measurement. For large-scale multi-qubit chips, employing an automated defect inspection tool can significantly
reduce labor costs and enhance inspection accuracy. Vision-based defect detection algorithms have been widely applied in the current semiconductor industry~\cite{Czimmermann2020VisualBased}, with applications in areas such as printed circuit board inspection~\cite{Zhou2023Review}. These techniques can be directly adapted for the inspection of quantum processor designs.

\subsection{Data-driven verification}
Extensive research has yielded a substantial collection of quantum processor designs validated by both simulation and experimental results. As previously discussed, simulating these devices is computationally intensive, making it highly advantageous to leverage this existing knowledge to accelerate the design workflow. For example, the SQuADDS~\cite{Shanto2024SQuADDS} dataset offers a collection of devices with simulation data benchmarked against published experiments. The dataset comprises diverse components, including various qubit types, 406 half-wave resonators, and 430 coupling capacitors. Furthermore, the SQuADDS can generate approximate device layouts based on user-specified parameters for qubits and resonators.

\subsection{Spectroscopy}
Spectroscopy is a fundamental experimental technique in quantum computing due to its simplicity and the comprehensive information it provides~\cite{Naghiloo2019Introduction}. In typical superconducting architectures, qubits do not output classical signals directly. Instead, measurements are performed by probing a readout resonator that is dispersively coupled to the qubit. Spectroscopy allows us to extract information from measurements before the signal can be classified as 0 or 1, and these kinds of results should be handled by the QDA workflow.
\begin{enumerate}
  \item One-tone spectroscopy: Drive the resonator alone and measure the transmission signal $S_{21}$ of the resonator versus the driving frequency or amplitude. When the driving is strong enough, the interaction overwhelms the qubit-resonator coupling, and the signal becomes irrelevant to the qubit. From this, we can obtain information about the bare resonator. When the driving is weak enough, the shift of the resonator by the qubit is observable, and one can see one or more peaks corresponding to the state $\ket{0},\ket{1}\cdots$ that depend on the thermal population. From one-tone spectroscopy, we can check whether the resonator works correctly, and determine the properties such as the resonator's $Q_i$, $Q_c$ and so on~\cite{Boehme2016Characterisation,Zmuidzinas2012Superconducting}.
  \item Two-tone spectroscopy: Drive the qubit and the resonator at the same time, and measure $S_{21}$ of the resonator. The resonator drive frequency is set to the value that is on resonance with the resonator when the qubit is $\ket{0}$. When the qubit drive is on resonance with the qubit, the qubit will leave $\ket{0}$, and the $S_{21}$ will change significantly. $\ket{1}-\ket{2}$ transition is also observable when the thermal population is large enough. From two-tone spectroscopy, we can determine the qubit frequencies, and further determine the qubit model by adjusting the qubit parameters. For example, the $E_C$, $E_J$, $E_L$ parameters of fluxonium qubits can be determined by two-tone spectroscopy with scanning $\Phiext$ for~\cite{Nguyen2019HighCoherence}.
\end{enumerate}
The spectroscopy analysis presents challenges, as spectra can be complex, potentially containing confounding signals such as sideband transitions~\cite{Wallraff2007Sideband} that complicate fitting algorithms. Furthermore, achieving required high-precision often demands a large number of data points. To enhance efficiency, adaptive measurement protocols can significantly reduce the data required and accelerate the measurement process~\cite{Nijholt2019textitAdaptive,Rol2016DiCarloLabDelftPycQEDpy3}. For the analysis, tools such as SuperGrad~\cite{Wang2025SuperGrad} accelerate the fitting procedure by implementing automatic differentiation in the computation of the objective function. Additionally, modern machine learning methods can be applied to rapidly provide highly accurate initial parameter estimates; for example, a model developed to estimate fluxonium $E_C/E_J/E_L$ parameters achieved an average accuracy of 95.6\%~\cite{Kung2025Automatic}.

\subsection{Pulse distortion correction}

As discussed in the cryogenic setup section, the multiple components in the input line from the electronics to the quantum processors can distort the pulse shape. Without correction, especially of the DC flux pulse, it will fail to produce the expected results~\cite{Rol2019Fast,Sung2021Realization}. Cryoscope~\cite{Rol2019Timedomain,Akesson2022Correcting} is an important technique to characterize the distortion. It utilizes Ramsey-style experiments on the qubit with non-linear flux dependence to map the phase at discrete time points throughout the pulse back to the pulse amplitudes. Once the pulse distortion is known, it is possible to apply infinite impulse response (IIR) filters and finite impulse response (FIR) filters to the flux pulse to correct the distortion. This procedure can be iterated to optimize the filters. With these real-time digital filters, we can perform fast and high-fidelity operations better on quantum processors with flux pulses. For flux pulses that are applied to qubits without readout resonators, which are often used as tunable couplers, we can measure their coupled neighbors with similar techniques to obtain the predistortion parameters for high-fidelity two-qubit gates~\cite{Li2025Highprecision}.

\subsection{Parameters correction}
Device parameters invariably exhibit fluctuations stemming from fabrication process variability. These fluctuations can include systematic errors, which are amenable to correction through measurement and compensation, often guided by empirical formulas.
For example, the capacitance per area of Josephson junctions (specific capacitance, $C_s$) is an important design parameter of superconducting quantum processors. The accurate circuit quantization process requires precise capacitance values of all components. The junction capacitance is not negligible, and a systematic error will be introduced if it is treated as zero. Furthermore, the Josephson junction capacitance per area is strongly related to the \ce{AlO_x} fabrication condition, as the \ce{AlO_x} structure of sub-nanometer thickness is very sensitive to the oxidation process. $C_s$ can be determined from the current-voltage relationship in low-temperature experiments and scanning electron microscope images of junctions. The large range of $C_s$ has been demonstrated in measurements, including $100\pm 25\text{fF/}\mu\text{m}^2$ for Al/\ce{AlO_x}/Al~\cite{Deppe2004Determination},  $46 \text{fF/}\mu{m}^2$~\cite{Xiong2017Development}, $33\sim 50\text{fF/}\mu{m}^2$~\cite{Xiong2018Measurement} and $40\sim 70\text{fF/}\mu{m}^2$ for Nb/Al-\ce{AlO_x}/Nb~\cite{Maezawa1995Specific}. Therefore, an iterative feedback loop between experimental characterization and design refinement is essential to calibrate the specific junction capacitance $C_s$. This process ensures that subsequent designs incorporate accurate qubit capacitance values.

\subsection{Noise extraction \label{subsec:noise_extraction}}
As discussed in \Cref{sec:decoherence}, the decoherence is one of the key factors in superconducting quantum computing, and accurate characterization of the decoherence effect is essential for improving the design of qubits. The most straightforward metric is the $T_1$ and $T_2$ of qubits, which are often reported in the characterization of large quantum processors~\cite{Arute2019Quantum, Acharya2024Quantum, Zhao2022Realization,Gao2025Establishing}. However, coherence times alone do not provide a complete description of decoherence. Given the complexity rooted in the nature of decoherence, numerous methods have been developed to discern the corresponding channels.

\subsubsection{Dielectric loss and TLSs}
TLSs play an important role in the decoherence of solid-state-based qubits, and extensive work has focused on their characterization~\cite{Muller2019Towards}. The interaction between the TLSs and the qubit is observable as an avoided crossing in the spectroscopy with tuning qubit frequency. The density and splitting of TLSs in the frequency domain reveal that they agree with the model that TLSs are impacted by the electric field of the qubit, suggesting that materials with smaller dielectric loss tangent are preferable for achieving high coherence time~\cite{Martinis2005Decoherence}. This model has been subsequently verified, establishing the reduction of dielectric loss and TLS effects (through material selection and design) as a central principle in qubit development. The frequency distribution of TLSs can also be probed by the all-microwave ac-Stark shifted technique on fixed frequency qubits, and it is shown to be directly correlated with the average $T_1$ over long time~\cite{Carroll2022Dynamics}. At low frequencies, TLS-qubit coupling is observed to weaken, which suggests a design pathway toward improving qubit stability~\cite{Sun2023Characterization}. The analysis of the dielectric loss allows us to optimize the fabrication process combining with other material characterization techniques~\cite{Chayanun2024Characterization} and the layout for electric field distribution~\cite{Deng2023Titanium,Ganjam2023Improving}.

The time fluctuation of TLSs in the frequency domain can be measured to probe their dynamics~\cite{Burnett2019Decoherence,Klimov2018Fluctuations,Bejanin2021Interacting}. From the dynamics, it can be seen that the fluctuation of $T_1$ of qubits is dominated by fluctuating TLSs. And TLSs can be classified into two categories, one with telegraphic spectral diffusion and another with diffusive spectral diffusion. Both can be understood in the interacting defect model. This information is useful to identify the qubit quality and guide the calibration and control. Some TLSs have also been correlated with trapped quasiparticles~\cite{deGraaf2020Twolevel}.

More detailed spacial distribution, energy distribution, electric dipole moments and coherence times of TLSs can be studied by applying an electric field on the qubits~\cite{Grabovskij2012Strain,Lisenfeld2016Decoherence,Lisenfeld2019Electric,Bilmes2020Resolving,Bilmes2022Probing,Lisenfeld2015Observation}. This reveals that the majority of the TLSs interacting with qubits are in the naturally formed \ce{Al2O3} layer on the metal-air interface and another large part comes from the junction oxides, which gives a good optimization target in the design concerning the electric field. 

\subsubsection{Flux noise}
The flux noise typically exhibits $1/f$ behavior in common qubits~\cite{Yoshihara2006Decoherence,Koch2007Model,Yan2012Spectroscopy,Anton2012Pure,Bylander2011Noise,Yan2013Rotatingframe,Quintana2017Observation,Braumuller2020Characterizing}. The PSD of frequency fluctuation contains rich information about the dephasing noise. Because flux noise is the dominant dephasing mechanism for flux-tunable qubits (when biased away from flux-frustration points), PSD analysis is a critical tool for its characterization. Techniques such as CPMG and spin-locking are widely used to measure the PSD in the MHz frequency range~\cite{Bylander2011Noise,Yan2013Rotatingframe}, while long-time sampling of the measurement results can be utilized for ultra-low frequencies (in the Hz range)~\cite{Quintana2017Observation}. These data allow for the extraction of the noise type, the $1/f$ noise amplitude, and the exponent $\alpha$ (in $1/f^{\alpha}$), facilitating an understanding of the noise's correlation with other factors. 

The flux noise also plays an important role in devices made from disordered superconducting materials~\cite{Grunhaupt2019Granular,Maleeva2018Circuit,Kristen2023Random,Kristen2023Observation,Rieger2023Granular,Hazard2019Nanowire}, and the study of the noise greatly helps to understand the potential of these new materials in superconducting qubits. For instance, in devices made from granular aluminum, the frequency fluctuation PSD exhibits a distinct Lorentzian shape, distinguishing it from the noise profile of conventional materials~\cite{Kristen2023Random}. Similarly, noise analysis in qubits fabricated from \ce{Ti_xAl_yN} indicates that the flux noise scales with the material's volume rather than its surface area~\cite{Gao2025The}.
Meta-analysis of multiple disordered superconducting materials suggests that the volume scaling law is universal and may have originated from additional non-equilibrium quasiparticles induced by disorder\cite{Charpentier2025Universal}.

\section{Quantum Instruction Set Design}\label{sec:qisa}

\subsection{Overview of the logic-level QDA}
In the previous sections, we have outlined a comprehensive workflow for designing quantum processors. This workflow integrates chip layout and control schemes---detailed in~\Cref{sec:chip_design} and~\Cref{sec:control} respectively---thereby enabling the high-fidelity execution of designed quantum operations. Yet, this does not mark the conclusion of our design narrative---rather, building upon these designed quantum operations, we now enter a new domain defined by immense design freedom: \textit{logic-level design}. This encompasses the development of quantum instruction sets, the formulation and implementation of quantum algorithms, the design of quantum error-correcting codes and fault-tolerant schemes, and, more broadly, methodologies for utilizing the quantum devices engineered using the approaches outlined in the preceding sections.

In this and subsequent sections, we will focus on explaining the general workflow and design principles at the logical level. These materials can function independently of physical-level materials, offering an overview of QDA from a computer science perspective. In fact, this is precisely the type of approach that most quantum design automation literature refers to in the computer science community~\cite{willeDesignAutomationQuantum2024}.

While the bottom-up approach employed to organize the material in this paper equips readers with sufficient information to implement quantum algorithms, directly exposing electronic control primitives to quantum processor users would be impractical for those without deep expertise in experimental physics or electronics. Instead, users with backgrounds in quantum computing or quantum algorithms would benefit from a more accessible framework---enabled through an abstraction layer called a \textit{quantum instruction set}---allowing them to implement quantum computing tasks via quantum circuits or high-level quantum programming languages.

From this perspective, we then construct several critical building blocks that form the logic-level QDA stack. These span a broad spectrum---from low-level hardware-software interfaces to high-level program transformation routines, to name just a few key components:

\begin{itemize}
    \item \textbf{Quantum instruction set architecture (Quantum ISA, or QISA)}. This part serves as the interface between the physical-level and logic-level halves of QDA: the physical-level QDA provides a way to execute operations defined in the quantum instruction set, while the logic-level QDA finds a way to transform into the chosen quantum instruction set. Typically, the quantum instruction set includes qubit initialization, a universal gate set, and measurement. The universal gate set is the key component of a quantum ISA that dominates hardware-implementation accuracy and cost, as well as software expressivity for quantum program execution. Thus, in the narrow sense, quantum ISA often refers to the natively supported universal gate set.
    \item \textbf{Quantum circuit synthesis}. This routine, also dubbed compilation, transforms a high-level description or building block of a quantum program to a lower-level one that can be composed of smaller building blocks, like primitive unitary gates or quantum operators, opening up opportunities for further analysis, optimization and implementation of the quantum program.
    \item \textbf{Quantum circuit optimization}. This routine performs compiler optimization on a quantum program, producing an equivalent quantum program that is ``better'' than the input program. The notion of ``better'' can be assessed in different ways: circuit depth, gate count, count of a specific kind of gate, etc.
    \item \textbf{Code generation}. This routine implements the optimized circuit with the quantum instruction set, notably:
    \begin{itemize}
        \item \textbf{Qubit mapping and routing}. Since we are mostly working with superconducting architecture throughout this paper, one natural requirement for defining the quantum ISA is to only allow operations on physically adjacent qubits, usually inducing an interaction graph between physical qubits. To resolve the backend topology constraint, there requires qubit mapping/placement that maps logical qubits to physical qubits, and qubit routing based on inserting SWAP gates that exchanges state subspaces of two operand qubits, such that non-adjacent logical qubit states can be moved next to each other. Qubit mapping and routing can be modeled uniformly as routing refers to dynamic qubit mapping via SWAP insertion that alerts intermediate logical-to-physical mapping.
        \item \textbf{ISA rebase}. This converts the quantum gates within an optimized quantum circuit to the natively implemented gate set. For instance, two-qubit gates (or two-qubit subcircuits) are decomposed into subcircuits consisting of native CZ/iSWAP gates and arbitrary single-qubit gates. Arbitrary single-qubit gates are further represented as a sequence of continuous phase-shifted virtual-$ Z $ operations and fixed $R_x$ rotations.
    \end{itemize}
    \item \textbf{Circuit simulation}. To check if a quantum algorithm is implemented correctly, it is useful to simulate the quantum computing task on a classical computer. While there is an exponential cost of simulating a general quantum program on a classical computer, it remains meaningful to reduce simulation overhead to an acceptable extent for simulation of larger small-scale quantum algorithms.
\end{itemize}

While some may contend that quantum circuit synthesis is unrelated to quantum chip design, we decide to include a detailed discussion of this topic for the following reasons:

\begin{itemize}
\item By far, practical quantum computing on realistic quantum hardware requires close connection and co-optimization between software and hardware: the quantum program needs to be tailored to a specific hardware, while the design of quantum hardware may still vary according to software need (for example, the quantum annealing machine, or quantum error correction codes).
\item Many researchers have recognized the quantum analog of logic synthesis---one of the core components in classical EDA---as a key challenge in advancing quantum EDA. This perspective holds merit, as quantum program synthesis, together with the subsequent transpilation process, forms a crucial layer that transforms an abstract quantum program to meet the physical constraints and architectural characteristics of a specific target quantum device.
\end{itemize}

It is worth noting that while the logic-level QDA routines outlined above provide a rough sketch of a workflow with a logical transformation sequence, real-world quantum program compilers often do not adhere strictly to this structure. Specifically, individual transformation routines may be reordered relative to one another, repeated at various stages, or merged with other transformations---all depending on the overarching design objectives. For example:
\begin{itemize}
    \item Multiple levels of intermediate representations (IRs) may be adopted during compilation; therefore, multiple rounds of synthesis-optimization cycles are required. A famous intermediate representation is the ZX-calculus~\cite{Kissinger2020Reducing}, where the circuit is first ``synthesized'' into a ZX graph, followed by ZX-calculus optimizations~\cite{Duncan2020Graphtheoretic}, and then ``extracted''~\cite{backensThereBackAgain2021} back into a circuit before gate-based optimizations. 
    \item Post-mapping optimization~\cite{wuQGoScalableQuantum2022} may require circuit-level or even pulse-level optimization passes to be inserted after the qubit mapping/routing and ISA rebase steps to reduce circuit cost as a final step.
\end{itemize}

\subsection{The role of quantum instruction sets}

We will begin our exploration of logic-level QDA by first introducing quantum instruction sets. In classical computing systems, the Instruction Set Architecture (ISA) functions as the interface between hardware and software: it defines a set of basic supported operations for computation; hardware is engineered to execute programs written in accordance with the ISA directly; and the software stack focuses on compiling computer algorithms and programs into ISA-compliant form.

Classical ISAs have been developed with varying design objectives, the most prominent distinction being between CISC (Complex Instruction Set Computer) and RISC (Reduced Instruction Set Computer) architectures. CISC ISA (such as x86) incorporates coarse-grained operations, which simplify manual ISA program writing and reduce the total number of operations---though this comes at the expense of more complex hardware circuitry. In contrast, RISC ISAs (including ARM and RISC-V) utilize finer-grained operations, enabling simpler hardware implementations at the cost of complicated compilation and optimization processes.

Similar to their classical counterparts, quantum computers also rely on an instruction set to define the fundamental operations performable on qubits.
Quantum hardware vendors provide access to this set, either through a cloud interface or a fully deployed device. This access typically comprises two main categories of operations. The first is a universal \textit{native gate set} of unitary operations, such as $R_z(\theta)$, $R_x(\frac{\pi}{2})$, and CZ (or iSWAP) gates, which are common for superconducting systems~\cite{Krantz2019A}. 
The second includes essential non-unitary operations such as Measurement and Reset.
The physical design and implementation of these operations are detailed in~\Cref{sec:control}. 

There is also a fundamental distinction between quantum and classical processors, which lies in the implementation of computational logic.
Classical CPU instruction sets are typically implemented as specialized, hardwired circuits on values copied from register files, whereas in quantum processors, qubits are integrated as the core physical units, while the logic layer---realized as quantum operations---is dynamically encoded through tunable control waveforms. These waveforms, parameterized by attributes such as amplitude, phase, and duration, enable precise modulation of quantum states, representing a paradigm shift in hardware-software co-design and dynamic configurability.

\subsection{Considerations in quantum instruction set design}

First and foremost, a quantum instruction set must enable universal quantum computing. As such, its design must comply with specific requirements underpinned by rigorous theoretical guaranties. By way of comparison, in the classical computing paradigm, the XOR and OR operations suffice to implement any form of binary logic. In contrast, general quantum operations are typically realized through a small set of precisely calibrated pulse sequences---often a fixed, minimal number of them.
For this reason, the concept of universality is more commonly articulated through the so-called Solovay-Kitaev theorem: this theorem states that, for a finite set of $m$ instructions from $\mathrm{SU}(d)$, if the instruction set can approximate any gate (i.e., generate a dense subset) in $\mathrm{SU}(d)$, the approximation can always be done efficiently: any unitary operator $U \in \mathrm{SU}(d)$ can be approximated to a specified desired accuracy $\epsilon$ by a finite sequence of $O\left(m\log^{3.97}\left(\frac{m}{\epsilon}\right)\right)$ instructions~\cite{dawson2005solovay}. Moreover, such a sequence can be generated by a classical computer in a running time that also scales as $O\left(m\log^{2.71}\left(\frac{m}{\epsilon}\right)\right)$. The exponent bound on the number of instructions has only recently been refined, improving from $\log_{\frac{3}{2}} 5 \approx 3.97$ to $\log_{\frac{1+\sqrt{5}}{2}} (2) \approx 1.44$~\cite{kuperberg2023breaking}.
The Solovay-Kitaev theorem underpins the practical feasibility of universal quantum computation by bridging abstract quantum operations and physically realizable gate sequences. Moreover, as quantum error-correcting codes and fault-tolerant schemes enable the scalability of encoding logical qubits with arbitrary precision by utilizing redundant physical qubits, they naturally align with the guaranties of the Solovay-Kitaev theorem, ensuring the capability to run any quantum program in a fault-tolerant manner with arbitrarily desired accuracy. 

We will delve further into this line of research in~\Cref{sec:qec} on the design of quantum error-correcting codes and fault-tolerant schemes. For the present, we focus on quantum instruction sets as the fundamental building blocks for quantum algorithms and error correction codes from a bare-metal perspective.
One major concern regarding quantum instruction set design is whether to support the exact implementation of continuous quantum gates. Given that any quantum instruction inherently carries inaccuracies stemming from design and fabrication, \textit{exact universality} at the logic level becomes particularly advantageous compared with approximating continuous gates with discrete ones.
While it is widely known that the continuous set of all single-qubit operations, when combined with any entangling two-qubit gate, suffices to achieve exact universality---that is, they can be used to implement any arbitrary $n$-qubit unitary transformation~\cite{harrow2008exact}, the scenario of exact universality is not straightforward. One might question: if calibrations are required for any designed instructions, how can we support a continuous instruction set such as all single-qubit rotations? Fortunately, the widely adopted virtual-$Z$ technique enables flexible adjustment of arbitrary Z-rotation angles; notably, the phase shifts of the microwave drives can thus be treated as having zero duration and arbitrary precision.

Secondly, quantum instruction sets, much like their classical equivalents, possess a dual nature. They strike a balance between high-precision physical execution and efficient representational abstraction---a consideration that is frequently overlooked in the context of quantum control, as examined in~\Cref{sec:control}. The design of more effective quantum instruction sets remains an active area of research, with several approaches explored: deploying continuous quantum instruction sets such as $\textrm{fSim}$~\cite{Foxen2020Demonstrating} or $\textrm{XY}$~\cite{Abrams2020Implementation} to enhance expressivity; replacing $\textrm{iSWAP}$ with its square root~\cite{Huang2023Quantum}, which offers higher precision and improved expressivity; and directly implementing $\textrm{SU}(4)$ modulo single-qubit rotations, a strategy formally referred to as the AshN scheme~\cite{Chen2024One}---this approach aims to achieve ultimate quantum gate expressivity while enhancing control precision.

Thirdly and importantly, quantum instruction set architectures---like their classical counterparts---are not merely design constructs between hardware and software; a quantum instruction set requires a supporting ecosystem around it, specifically, the pipeline we refer to as \textbf{4C}, encompassing \underline{C}ontrol, \underline{C}alibration, \underline{C}haracterization, and \underline{C}ompilation. Each phase in the \textbf{4C} pipeline demands careful design of the scheme. For instance, just a few years ago, the most widely adopted characterization scheme was randomized benchmarking, which is only capable of handling Clifford operators. However, as noted in~\cite{harrow2008exact}, non-Clifford yet entangling operations---though unsupported by randomized benchmarking---can still enable universal computing when supplemented with single-qubit rotations.
As a result, in earlier days, the exploration of non-Clifford quantum instructions faced a key challenge: their fidelity could not be characterized truthfully. Instead, researchers typically relied on tomography-based approaches, which are sensitive to SPAM (state preparation and measurement) errors. It is only recently that cross-entropy benchmarking has been introduced, emerging as a new mainstream approach and accelerating the development of novel quantum instruction sets.

\subsection{Quantum control instruction sets}\label{subsec:qcis}

In this paper, we use the term ``quantum instruction set'' to refer to the commands executed on quantum processors---direct analogs of classical machine instructions.

However, as noted earlier, quantum processors are typically integrated with or controlled by classical computers via specialized control electronics. Scenarios such as variational quantum algorithms, quantum error correction, and numerous others involve not only quantum operations but also classical operations---often with tight interdependencies between the two. Furthermore, to seamlessly integrate quantum processors with classical systems (CPU, FPGA, or even GPUs), a growing trend has emerged~\cite{fuExperimentalMicroarchitectureSuperconducting2017,fuEQASMExecutableQuantum2019a,zhouHiMAHierarchicalQuantum2024,zhangClassicalArchitectureDigital2024,zettles262DesignConsiderations2022}: replacing standalone classical computers with classical coprocessors. The dedicated instruction sets of these coprocessors demand careful design to deliver meaningful performance gains; we refer to the instructions executed on such classical coprocessors as a ``quantum control instruction set''. 

It should be noted that a discerning reader will recognize that the term ``quantum instruction set'' in the literature can encompass different concepts~\cite{fuExperimentalMicroarchitectureSuperconducting2017,fuEQASMExecutableQuantum2019a,zhouHiMAHierarchicalQuantum2024,zhangClassicalArchitectureDigital2024,zettles262DesignConsiderations2022}: these pieces of literature refer to the instructions executed on classical coprocessors belonging to a ``quantum instruction set'', which is, in fact, closer to the ``quantum control instruction set'' by our definition.

Here, we wish to emphasize that the quantum control instruction set (QCIS) is highly dependent on three core factors: the design of the quantum chip itself (a single chip or interconnected chips), the integration scheme of control electronics with the quantum chip, and the interface with the classical coprocessor. A key challenge in implementing the microarchitecture for a QCIS lies in its scalability: a single controller will likely struggle to manage up to millions of qubits---an limitation that becomes even more pronounced when accounting for the demands of both near-future and long-term quantum computing scenarios. Specifically, these scenarios will require syndrome measurements (a critical component in fault-tolerant quantum computing, which we will discuss in~\Cref{sec:qec}), where two resource-intensive operations are prevalent: (1) gate operations that act on all qubits (global gates) and (2) classical feedback control, dynamically driven by measurement outcomes.

The design of the quantum system---specifically, the classical architecture, which refers to how classical components are organized within a quantum computer (to be distinguished from the quantum architecture, which governs the organization of quantum chips)---is another critical area; yet, it remains insufficiently mature to be addressed in a systematic manner within this paper. Here, we merely highlight a few notable efforts in this domain.

eQASM represents one of the earliest attempts to develop a coprocessor tightly integrated with a quantum system~\cite{fuExperimentalMicroarchitectureSuperconducting2017, fuEQASMExecutableQuantum2019a}. While its design primarily targets scenarios of the NISQ era, the core concept behind it has since evolved into a mainstream approach in the field.

IBM has adopted a distributed architecture for quantum control, as outlined in~\cite{zettles262DesignConsiderations2022}. This framework consists of a central hub node and multiple dedicated qubit controller nodes, with a clear division of responsibilities:
\begin{enumerate}
\item  The central hub, powered by a general-purpose PowerPC processor, acts as the computation coordinator. Its key functions include receiving qubit measurement outcomes, executing classical computing tasks, and broadcasting control results to all qubit controller nodes.
\item Each qubit controller node directly manages a subset of qubits. Beyond sending measurement data to the central hub and receiving classical control information in return, these nodes are equipped with a specialized processor. This processor features a custom ISA that incorporates a comprehensive suite of arithmetic, branch, and storage instructions---enabling it to handle complex control flow for quantum operations.
\end{enumerate}

Alibaba~\cite{zhangClassicalArchitectureDigital2024} has adopted a comparable distributed architecture, with a primary focus on the central hub---a RISC-V processor tasked with executing classical instructions (such as syndrome decoding logic) and broadcasting pulse-level commands to PXIe-connected in-house control electronics. These control electronics directly control the same quantum chip, forming a streamlined control chain. A key advantage of this architecture is its ability to reduce instruction issuance and transmission costs to constant values, meaning these costs do not scale with the number of qubits. Notably, this efficiency is achieved without introducing additional overhead in decoding or dispatching processes. To enhance performance further, the multi-core RISC-V processor is dedicated to processing syndrome information via a sliding-window approach, thereby providing sufficient decoding capacity to support logical operations.

All these efforts are built on room-temperature control electronics, and while they represent early steps in quantum control instruction set design, they currently operate at the scale of only dozens of qubits. As quantum systems grow in size, we can anticipate that the associated challenges will become increasingly pronounced. Meanwhile, it is evident that current room-temperature control electronics may emerge as a critical bottleneck as system scales expand. For large-scale quantum systems, Single-Flux-Quantum (SFQ) technology~\cite{mukhanov2011energy,mcdermott2018quantum} or cryo-CMOS~\cite{charbon2016cryo,patra2017cryo,sebastiano2017cryo} technology is likely to be employed in the near future. Designing architectures that accommodate the stringent cooling power constraints of these new technologies will undoubtedly open an exciting new frontier in research.

\subsection{What's next}

Once a quantum device is equipped with a dedicated instruction set, it becomes capable of executing quantum computing tasks. Yet, a critical gap remains between quantum computing tasks articulated in the quantum ISA and those written by users---analogous to the divide between a classical CPU's machine code and the high-level programs crafted by users. To bridge this gap, a sequence of steps involving synthesis, optimization, and code generation is essential. These processes transform the user's program into a format that the quantum microarchitecture can both recognize and execute with optimal efficiency.

We aim to illustrate, through the work presented in this section, that significant flexibility exists in instruction set design during quantum computer development---and that with robust ecosystem support, this flexibility can yield substantial benefits.
Considerable efforts have been dedicated to exploring non-traditional quantum instruction sets. Beyond the anticipated advantages for Noisy Intermediate-Scale Quantum (NISQ) applications, recent proposals demonstrate benefits across multiple domains: iSWAP-based surface code implementations~\cite{eickbusch2024demonstrating}, quantum low-density parity-check (qLDPC) codes~\cite{geher2025directional}, and Complex Instruction Set Computing (CISC)-style approaches that combine CX and iSWAP-based surface code implementations~\cite{zhou2024halmaroutingbasedtechniquedefect, geher2025directional}. The exploration of novel quantum instruction sets---along with their contributions to enhancing system performance---will, in turn, sustain and advance this momentum.

\section{Quantum Circuit Synthesis and Optimization}\label{sec:circuit}

In classical Electronic Design Automation (EDA), logic synthesis is the pivotal step that translates high-level hardware descriptions (e.g., Verilog) into low-level physical implementations (e.g., ASICs or FPGAs). Its core mission is to optimize the resulting circuit against a spectrum of real-world constraints, including logic complexity, timing, power consumption, and manufacturability.

The QDA community has explored a similar synthesis concept, but its applicability is not straightforward due to fundamental differences between classical and quantum hardware. Unlike classical logic, which is hardwired, the quantum instruction sets on platforms like superconducting qubits are realized via dynamic microwave pulses. This removes many of the spatial constraints of classical hardware but introduces a far more critical limitation: extreme sensitivity to noise. Consequently, minimizing the length of gate sequences to reduce error accumulation becomes the paramount concern for quantum synthesis.

Against this backdrop, we define quantum circuit synthesis as the process of translating abstract quantum algorithms into concrete, hardware-executable gate sequences. It serves as the essential bridge between theoretical algorithms and their physical realization. Leveraging the native primitives of a given quantum device, synthesis aims to construct circuits that are not only logically correct but also optimized for noise resilience---primarily by minimizing circuit depth and gate count to mitigate the effects of decoherence.

\textbf{Quantum circuit synthesis and optimization} encompass all transformations that effectively convert higher-level representations of quantum algorithms into lower-level hardware primitives. It is important to note that, based on different design objectives, there is significant flexibility in selecting both the higher-level and lower-level descriptions.

\begin{enumerate}
\item \textit{High-level Quantum Program Representations} are the inputs to the synthesis process. While they can originate from high-level languages, from a QDA perspective, the primary input is a quantum circuit that is not yet optimized for a specific hardware target. This can include:
\begin{enumerate}
    \item \textit{Logical Quantum Circuit Fragments}: The most common input is a logical circuit composed of gates that are not directly supported by the target hardware. These circuits are often inefficiently structured and can feature complex operations, ranging from multi-qubit gates such as Toffoli and CSWAP to nested constructs, including controlled or adjoint sub-circuits.
    \item \textit{Abstract Unitary Descriptions}: In some cases, a circuit block may be described by its mathematical function, such as a target unitary matrix or a Hamiltonian evolution, which must be synthesized into a concrete gate sequence.
\end{enumerate}

\item \textit{Low-level Hardware Primitives} are the fundamental building blocks that form the ``machine code'' of quantum circuits, varying based on the target hardware or design constraints. Examples include:
\begin{enumerate}
    \item \textit{Hardware-specific controls} are pulses or electronic control operations directly implemented in the quantum-classical control interfaces, that is, components of the quantum control instruction set introduced in~\Cref{subsec:qcis}.
    \item \textit{Native quantum gates} include single-qubit gates (Pauli-X, Y, Z, phase gates) and two-qubit gates (CZ, iSWAP) that are directly engineered and calibrated on a quantum processor.
    \item \textit{FTQC primitives} are those tailored to fault-tolerant architectures, such as magic-state distillation components or lattice surgery\cite{Litinski2019gameofsurfacecodes}. Terminating compilation at the FTQC primitive layer opens up freedom for the underlying design of QEC and FTQC schemes\cite{Watkins2024highperformance, beverlandAssessingRequirementsScale2022}.
\end{enumerate}
\end{enumerate} 

\begin{figure}
    \centering
    \includegraphics[width=\linewidth]{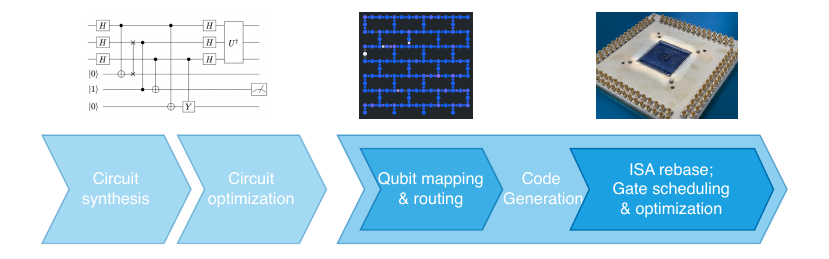}
    \caption{A typical Workflow for Quantum Circuit Synthesis and Transpilation. The circuit synthesis and optimization stages aim to decompose circuits into 1Q and 2Q gates, with the goal of minimizing either the total gate count or circuit depth. The qubit mapping and routing stage bears similarities to register allocation and instruction scheduling in traditional computing; its core objective is to reduce routing overhead---specifically, the number of SWAP gates inserted. The final stage, known as the ISA rebase, gate scheduling, and optimization phase, involves translating the processed circuit into pulse sequences. Here, the primary desired outcome is minimizing the total pulse duration. Note that in some quantum SDKs such as Qiskit, the synthesis procedure may be considered as part of the transpilation process.}
    \label{fig:compilation-workflow}
\end{figure}

We will introduce quantum circuit synthesis from three key aspects, and the details of these aspects are illustrated in~\Cref{fig:compilation-workflow}:
\begin{enumerate}
  \item \textbf{Circuit synthesis}, where higher-level constructs in the quantum algorithm are synthesized into a lower-level gate-based quantum circuit.\footnote{Another area of research centers on the synthesis of quantum state preparation. However, we do not survey this topic in the present paper, as our primary goal is to illustrate a comprehensive workflow rather than provide an overview of exhaustive research themes.}
  \item \textbf{Circuit optimization}, wherein the aforementioned quantum circuit is further transformed and optimized to reduce execution overhead.
  \item \textbf{Code generation}, in which the quantum circuit is adapted to a given quantum hardware architecture, considers the constraints of physical-qubit connectivity and the native gate set.
\end{enumerate}

It is important to note that:

\begin{itemize}
  \item While we divide quantum circuit synthesis into three top-down steps, in real quantum compilers, these steps may interleave freely. For example, in~\Cref{subsubsec:beyond} we will show optimizations based on aggregating gates and performing resynthesis; moreover, it is natural to insert an optimization pass before and after mapping to hardware (rebasing and post-mapping optimization, correspondingly), as long as the circuit is still valid on hardware after optimization.
  \item Determining the optimal implementation for an arbitrary quantum circuit is a QMA-hard problem~\cite{Janzing2005}; this complexity classification persists even when the synthesis target is extended to arbitrary quantum gates. This computational hardness carries a critical implication: in the vast majority of practical scenarios, identifying the optimal solution for quantum circuit synthesis remains computationally intractable.
  \item Identifying robust higher-level quantum descriptions that balance human interpretability and syntactic tractability remains a non-trivial challenge.
\end{itemize}

Against this backdrop, we will discuss a collection of quantum circuit synthesis and transformation approaches tailored for distinct abstraction levels, whose core value and applicable logic have only gradually been understood and clarified within the quantum computing community over the past decade.

Notably, while both quantum compiler and QDA operate on quantum circuits, their scopes differ fundamentally. QDA sits at the interface between compilers and physical realization, focusing primarily on circuit-level design automation tasks such as synthesis, optimization, layout, and verification. These processes treat the quantum circuit as the design unit---analogous to the gate-level representation in classical EDA---and aim to optimize metrics such as depth, gate count, and fidelity under hardware constraints. In contrast, quantum compilers span higher abstraction layers, bridging programming languages, intermediate representations (IRs), and circuits. Compiler optimizations typically address high-level program semantics (e.g., loop transformations, uncomputation strategies, or symbolic simplifications) that precede circuit-level synthesis. Accordingly, this survey centers on synthesis and optimization within QDA, where the focus is on structural and physical transformations of quantum circuits, leaving programming-language-level compilation techniques beyond its intended scope.

\subsection{Circuit synthesis}

An operation synthesis pass transforms higher-level descriptions of a quantum subroutine into a sequence of basic quantum gates.

\subsubsection{Unitary synthesis}\label{sec:unitary_synthesis}

\begin{table}[htbp]
\centering
\caption{Taxonomy of unitary synthesis methods.}
\begin{small}
\begin{tblr}{
  width = \linewidth, 
  colspec = {|X[l]|X[l]|X[l]|X[l]|X[l]|}, 
  hlines, 
  row{1} = {font=\bfseries}, 
  column{1} = {font=\bfseries}, 
}
Method & Analytical (Closed-form) & Constructive (Recursive) & Approximate synthesis & Fault-Tolerant (Discrete) \\
Representative methods & Euler, KAK & QR, CSD, QSD & Madden et al. & Solovay-Kitaev, Ross-Selinger \\
Gate set & Continuous & Continuous & Continuous & Discrete (Clifford+$T$) \\
Goal & Symbolic exactness & Structured decomposition & Near-optimal approximation; trade-off & $\epsilon$-approximation with minimal $T$-count \\
Scalability & Low & Exponential & Low–medium & High (polylog) \\
Applicability & 1-2 qubits & $n$ qubits & $\leq$4-5 qubits & $n$ qubits \\
Two-qubit cost & Minimal (analytical) & Tight upper bounds & Empirical optimal & Depends on T-depth \\
\end{tblr}
\end{small}
\label{tab:synth_taxonomy}
\end{table}

Unitary synthesis refers to the process of decomposing a given unitary operation into a sequence of quantum gates drawn from a specified instruction set. This process is fundamental in quantum computing, as it enables the implementation of complex quantum algorithms on hardware that supports only a limited set of native gates. \Cref{tab:synth_taxonomy} presents a taxonomy of unitary synthesis methods, categorizing them based on their approach, target gate set, goals, scalability, applicability, and synthesis cost.

\paragraph{Analytical synthesis with closed-form solutions}
The simplest example for $\mathrm{SU}(2)$ unitary synthesis, i.e., single-qubit gate decomposition, is the famous Euler decomposition with that a general single-qubit gate represented as $U3$ gate can be decomposed into a sequence of $R_z$-$R_y$-$R_z$ rotations~\cite{Nielsen2010Quantum} as 
\begin{align}
U3(\theta, \phi, \lambda)
: = 
    \begin{pmatrix}
        \cos\frac{\theta}{2}          & -e^{i\lambda}\sin\frac{\theta}{2} \\
        e^{i\phi}\sin\frac{\theta}{2} & e^{i(\phi+\lambda)}\cos\frac{\theta}{2}
        \end{pmatrix}
= e^{i\frac{\phi+\lambda}{2}}R_z(\phi)R_y(\theta)R_z(\lambda).
\label{eq:euler_decomposition}
\end{align}
The derived ABC decomposition building on~\Cref{eq:euler_decomposition} can decompose any controlled-unitary gate into CNOTs and single-qubit rotations as
\begin{align}
  U3(\theta, \phi, \lambda) &= e^{\alpha} R_z\left(\phi\right) R_y\left(\theta\right) R_z\left(\lambda\right)\\
&= e^{\alpha} \left[R_z\left(\phi\right)R_y\left(\frac{\theta}{2}\right)\right]X\left[R_y\left(-\frac{\theta}{2}\right)R_z\left(-\frac{\phi+\lambda}{2}\right)\right]X\left[R_z\left(\frac{\lambda-\phi}{2}\right)\right]\\
&= e^{\alpha} AXBXC,
\end{align}
such that the decomposition rule for any two-qubit controlled-unitary gate can be illustrated as
\begin{small}
\begin{center}
\begin{quantikz}[row sep=0.3cm, column sep=0.2cm, align equals at=1.5]
    & \ctrl{1} & \ghost{X}\qw \\
    & \gate{\mathrm{U3}} & \ghost{X}\qw
\end{quantikz} = \begin{quantikz}[row sep=0.3cm, column sep=0.2cm, align equals at=1.5]
    & \ghost{X}\qw & \ctrl{1} & \qw & \ctrl{1} & \qw & \ctrl{1} & \qw \\
    & \gate{C} & \targ{} & \gate{B} & \targ{} & \gate{A} & \gate{\mathrm{Ph}(\alpha)} & \qw
\end{quantikz} $\sim$ \begin{quantikz}[row sep=0.3cm, column sep=0.2cm, align equals at=1.5]
    & \ghost{X}\qw & \ctrl{1} & \qw & \ctrl{1} & \gate{R_z(\alpha)} & \qw \\
    & \gate{C} & \targ{} & \gate{B} & \targ{} & \gate{A} & \qw
\end{quantikz}  
\end{center}
\end{small}

where the relation controlled-$\mathrm{Ph}(\alpha) = \begin{pmatrix}
  1 & 0 \\ 0 & e^{i\alpha}
\end{pmatrix} \otimes \begin{pmatrix}
  1 & 0 \\ 0 & 1
\end{pmatrix} \sim R_z(\alpha)\otimes I$ is applied.

In~\Cref{sec:qisa}, we discussed various options for defining quantum instruction sets. Among these, the de facto standard comprises CNOT/CZ gates paired with all single-qubit rotations. Notably, CNOT can be substituted with one or more physically realizable entangling two-qubit gates to construct alternative QISAs~\cite{harrow2008exact}---with corresponding synthesis passes then developed for these customized instruction sets.  

For the simplest synthesis problem---i.e., how to implement a given two-qubit unitary operator using the minimum number of CNOT gates or other vendor-specified quantum instructions---the well-known KAK decomposition~\cite{Khaneja2001Cartan} greatly simplifies the analysis of general two-qubit gates. Instead of addressing the $15$ real parameters required to characterize a general two-qubit gate, one only needs to work with a family of $3$ real parameters. More specifically, any unitary $U \in \mathrm{SU}(4)$ can be expressed as $U = K_1 A K_2$, where $K_1$ and $K_2$ are tensor products of single-qubit unitaties, and $A = e^{-i\sum_{i=1}^3 \lambda_i\sigma_i\otimes\sigma_i}$ is the canonical two-qubit operation. Building on this decomposition, it is straightforward to verify that at most three CNOT gates are sufficient for implementing any arbitrary two-qubit gate~\cite{Vidal2004Universal}. As we briefly mentioned in~\Cref{sec:qisa}, the AshN scheme~\cite{Chen2024One} can directly implement any two-qubit gate up to single-qubit rotations for a wide range of architectures, including the mainstream frequency-tunable transmon qubits.

\paragraph{Constructive algorithms}
A more general scenario involves decomposing a given $n$-qubit unitary gate into a vendor-specified quantum instruction set---for instance, the combination of $\mathrm{SU}(2)$ gates and CNOT gates. To address this decomposition challenge, several sophisticated elimination-based methods have been developed, including QR decomposition~\cite{Cybenko2001Reducing}, cosine-sine decomposition~\cite{Mottonen2004Quantum}, and most notably, quantum Shannon decomposition (QSD)~\cite{Shende2006Synthesis}. QSD remains the most pragmatic algorithm for unitary synthesis to date and has been integrated into major quantum compiling toolkits such as Qiskit~\cite{Wille2019IBMs} and Cirq~\cite{Developers2024Cirq}. Additionally, for nearly two decades, QSD has provided the tightest available upper bound $\lfloor \frac{23}{48}4^n - \frac{3}{2}2^n + \frac{4}{3} \rfloor$ on the number of CNOT gates required to implement an arbitrary $n$-qubit unitary. Until very recently, this long-standing bound was improved to: $\lfloor \frac{22}{48}4^n - \frac{3}{2}2^n + \frac{5}{3}\rfloor$~\cite{krol2024beyond}. However, such tight upper bound results achieved by these constructive algorithms still remains far from the theoretical lower bound $\lceil \frac{1}{4}(4^n - 3n - 1) \rceil$, which can be calculated by counting the degrees of freedom of an $n$-qubit unitary operator~\cite{Shende2004Minimal}. 

Similar lower and upper bounds have also been established for the entire $\mathrm{SU}(4)$ instruction set, which can be physically realized via the AshN scheme~\cite{Chen2024One}, with the lower bound given by $\lceil \frac{1}{9}(4^n - 3n - 1) \rceil$ and the upper bound by $\lfloor \frac{23}{64}4^n - \frac{3}{2}2^n\rfloor$.

\paragraph{Approximate synthesis}
When the system size is small, a broader range of synthesis options becomes available. For instance, approximate synthesis~\cite{madden2022best,Khatri2019quantumassisted,davis_heuristics_2019}---a technique capable of constructing near-optimal quantum circuits to approximate target unitary operations---balances two core objectives: minimizing circuit complexity (e.g., gate count, depth) and preserving sufficient fidelity with the target. Unlike analytically exact synthesis, which aims for provably optimal solutions but becomes computationally intractable for larger systems, approximate synthesis enables the practical implementation of quantum algorithms by leveraging controlled, bounded errors. This makes it indispensable for scaling quantum hardware beyond small qubit counts, as its performance relies on navigating tradeoffs between approximation accuracy and resource efficiency, with methods tailored to specific gate sets. Notably, although approximate synthesis can achieve efficient and even optimal synthesis schemes, it is still not scalable for circuits or unitaries with a qubit number beyond $4$ or $5$. Moreover, while not yet rigorously proven, the approximate synthesis result---characterized by an extremely low bounded error---may be interpreted by some as a suggestive clue that exact synthesis could potentially be achievable. 

Through the adoption of approximate synthesis, we can randomly sample $n$-qubit unitary operators according to the Haar measure and then numerically determine the minimum number of two-qubit instructions required to ensure that circuits containing this number of two-qubit operations can adequately approximate all the sampled $n$-qubit unitary operators. According to~\cite{Chen2024One}, 
\begin{enumerate}
\item for the case of CNOT and single-qubit rotations as instructions, $3$, $14$, and $61$ CNOT gates are needed to approximate randomly sampled $2$-, $3$-, and $4$-qubit unitary operators;
\item for the case of $\mathrm{SU}(4)$ and single-qubit rotations as instructions, $1$, $6$, and $27$ $\mathrm{SU}(4)$ gates are needed to approximate randomly sampled $2$-, $3$-, and $4$-qubit unitary operators.
\end{enumerate}
Note that for $n = 2, 3, 4$, these values coincide exactly with the analytical lower bounds mentioned above! This observation motivates the following conjecture:
\begin{conjecture}
Any $n$-qubit unitary operator can be synthesized by $\left\lceil \frac{4^n - 3n - 1}{4} \right\rceil$ CNOT gates (or $\left\lceil \frac{4^n - 3n - 1}{9} \right\rceil$ general two-qubit gates), with the assistance of single-qubit rotations.
\end{conjecture}

\paragraph{Discrete synthesis particularly for fault-tolerant architecture}
Some readers may have noticed that in the literature, one often explores a discrete gate set such as Clifford+T, where the Clifford group is generated by $\{H, S, \text{CNOT}\}$. This corresponds to the framework of fault-tolerant quantum computing (FTQC), which we will elaborate on in~\Cref{sec:qec}. For circuit synthesis in FTQC, the primary objective is typically to minimize the count or depth of more expensive non-Clifford gates, such as $T$ gates. It is important to note that this stands in contrast to bare-metal (non-fault-tolerant) execution, where two-qubit instructions generally incur higher noise levels than single-qubit ones. In FTQC, by contrast, the single-qubit $T$ gate is more costly than two-qubit gates selected from the Clifford group. Another important point is that the relative cost of different instructions depends heavily on the implementation scheme. For instance, substantial efforts are being directed toward developing cheaper $T$ logical instructions. Notably, this relative cost hierarchy may shift as the technology advances.

\subsubsection{Synthesis with dynamic circuits}

Beyond the conventional ``static'' quantum circuits, which are defined by fixed, preprogrammed gate sequences, dynamic circuits have been introduced to enable time-dependent, adaptive, or interactive behavior. These dynamic architectures incorporate mid-circuit measurements and feedback control~\cite{Corcoles2021Exploiting}, and they mirror how classical dynamic circuits process time-varying signals. 

Over a decade ago, dynamic circuits were first adopted in quantum error correction via the so-called repeat-until-success (RUS) strategy~\cite{Paetznick2014RepeatUntilSuccess}. For instance, the magic state distillation protocol---which consumes multiple ``magic states'' to generate a high-fidelity magic state---can be implemented using this RUS approach: the protocol first accepts a handful of low-fidelity magic states, then executes Clifford gates and measurements; if the target measurement outcome is obtained, the desired output state is successfully generated, and if not, the protocol restarts and repeats the process until a successful run is achieved. Since then, considerable efforts have explored whether RUS can be directly leveraged to synthesize unitary gates. For single-qubit gate synthesis, Paetznick et al.~\cite{Paetznick2014RepeatUntilSuccess} proposed a direct-search algorithm that achieves nearly a $2\times$ performance improvement over gate decomposition methods. However, this approach relies on exhaustive search, which incurs exponential time complexity. To address this inefficiency, Bocharov et al.~\cite{Bocharov2015Efficient} developed a probabilistic search method, reducing the synthesis time for ``axial'' gates (i.e., $R_z(\theta)$) to polynomial complexity. Separately, Brown et al.~\cite{Brown2023Advances} provided a practical running example: they demonstrated the compilation of an RUS-based Q\# program onto Quantinuum's quantum hardware, bridging theoretical RUS design with real-world hardware deployment. Bocharov et al.~\cite{Bocharov2015Efficient} further expanded on RUS by proposing a variant called probabilistic quantum circuits with fallback, where the infinite retries inherent to traditional RUS are replaced with a resource-intensive---but finite---``fallback'' operation.

Though originally motivated by quantum error correction, dynamic circuits have recently expanded their utility: they are now employed for quantum state preparation, the implementation of long-range entangling gates, and the reduction of resource overhead in core quantum algorithms. To advance their practical deployment, several dynamic circuit synthesis approaches have been proposed---these methods leverage qubit reuse to cut down on overall resource overhead, including key metrics such as total qubit count and circuit depth~\cite{niu2024ac,fang2023dynamic}.

It is worth noting that synthesis with dynamic circuits can be viewed as introducing classical conditional branching instructions. Expanding the instruction set in this manner often enables more efficient implementations, provided that the introduced additional instructions themselves are efficient. What makes dynamic circuits particularly compelling in certain scenarios, therefore, is the efficient implementation of conditional branching---a capability that fundamentally relies on the tight integration of a classical coprocessor with the quantum processor, rather than dependence on a separate host computer. This integration is critical for minimizing communication latency. For additional architectural details, please refer to~\Cref{subsec:qcis}.

\subsubsection{Synthesis of higher-level constructs}\label{subsubsec:synth_highlevel}
In the previous section, we discussed the synthesis of a general unitary. In this section, we will focus on the synthesis of unitaries with specific structures, as well as how these structural features can facilitate the synthesis process.

\paragraph{Clifford circuit synthesis}
For $n$-qubit Clifford operations, instead of using unwieldy $2^n \times 2^n$ unitary matrices, a more efficient alternative is the stabilizer tableau---a compact, classical representation with a size of $2n \times (2n+1)$. This structured format captures the stabilizer state generated by a Clifford circuit or the stabilizer group preserved by a Clifford operation.  Specifically, the tableau employs a $2n \times (n+1)$ binary matrix to systematically track both stabilizers (operators that commute with the circuit's evolution and stabilize its output state) and destabilizers (their complementary counterparts). This compact representation avoids the exponential scaling of full unitary matrices, enabling efficient classical simulation and manipulation of Clifford operations---even for larger $n$. 

Thanks to the stabilizer tableau, it has been shown that any $ n $-qubit Clifford circuit can be synthesized into a ``canonical form'' using at most $O(\frac{n^2}{\log n})$ gates---a result dating back over two decades~\cite{PhysRevA.70.052328}. This canonical form is represented by the sequence $H$-$C$-$P$-$C$-$P$-$C$-$H$-$P$-$C$-$P$-$C$, where $H$, $C$, and $P$ denote Hadamard, CNOT, and Phase gates, respectively.

The simplicity of Clifford tableaus also facilitates the application of classical approaches to Clifford synthesis. For instance, Schneider et al.~\cite{Schneider2023SATClifford} developed a SAT encoding scheme rooted in stabilizer simulation, which enables the discovery of a Clifford synthesis with the minimum number of CNOT gates. Meanwhile, Webster et al.~\cite{websterHeuristicOptimalSynthesis2025} proposed A* search based greedy algorithms to tackle the Clifford synthesis problem. On the practical implementation front, IBM's Qiskit Transpiler Service~\cite{kremerPracticalEfficientQuantum2024} leverages reinforcement learning techniques to generate efficient synthesis of Clifford tableaus, with particular optimization for devices constrained by topological limitations.

\paragraph{Oracle synthesis}
Applications such as Grover's algorithm rely on an oracle---the implementation of a target classical function within a quantum circuit. Specifically, for a classical function $f: \{0, 1\}^m \rightarrow \{0, 1\}^n$, we need to synthesize a unitary operator $U_f$ that acts on $(m+n)$ qubits and satisfies the property $U_f\ket{x}\ket{y} = \ket{x}\ket{y \oplus f(x)}$  where $\oplus$ denotes the bitwise XOR operation.  

Notably, most classical functions $f$ are irreversible, which means we cannot directly convert their classical circuits into the ideal unitary $U_f$. Instead, we first construct a ``dirty'' unitary $U_f'$ such that $U_f'\ket{x}\ket{0} = \ket{f(x)}\ket{\text{garbage}}$,  
where $\ket{\text{garbage}}$ represents unintended auxiliary states introduced during the computation. To eliminate this garbage and recover the ideal $U_f$, we perform uncomputation of $U_f'$---a step that requires first copying the desired $\ket{f(x)}$ state to preserve it.  

If $f$ is inherently reversible (i.e., it maps distinct inputs to distinct outputs), Bennett~\cite{bennettLogicalReversibilityComputation1973} provided a direct construction for $U_f$, where the unitary acts on the input qubits alone: $U_f\ket{x} = \ket{f(x)}$.  

A common strategy for oracle synthesis is to first restructure the classical circuit of \( f \) into a form that is easier to translate into quantum gates. Two prominent approaches include:  
\begin{enumerate}
\item ESOP Conversion: Sanaee et al.~\cite{sanaeeESOPBasedToffoliNetwork2010} propose rewriting $f$ as an Exclusive-Or Sum of Products (ESOP)---a representation consisting solely of XOR operations applied to AND terms. This ESOP form can then be directly mapped to a quantum circuit using multi-Toffoli gates (for implementing AND terms) followed by a sequence of CNOT gates (for implementing XOR operations).  
\item XAG Conversion: Meuli et al.~\cite{meuliXorAndInverterGraphsQuantum2022} advocate converting $f$ into an XOR-And-Inverter Graph (XAG)---a computational network restricted to XOR, NOT, and AND operations. Using a technique called pebbling (a method to manage auxiliary resources), XAGs can be translated directly into quantum gates: multi-Toffoli gates and CNOT (as with ESOP conversion) plus X gates for implementing NOT operations.
\end{enumerate}

Another influential method is the LHRS framework~\cite{soekenLUTBasedHierarchicalReversible2019}, which leverages Look-Up Table (LUT) mapping---a standard technique in classical EDA---to enable hierarchical oracle synthesis. 

\iffalse
The process works as follows:  
\begin{enumerate}
\item First, the classical circuit of $f$ is synthesized into a $k$-feasible network, composed of $k$-LUTs (each LUT takes at most $k$ input bits and produces $1$ output bit).  
\item Each $k$-LUT can be implemented as a standalone target unitary $U_f$, but this requires auxiliary qubits initialized to $\ket{0}$ to support computation.  
\item The number of auxiliary qubits used and the overall computational complexity of the oracle are balanced via a trade-off determined by the order of uncomputation in a pebble game~\cite{meuliReversiblePebblingGame2019,zhangDivideAndConquerPebblingStrategy2025}.  
\item For efficiency, the optimal implementation of each $k$-LUT in the Clifford+$T$ gate set can be precomputed and stored in a database for quick retrieval.  
\end{enumerate}
\fi

In practice, tools like IBM's Qiskit Transpiler Service~\cite{kremerPracticalEfficientQuantum2024} streamline oracle integration by natively supporting two common types of classical oracles:  
\begin{enumerate}
\item Linear Functions: Functions that can be represented as linear transformations over the binary field (e.g., parity checks) and map to simple quantum circuits using only CNOT and X gates.  
\item Permutation Circuits: Circuits that implement reversible permutations of input bits (e.g., bitwise rotations), which translate to unitaries without requiring garbage states or uncomputation.
\end{enumerate}

\subsection{Quantum circuit optimization}

The results derived from operation synthesis can be further refined through optimization techniques---a process commonly recognized as quantum circuit optimization.

Optimization algorithms seek to identify functionally equivalent implementations of the original quantum circuit, with the core goal of minimizing overhead. This minimization is guided by specific objectives---such as reducing circuit depth, shrinking circuit size, or lowering total error---while ensuring the compilation time remains reasonable. In essence, quantum circuit optimization serves as the quantum computing analog to compiler optimization in classical computing, fulfilling a parallel role in enhancing computational efficiency across the two paradigms.

The boundary between circuit synthesis and circuit optimization is somewhat blurred, as circuit synthesis often incorporates implicit objectives---such as minimizing circuit length---that align closely with the core goals of quantum circuit optimization.

\subsubsection{Pattern matching and circuit rewriting}

One natural approach to optimizing large quantum circuits is to follow the paradigm of traditional compilers: analyzing one local subcircuit at a time and applying incremental, step-by-step transformations to the original program to ultimately yield an improved version~\cite{10.5555/1177220, sarkarNanopassInfrastructureCompiler, mckeemanPeepholeOptimization1965}.

However, challenges emerge as the problem scales:
\begin{enumerate}
\item Hand-crafted rewrite rules may fail to cover all desired optimizations;
\item Managing numerous rewrite patterns simultaneously suffers from two critical issues: efficiency bottlenecks and the phase ordering problem---where hasty application of a rewrite step can eliminate opportunities for more impactful optimizations later;
\item A beneficial cost-reducing rewrite may only become accessible after traversing a ``plateau'' of multiple non-cost-reducing rewrites~\cite{liQuarlLearningBasedQuantum2023}.
\end{enumerate}

We will briefly overview several attempts to address these challenges.

\paragraph{Selecting Patterns for Circuit Rewriting.} The first core challenge in pattern rewriting is determining which patterns to use for circuit transformation. Early solutions relied on sets of hand-crafted circuit rewrite rules; notably, Nam et al.~\cite{Nam2018Automated} defined a widely adopted suite of circuit rewriting, commutation, and conjugation rules. A more general alternative avoids human-specified optimization rules, instead leveraging machine-generated circuit identities. Key examples of this approach include: (1) Quanto\cite{Pointing2021Quanto} establishes a database of quantum circuit identities, where circuits are indexed by floating-point hashes of their unitary matrices. To prevent database explosion with an increasing qubit count, it only constructs entries for small-sized circuit ``tiles'' and applies tile-based rewriting via a cost-driven search.
(2) Quartz~\cite{Xu2022Quartz} generalizes Quanto by extending rewriting support from fixed unitary matrices to parametric quantum circuits. It verifies circuit equivalence using a Satisfiability Modulo Theories (SMT) solver.
(3) QUESO~\cite{xuSynthesizingQuantumCircuitOptimizers2023} generates circuit identities via path summation. To accelerate quantum circuit identity testing, they introduce a data structure called the polynomial identity filter, which groups circuits that are likely to be equivalent.

\paragraph{Detecting Matching Patterns in the Circuit.} The second challenge involves identifying patterns in the target circuit that match those stored in the database. Two notable contributions address this problem: Iten et al.~\cite{itenExactPracticalPattern2022} focus on finding a maximal matching---a set of disjoint matches that cannot be extended by adding another match---for a given pattern in the circuit. This matching accounts for gates that can be commuted (i.e., reordered without altering the circuit's functionality). Mondada et al.~\cite{mondadaScalablePatternMatching2025} tackle the scalability of matching large numbers of small patterns. Through preprocessing, their method ensures that the complexity of matching all patterns against the input circuit is independent of the number of patterns considered---a critical advantage for large pattern libraries.

\paragraph{Ordering the Application of Rewrite Rules.} The third challenge concerns the order in which pattern rewrite rules are applied. Importantly, rewrites that do not directly reduce circuit cost have been shown to play a key role in the overall optimization of quantum circuits~\cite{liQuarlLearningBasedQuantum2023, xuSynthesizingQuantumCircuitOptimizers2023}. Key approaches to addressing this challenge include: Xu et al.~\cite{xuSynthesizingQuantumCircuitOptimizers2023} employ beam search: they maintain a priority queue of the top-$k$ best-known circuits (for a fixed k). Each time a circuit is extracted from the queue, a maximal match of a given pattern is applied to generate potential improved candidates---similar to the maximal matching strategy in Iten et al.~\cite{itenExactPracticalPattern2022}. In recent years, machine learning (ML) and reinforcement learning (RL) have been proposed to guide the selection of which pattern rewrite rule to apply next~\cite{Fosel2021Quantum, liQuarlLearningBasedQuantum2023}.

\paragraph*{}We close this subsection with a discussion of OAC~\cite{aroraLocalOptimizationQuantum2025}, a meta-optimizer that treats other quantum optimization methods as expensive ``rewrite rules'' for circuit tiles with a maximum depth of $2\Omega$. While the computational cost of the underlying optimization oracles typically grows with circuit size, OAC bounds the size of oracle calls using a divide-and-conquer strategy. It splits the circuit into two segments; it optimizes each segment recursively; and ``melds'' the seam between the two segments by recursively optimizing the subcircuit spanning this seam. Since the process terminates only when no optimizable seams remain, the output of OAC is locally optimal: no $\Omega$-local segment (i.e., a segment involving $\Omega$ or fewer qubits/gates) in the final circuit can be further optimized by the underlying oracle.

\subsubsection{From local patterns to beyond}
\label{subsubsec:beyond}

% In principle, two quantum circuits are equivalent if and only if their matrix representations differ merely by a global phase. While such equivalence can, in practice, be derived from small-scale equivalences that are commonly referred to as ``rewriting rules'', these ``local'' rewriting rules may not be sufficient to capture all possible equivalences, especially those between large and complicated quantum circuits. We discussed circuit optimization via local patterns in the previous subsection; now, we will explore how to overcome the limitations of local optimization and leverage opportunities to achieve more profound optimization.

In principle, equivalence between two quantum circuits is a global property --- two quantum circuits are equivalent if and only if their matrix representations differ merely by a global phase. While some optimization opportunities can be readily captured via local rewriting rules applied directly to the circuit, others are more easily discovered by converting the circuit (or at least a part of it) into another representation and potentially looking at it from a more global perspective.

\paragraph{Resynthesis-based Optimization.} Since we have demonstrated several gate synthesis techniques, it is natural to consider whether we can use a high-quality gate synthesis pass to optimize a quantum circuit: we simply take the circuit description, feed it into a gate synthesis algorithm, and obtain a potentially improved circuit. This approach corresponds to the superoptimization technique in classical program compilation, where the compiler generates an optimal equivalent program. Unfortunately, as we explained in the previous section, typical unitary synthesis methods produce constructive results but rarely yield highly optimized circuits. Instead, approximate synthesis may be better suited to this scenario. Another example of resynthesis involves Clifford subcircuits: in a Clifford+T context, we aggregate Clifford gates, represent the accumulated operations via a Clifford tableau, and resynthesize the tableau using the approach described in~\Cref{subsubsec:synth_highlevel}.

\begin{figure}
\centering

\begin{subfigure}[b]{\textwidth}
    \includegraphics[clip, trim=5 5 5 5,width=\textwidth]{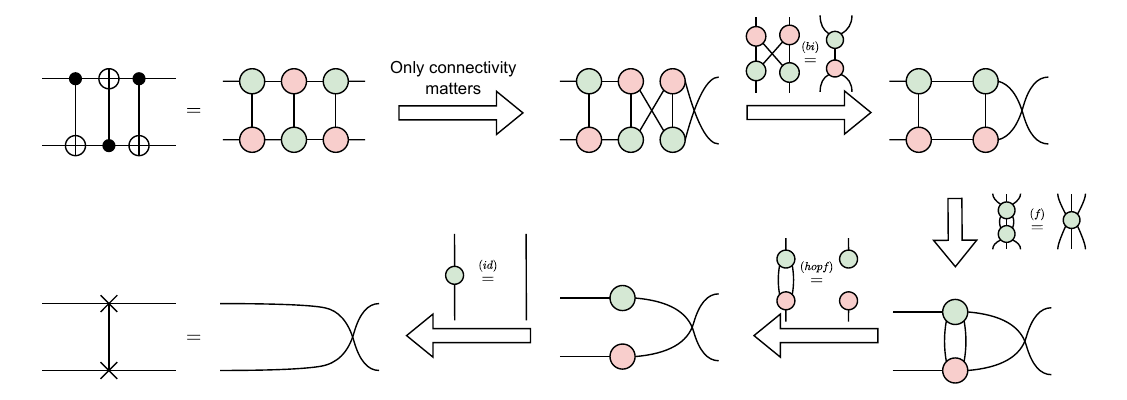}
    \caption{}
    \label{fig:ZX_example}
\end{subfigure}
\begin{subfigure}[b]{0.4\textwidth}
    \includegraphics[clip, trim=20 5 20 5,width=\textwidth]{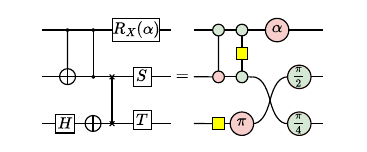}
    \caption{}
    \label{fig:ZX_circuit_to_diagram}
\end{subfigure}
\begin{minipage}[b]{0.5\textwidth}
\centering
\begin{subfigure}[b]{0.8\textwidth}
    \includegraphics[clip, trim=20 3 20 5,width=\textwidth]{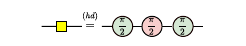}
    \caption{}
    \label{fig:ZX_rule_hd}
\end{subfigure}
\begin{subfigure}[b]{\textwidth}
    \includegraphics[clip, trim=15 3 15 5,width=\textwidth]{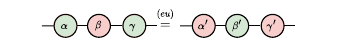}
    \caption{}
    \label{fig:ZX_rule_eu}
\end{subfigure}
\end{minipage}

\caption{Examples of ZX calculus. 
(a) An example adapted from \url{https://zxcalculus.com/}, demonstrating rewriting a quantum circuit with ZX-calculus rewriting rules. A green node with label $\alpha$ (omitted when $\alpha = 0$) represents a $Z$ spider $\ket{0}^{\otimes n} + e^{i\alpha} \ket{1}^{\otimes n}$, and a red node represents a $X$ spider $\ket{+}^{\otimes n} + e^{i\alpha} \ket{-}^{\otimes n}$; the direction of ket and bra does not matter and can be changed trivially by ``bending'' legs of the spider; composition of spiders obeys the semantics of tensor networks. Rewriting rules used in the transformation are the \textit{bialgebra} rule, same-color spider \textit{fusion}, the \textit{Hopf} rule and the \textit{identity} rule; soundness of these rewriting rules can be checked by hand. All rewriting rules hold with color flipped.
(b) More examples of converting a quantum circuit to ZX-diagram. The global scalar factor is ignored.
(c) Euler decomposition rewrite for Hadamard gate.
(d) Color-swap Euler decomposition rewrite. Parameters $\alpha, \beta, \gamma$ and $\alpha', \beta', \gamma'$ satisfies the non-linear constraint defined in~\cite{vilmartNearMinimalAxiomatisationZXCalculus2019a}.
}
\label{fig:ZX_calculus}
\end{figure}

\paragraph{ZX-calculus.} The ZX-calculus~\cite{weteringZXcalculusWorkingQuantum2020} creates a transformative new avenue for moving beyond the ``rigid'' structure of traditional quantum circuits. Unlike conventional circuits---where gates are fixed to discrete ``qubit wires''---the ZX-calculus represents quantum processes using \textit{ZX-diagrams}: flexible structures composed of freely connected Z-spiders and X-spiders. As long as the underlying connectivity of these components is preserved, ZX-diagrams can be deformed arbitrarily, behaving like general string diagrams~\cite{coeckePicturingQuantumProcesses2017}. This flexibility unlocks unique opportunities: for instance, \Cref{fig:ZX_example} demonstrates converting a quantum circuit to a ZX-diagram and transforming it. Note that while converting a circuit to a ZX-diagram is usually trivial and preserves the circuit structure (``qubit wires'' and ``gate wires'' can still be distinguished from diagram, e.g.~\Cref{fig:ZX_example} and~\Cref{fig:ZX_circuit_to_diagram}), application of rewriting rules depends only on connectivity between spiders, regardless of the wires originally come from, enabling applying unified rewrite rules spanning across qubits and gates.

A critical theoretical interest for any new circuit rewriting system is \textit{completeness}---the guarantee that if two ZX-diagrams represent two equivalent quantum circuits, there exists a sequence of rewrites to convert one into the other. Several milestones of completeness results are:
\begin{enumerate}
\item \textbf{Clifford completeness}: The $\frac{\pi}{2}$-fragment of the ZX-calculus, where all angles in the ZX-diagram are multiples of $\frac{\pi}{2}$, is universal for Clifford circuits.
The initial version of the ZX-calculus rewriting rules proposed in~\cite{coeckeInteractingQuantumObservables2008} does not include the rule of Euler decomposition for the Hadamard gate (see~\Cref{fig:ZX_rule_hd}). It is pointed out in~\cite{duncanGraphStatesNecessity2009} that the ~\Cref{fig:ZX_rule_hd} rule cannot be derived from the initial rules. \cite{Backens_2014} proves that adding the rule makes the $\frac{\pi}{2}$-fragment of the ZX-calculus complete for Clifford circuits.
\cite{schroderdewittZXcalculusIncompleteQuantum2014} points out that the set is still incomplete for general quantum circuits.
\item \textbf{Clifford+T completeness}: Like the case for Clifford circuits, the $\frac{\pi}{4}$-fragment of ZX-calculus is universal for Clifford+$T$ circuits, which is approximately universal for quantum circuits. \cite{hadzihasanovicTwoCompleteAxiomatisations2018} and~\cite{jeandelCompletenessZXCalculus2020} both propose a set of extra rules to make the ZX-calculus Clifford+$T$ complete, based on back-and-forth translation between another graphical language named the ZW-calculus~\cite{coeckeCompositionalStructureMultipartite2010} known to be complete~\cite{hadzihasanovicDiagrammaticAxiomatisationQubit2015}.
\item \textbf{General completeness}: A generally complete set of ZX-calculus rewriting rules can be obtained by adding one axiom~\cite{jeandelCompletenessZXCalculus2020} to the Clifford+$T$ complete rewriting rules. Furthermore, ~\cite{vilmartNearMinimalAxiomatisationZXCalculus2019a} gave a minimal axiomization that only adds an axiom of converting between ZXZ and XZX decompositions of a single-qubit gate (as guessed in~\cite{schroderdewittZXcalculusIncompleteQuantum2014}, see~\Cref{fig:ZX_rule_eu}) with a clear physical interpretation.
\end{enumerate}
We refer readers to~\cite{weteringZXcalculusWorkingQuantum2020} for a more detailed history of the completeness of the ZX-calculus.

A major challenge in adapting the ZX-calculus for practical circuit transformation arises from the flexibility of ZX-diagrams themselves: unlike the ideal circuit-like situation in~\Cref{fig:ZX_circuit_to_diagram}, an arbitrary ZX-diagram after heavy optimization may not resemble a traditional quantum circuit at all. To resolve this, we require a procedure to convert optimized ZX-diagrams back into executable quantum circuits---a step known as \textit{circuit extraction}~\cite{Duncan2020Graphtheoretic, backensThereBackAgain2021}. Notably, \cite{beaudrapCircuitExtractionZXdiagrams2022} proves that extracting a circuit from a general ZX-diagram is \(\#P\)-hard---even when the diagram is guaranteed to represent a unitary operator.

To circumvent this hardness, ZX-based circuit transformations and optimizations should not allow arbitrary rewrites of ZX-diagrams. Instead, rewrites should be restricted to preserve some critical \textit{invariant} of the ZX-diagram: a simple circuit extraction routine should remain feasible after every rewriting step.
A prominent example is Quantomatic~\cite{faganOptimisingCliffordCircuits2019}, which rewrites ZX-diagrams while preserving a graph invariant called \textit{causal flow} (or \texttt{cflow}). Causal flow is an auxiliary invariant that intuitively ``marks'' which edges correspond to ``qubit wires'' and which for CNOT gates. If the \texttt{cflow} invariant is held for the initial ZX-diagram (like the one in~\Cref{fig:ZX_circuit_to_diagram}) and upon every rewriting step, a simple circuit extraction routine would suffice to convert an optimized ZX-diagram back to a circuit.

An alternative invariant-based approach leverages another invariant named \textit{generalized flow} (or \texttt{gflow})---a concept originally developed for one-way quantum computing~\cite{mcelvanneyFlowpreservingZXcalculusRewrite2023, backensThereBackAgain2021, Duncan2020Graphtheoretic, Kissinger2020Reducing}. This method allows rewrites on ``graph-like'' ZX-diagrams while preserving \texttt{gflow}, a more relaxed invariant that allows the ZX-diagram to deviate further from circuit-like structures while ensuring polynomial-time circuit extraction.

For a comprehensive review of ZX-calculus-based quantum circuit optimization, we refer readers to~\cite{fischbachReviewQuantumCircuit2025}.

Quantum circuit resynthesis via ZX-diagram representations is a key technique employed by various compilers, including TKet~\cite{sivarajah2020t, cowtanPhaseGadgetSynthesis2020}, PAULIOPT~\cite{gogiosoAnnealingOptimisationMixed2023}, and Pennylane~\cite{bergholmPennyLaneAutomaticDifferentiation2022}.

The ZX-calculus's diagrammatic structure provides a visual framework for formalizing and streamlining the commutation behaviors and phase dynamics inherent to Pauli operators. Through algebraic manipulations and graphical transformations, they facilitate minimizing both circuit depth and gate count. While effective at handling Pauli gadget resynthesis, the commutation principles they rely on are inherently localized, posing challenges for hardware-adaptive optimization strategies.

\paragraph{Phase polynomial.} Whereas ZX-calculus-based optimization still relies on local rewriting rules, the phase polynomial is an example of a representation that can capture more global information, albeit only for a limited class of circuits. In contrast to general Clifford+$T$ circuits, CNOT+$T$ circuits exhibit very simple behavior. They always map computational basis states to computational basis states according to an affine transformation, with a possible phase shift of a multiple of $\pi/4$. This means that optimization opportunities are often found in CNOT+$T$ subcircuits, especially opportunities to reduce $T$ gates. The phase polynomial stands out as an effective representation for CNOT+$T$ circuits, equipping compilers to identify precisely these types of optimization opportunities.

\begin{figure}
\centering

\begin{subfigure}[t]{0.45\textwidth}
    \includegraphics[clip, trim=25 10 20 5,width=\textwidth]{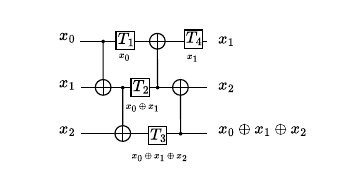}
    \caption{}
    \label{fig:Phase_polynomial_example}
\end{subfigure}
\begin{subfigure}[t]{0.5\textwidth}
    \raisebox{1cm}{
    \includegraphics[clip, trim=35 0 20 0,width=\textwidth]{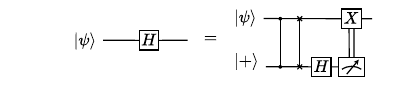}
    }
    \caption{}
    \label{fig:Phase_polynomial_hadamard_gadgetization}
\end{subfigure}
\caption{Examples for phase polynomial.
(a) A CNOT+$T$ circuit with phase polynomial $\varphi(x_2, x_1, x_0) = x_0 + (x_0\oplus x_1) + (x_0 \oplus x_1 \oplus x_2) + x_1 \mod 8$ and $f(x_2, x_1, x_0) = (x_1, x_2, x_0\oplus x_1\oplus x_2)$.
(b) Hadamard gadgetization in \cite{Heyfron2018An}, implementing a Hadamard gate with a Hadamard-free circuit and feedback control. The classically-controlled $X$ can be conjugated to end of the circuit into a classically-controlled Clifford at the cost of zero extra $T$ gates.
}
\label{fig:Phase_polynomial}
\end{figure}

The effect of a CNOT+$T$ circuit $V$ on a computational basis state $\ket{x}$ can be written as 
\begin{equation}
  V\ket{x} = \exp(\frac{i\pi}{4} \varphi(x))\ket{f(x)},
\end{equation}
where $f(x)$ is a permutation over the computational basis, and $\varphi(x)$ is in the form:
\begin{equation}
\label{eq:phase_polynomial_gate}
\varphi(x_{n-1}, \cdots, x_0) = \sum_{0\leq i < r} \left(\bigoplus_{0 \leq j < n} \chi_{ji}x_i\right) \mod 8, \chi_{ji}\in \{0, 1\}, x_i\in \{0, 1\}.
\end{equation}
The phase polynomial $\varphi(x)$ accumulates phase shifts from all $r$ $T$ gates in the CNOT+$T$ circuit: the contribution of every $T$ gate is $\frac{\pi}{4}$ multiplies a linear Boolean function (XOR sum indicated by $\{\chi_{ji}\}$) of input bits $\{x_i\}$ determined by the CNOT network before the $T$ gate. An example of a phase polynomial of a CNOT+$T$ circuit is given in~\Cref{fig:Phase_polynomial_example}.
For the purpose of reducing $T$ gates, $f(x)$ can be set to $f(x)=x$ by appending a reversed CNOT network to the circuit; thus, we can focus only on $\varphi(x)$. 

Expanding the XOR sum into polynomial form over $\{x_i\}$ by $x_i\oplus x_j = x_i + x_j - 2x_ix_j$ gives:
\begin{equation}
\label{eq:phase_polynomial}
  \varphi(x) = \sum_i a_i x_i + 2 \sum_{ij} b_{ij}x_ix_j + 4\sum_{ijk} c_{ijk}x_ix_jx_k \mod 8,
\end{equation}
where $a_i \in \{0, 1, \cdots 7\}, b_{ij}\in \{0, 1, 2, 3\}, c_{ijk}\in \{0, 1\}$.
Note that:
\begin{itemize}
  \item One direct circuit implementation of the phase polynomial, i.e.~\Cref{eq:phase_polynomial}, would be: applying $T^{a_i}$ to each qubit $i$, $CS^{b_{ij}}$ to qubits $i$ and $j$, and $CCZ^{c_{ijk}}$ to qubits $i$, $j$, and $k$.
  \item Since $T^2=S$ and $CS^2=CZ$, two phase polynomials are Clifford equivalent if all their coefficients ($a_i$, $b_{ij}$, $c_{ijk}$) have the same parity.
\end{itemize}

Considering only the parity of $a_i, b_{ij}, c_{ijk}$ allows representing $\varphi$ as a Boolean \textbf{signature tensor} $\mathcal{T}(\varphi)$, a symmetric order-3 tensor defined as:
\begin{align}
  &\mathcal{T}(\varphi)_{i, i, i} = a_i\mod 2 &\\
  &\mathcal{T}(\varphi)_{i, i, j} = b_{ij}\mod 2 &(i\neq j)\\
  &\mathcal{T}(\varphi)_{i, j, k} = c_{ijk}\mod 2 &(i, j, k \text{ distinct})
\end{align}

Most notably, a signature tensor with rank $1$, i.e. one that can be written in the form:
\begin{equation}
    \mathcal{T} = u \otimes u \otimes u, u = (u_{0}, \cdots, u_{{n-1}})
\end{equation}
always corresponds to the phase polynomial of a CNOT+$T$ circuit with exactly one $T$ gate:
\begin{equation}
    \varphi(x) = \bigoplus_{0\leq j < n} u_jx_j.
\end{equation}
Conversely, \eqref{eq:phase_polynomial_gate} provides a decomposition of the signature tensor $\mathcal{T}(\mathcal{\varphi})$ into a sum of $r$ rank-$1$ tensors:
\begin{equation}
\label{eq:waring_decomp}
    \mathcal{T}(\varphi) = \sum_{0\leq i<r} \boldsymbol{\chi}_j\otimes \boldsymbol{\chi}_j\otimes \boldsymbol{\chi}_j, \boldsymbol{\chi}_j = \{\chi_{ji}\}.
\end{equation}

Therefore, the problem of finding the minimum number of $T$ gates reduces to finding a decomposition \eqref{eq:waring_decomp} of $\mathcal{T}(\varphi)$ with the minimum possible $r$, known as a Waring decomposition.
\cite{Heyfron2018An} extends Lempel's factoring algorithm to find a low-rank decomposition, while AlphaTensor-Quantum~\cite{Ruiz2024Quantum} further leverages reinforcement learning to identify such a low-rank decomposition.

To apply this optimization technique to a general Clifford+$T$ circuit, the simplest way is to split the circuit using Hadamard gates. However, when the CNOT+$T$ subcircuits are short, optimization opportunities become limited. An alternative method is to implement all Hadamard gates through \textit{Hadamard gadgetization}~\cite{Heyfron2018An} (see~\Cref{fig:Phase_polynomial_hadamard_gadgetization}) at the cost of extra ancilla qubits and measurements.

Besides the Waring decomposition approach, the phase polynomial is also used in other resynthesis techniques.
\cite{Nam2018Automated} uses phase polynomials to detect if two $R_Z$ gates are performed on the same XOR sum and merges them if possible, enabling non-local merging of rotation gates across many CNOT gates.
PAULIOPT~\cite{gogiosoAnnealingOptimisationMixed2023} uses simulated annealing when compiling a circuit consisting of phase polynomials to minimize the CNOT count on a given nearest-neighbor architecture.

\paragraph{Pauli exponentiations.} 
% While all the aforementioned optimization techniques are, in principle, limited to local patterns and rewrites, a critical question persists: how might we capitalize on global optimization opportunities? Notably, this area has garnered considerably less attention in existing literature. To address this gap, we employ Pauli exponentiation (i.e., Pauli strings with coefficients)~\cite{yang2025phoenix} as a concrete example, demonstrating how such structures can be harnessed to explore these global optimization opportunities. 

An important idea of the phase polynomial method is to decompose the circuit into a sequence of \textit{Pauli exponentiations}, i.e., factors of the form $\exp(-i\theta_j P_j/2)$ where $P$ is a Pauli operator. The phase polynomial method applies to the special case where all $\theta_j = \pi/4$ and all $P_j$ commute with each other. \cite{yang2025phoenix} explores this idea further and finds that the structure of Pauli exponentiations actually occurs naturally in many practical applications of quantum computation.

Specifically, Pauli exponentiations can function as a high-level IR particularly for Hamiltonian simulation programs, where each multi-qubit Pauli rotation is mathematically expressed as the Pauli exponentiation according to
\begin{align}
  e^{-iHt} = e^{-i \sum_j \alpha_j P_j t} \approx \prod\nolimits_j e^{-i \alpha_j P_j t}.
  \label{eq:pauli_exp}
\end{align}
\Cref{eq:pauli_exp} emphasizes that when the coefficients $\{\alpha_j\}$ are small, the overall evolution governed by the Hamiltonian $ H $ can be approximated by the product of individual Pauli exponentiations~\cite{dalzell_quantum_2023}.
Beyond Hamiltonian simulation, Pauli exponentiations also naturally arise in variational quantum algorithms (VQAs)~\cite{Cerezo2021}, where parameterized quantum circuits are constructed using layers of Pauli rotations. In this context, optimizing groups of Pauli exponentiations can lead to significant reductions in circuit depth and gate count, thereby enhancing the performance of VQAs on near-term quantum hardware.

Existing works based on the representation of Pauli exponentiations can be divided into two categories:
\begin{itemize}
  \item \textbf{Gate cancelation via IR synthesis variations.}

  Each weight-$n$ Pauli exponentiation can be na\"ively synthesized by $R_z$ gates sandwiched between a pair of $n-1$ CNOT trees and single-qubit Clifford gates. Since the synthesized gate arrangement is not unique, Paulihedral~\cite{li2022paulihedral} and Tetris~\cite{jin2024tetris} identify gate cancelation opportunities between nearest-neighbor IRs, exposed by the variants of their synthesis schemes based on CNOT-tree unrolling. They proposed sophisticated co-optimization techniques that minimize the CNOT gate for both logic-level synthesis and SWAP-based routing, achieving good performance, especially on limited-topology NISQ devices. However, their optimization scope is limited to a finite set of local subcircuit patterns, and they rely solely on the CNOT-based ISA.

  \item \textbf{Tableau-based synthesis.}

  Tableau-based synthesis methods based on the binary symplectic form (BSF)---formal description of Pauli exponentiations exhibits highly-effective optimization to a more global extent. By representing the Pauli strings using the BSF, the complex problem of circuit synthesis is elegantly reformulated as the simplification of the BSF. The objective becomes to reduce the column weights of the BSF by strategically applying Clifford transformations. Each such transformation, while appearing as a simple algebraic manipulation on the BSF, corresponds to a non-local transformation on the quantum circuit. This method allows for the simultaneous simplification of multiple Pauli strings, a significant departure from the one-by-one or locally-focused optimizations found in other compilers. Rustiq~\cite{goubaultdebrugiere2024faster} and PHOENIX~\cite{yang2025phoenix} are two representative works in this category, both demonstrating substantial reductions in gate count and circuit depth across various quantum hardware architectures. Especially, PHOENIX exhibits quantum ISA-independent synthesis and optimization, making it broadly applicable across different quantum computing platforms. However, these methods are not totally topology-aware, requiring downstream qubit routing passes to adapt the synthesized circuits to specific hardware constraints.

\end{itemize}

\subsection{Code generation}

Once a quantum circuit has been logically synthesized and optimized, attention turns to its practical implementation on the underlying hardware. We will focus on backend transformations that are closely tied to the architectural features, especially regarding the qubit topology and the supported gate set. Correspondingly, the qubit mapping, routing, and ISA rebase passes are two critical components that adapt the logical quantum circuit to the constraints. The code generation workflow may differ for fault-tolerant quantum computation, which we will discuss in the next section. We will leave further exploration of these differences for that discussion.

\subsubsection{Qubit mapping and routing}

Qubit mapping and routing~\cite{zhu_quantum_2025} is one of the most well-explored topics in design automation and compiler research, as it shares similar methodologies with instruction scheduling and register allocation in classical computing.

Solving the backend topology constraint to map logical circuits onto physical devices is naturally modeled as a graph problem, where a logical circuit is represented as a directed acyclic graph (DAG) of gates, and the physical device is represented as a coupling graph of qubits. In most cases, even an optimal initial mapping cannot guarantee that all logical 2Q gates are mapped onto physically connected qubit pairs. The common solution is to dynamically change logical-to-physical qubit mappings by inserting SWAP gates, as a SWAP gate exchanges the state subspaces of two operand qubits, allowing non-adjacent logical qubit states to be moved next to each other. Typically, there are two kinds of approaches to this problem: (1) solver-based methods and (2) heuristic algorithms for scalable qubit routing.

\paragraph{Solver-based methods}

Solver-based methods address the qubit mapping and routing problem by first modeling it mathematically and then employing computational solvers to find a solution. This approach can yield optimal or near-optimal results, but its primary drawback is a lack of scalability. Since qubit mapping/routing is NP-hard in most scenarios, these exact methods typically suffer from exponential computational complexity, limiting their practical use to smaller circuits.

Below are prominent examples of solver-based strategies.
\begin{itemize}
  \item \textbf{Boolean Satisfiability (SAT) Solvers}

    SAT solvers identify solutions by finding satisfying assignments for Boolean variables. In quantum computing, they have been used to:
    \begin{itemize}
      \item Optimize the placement of adjacent qubits in sub-circuits~\cite{Hattori2018GateOrder2DNN}.
      \item Minimize T-gate and CNOT-gate counts~\cite{Meuli2018SATCNOT}.
      \item Synthesize optimal Clifford circuits~\cite{Schneider2023SATClifford}.
      \item Find near-optimal qubit mappings by navigating an exponentially large search space~\cite{willeMappingQuantumCircuits2019}.
    \end{itemize}
  \item \textbf{Satisfiability Modulo Theories (SMT) Solvers}

    SMT solvers generalize SAT by handling a broader range of variable types and constraints. They have been applied to:
    \begin{itemize}
      \item Layout synthesis aims to maximize circuit reliability or minimize duration \cite{Tan2020Optimal,Murali2019NoiseAdaptive}.
      \item Developing more succinct problem formulations and improved encoding for scalability \cite{Lin2022ScalableOptimalLayout}.
    \end{itemize}
  \item \textbf{Integer Linear Programming (ILP)}

    ILP is another exact-solver technique that formulates problems using integer variables within linear objective functions and constraints. It has been effectively applied to minimize quantum circuit depth while adhering to nearest-neighbor hardware constraints \cite{Bhattacharjee2019MUQUT,Bhattacharjee2017DepthOptimalPlacement}.
\end{itemize}

\paragraph{Heuristic algorithms}
Heuristic algorithms provide a scalable approach to qubit routing. An early example from Zulehner et al.~\cite{Zulehner2018Efficient} uses an A*-based search to minimize SWAP gate overhead by partitioning the circuit into concurrent gate layers. Adopting a similar layered approach, Li et al.~\cite{Li2019Tackling} introduced SABRE (SWAP-based bidirectional heuristic search), a bidirectional routing procedure that finds a better initial mapping to reduce the total number of inserted SWAP gates.

SABRE~\cite{Li2019Tackling} remains the de facto standard due to its balance between efficiency and mapping quality. SABRE adopts a greedy heuristic strategy to insert SWAP gates, maintaining a frontier set of two-qubit gates ready for execution but blocked by hardware connectivity. Candidate SWAPs that involve qubits in the frontier are evaluated using a distance-based cost function with a lookahead term to account for future gates. To further improve initialization, SABRE introduces a bidirectional optimization---it runs once in the forward direction to obtain a final mapping and then again on the reversed circuit to refine the initial layout. This combination of local heuristic evaluation and bidirectional refinement enables SABRE to achieve high-quality mappings with low computational overhead, making it a widely adopted baseline for subsequent routing algorithms, such as the current LightSABRE in Qiskit~\cite{Zou2024LightSABRE}.

While SABRE acknowledged the trade-off between gate count and circuit depth, it did not prioritize depth optimization. Subsequent research has increasingly focused on improving circuit depth and parallelism, using either SABRE-like heuristics~\cite{lao2022timing,ddroute2025,Zou2024LightSABRE} or graph matching techniques~\cite{childs2019circuit}. A notable contribution is TOQM, an A*-based method from Zhang et al.~\cite{Zhang2021Timeoptimal} that systematically targets depth-optimality and reports superior performance over prior solver-based approaches~\cite{Tan2020Optimal}. Nevertheless, achieving holistic routing optimality is rarely guaranteed by theoretical bounds, as it heavily depends on the specific ISA, device topology, and circuit cost model.

The recent emergence of advanced quantum ISAs---such as superconducting fractional gates, ion-trap partial entangling gates~\cite{yale2025realization}, and AshN gates~\cite{Chen2024One}---has spurred efforts to co-design compilation and routing to better leverage hardware-specific capabilities. For instance, McKinney et al.~\cite{mckinney2024mirage} investigated the practical performance of the $ \sqrt{\mathrm{iSWAP}} $ basis gate and its mirror, proposing a modified SABRE algorithm that combines gate decomposition with qubit routing. However, this work represents a preliminary attempt, with a limited scope for optimization and a lack of sophisticated algorithmic design.

\paragraph*{}There has also been extensive research on the qubit mapping problem beyond the standard setting. Noise-aware qubit mapping~\cite{Murali2020Software, Murali2019NoiseAdaptive, Tannu2019Not, Tannu2019Ensemble} takes the hardware noise model into consideration to find a qubit mapping and routing strategy that could mitigate hardware noise. Duckering et al.~\cite{Duckering2021Orchestrated} introduce Toffoli in the circuit synthesis step, running qubit mapping and routing directly on circuits with Toffoli gates and decomposing Toffoli gates according to the connectivity of the Toffoli operands. Hillmich et al.~\cite{Hillmich2021Exploiting} exploit dynamic circuits to create entangled Bell pairs on ancilla qubits in the device and use quantum teleportation to transfer a qubit far-away instead of using many SWAP gates.

\subsubsection{ISA rebase}

The final synthesis stage is the ISA rebase pass, which translates all abstract quantum gates into the hardware's native instruction set. On superconducting platforms, any single-qubit gate can be synthesized with minimal overhead using an exact sequence of virtual-Z and fixed-angle rotations. The primary challenge, therefore, lies in the synthesis of two-qubit gates. While two-qubit rebasing is a subset of the general unitary synthesis methods discussed in~\Cref{sec:unitary_synthesis}, it requires special consideration, as it must target a specific and sometimes complex set of hardware-native primitives. This task is typically approached using either analytical or numerical methods, each with distinct applicability.

\paragraph{Analytical rebase}
Analytical rebase refers to the synthesis of two-qubit unitaries using closed-form solutions that achieve the optimal two-qubit gate count. Such analytical solutions are available only for a limited set of basis gates. For example, any two-qubit unitary can be synthesized using at most three CNOT gates \cite{Vidal2004Universal}, two or three $\sqrt{\mathrm{iSWAP}}$ gates \cite{Huang2023Quantum}, or two B gates \cite{zhang2004minimum}. Peterson et al.~\cite{Peterson2022OptimalSynthesis} recently proposed optimal synthesis procedures for selected fractional $\mathrm{XX}$ rotation gates (e.g., ${\sqrt[3]{\mathrm{CNOT}},,\sqrt{\mathrm{CNOT}},,\mathrm{CNOT}}$). For other basis gates or heterogeneous ISAs, analytically optimal synthesis schemes remain unknown, necessitating the use of numerical methods.

\paragraph{Numerical rebase}
A brute-force numerical approach would involve a combined structural search for the two-qubit gate arrangement and numerical optimization for the single-qubit parameters. This is computationally expensive and struggles to guarantee optimality.
A more effective strategy leverages monodromy polytope theory~\cite{Peterson2020FixedDepthTwoQubitCircuits}. This framework can precisely determine the synthesis coverage---the set of achievable unitaries---for any circuit template constructed from a fixed number of basis gates. This transforms the problem: instead of a costly search, one can first select the optimal fixed-depth template for the target two-qubit gate. Subsequently, a much more tractable numerical optimization is performed only on the surrounding single-qubit gate parameters to find the final solution.

\subsection{What's next}
As noted, the multiplicity of compilation passes and the flexibility in their sequencing, coupled with the expansive design space of instruction sets, introduce considerable latitude in formulating quantum circuit synthesis and optimization problems---far beyond the conventional paradigm of compiling circuits into CNOT gates with single-qubit rotations. To thoroughly explore this research avenue, artificial intelligence-assisted approaches hold significant promise. Reinforcement learning, for instance, can navigate the complex landscape of synthesis pathways to identify optimal sequences, while deep learning models could enable rapid performance prediction across diverse synthesis schemes, collectively enhancing our ability to exploit the full potential of quantum circuit design.

In this section, our focus lies on the compilation for the bare-metal mode. While our ultimate goal is to advance toward large-scale fault-tolerant quantum computing—a realm where quantum algorithms will first be synthesized and optimized at the logical operation level, then compiled via logical operation schemes such as lattice surgery—we have not elaborated on this avenue here. This omission stems partly from our intention to reserve detailed discussion for the next section, which will cover the design of quantum error-correcting codes and fault-tolerant quantum computing. Additionally, this field is evolving at a rapid pace, making it prudent to address it with the most current developments in mind.

\section{Design in Quantum Error Correction and Fault-tolerant Quantum Computing}\label{sec:qec}

The theoretical advantages of quantum computing suggest that a perfect quantum system on the order of one thousand could break RSA-2048 encryption. However, at the physical implementation level, qubits across all technical pathways are vulnerable to environmental noise---their fragility is comparable to that of vacuum tubes in the mid-20th century, where even minor vibrations or temperature fluctuations are sufficient to disrupt their stability. Historically, the invention of the transistor completely replaced vacuum tubes due to its overwhelming advantages in energy efficiency and reliability, laying the cornerstone for modern computers; yet today's quantum computing still lacks a revolutionary carrier analogous to the transistor.

In the previous sections, we have discussed the design of quantum chips in~\Cref{sec:chip_design}, quantum instruction sets in~\Cref{sec:qisa}, and quantum circuits in~\Cref{sec:circuit}. All of these topics primarily focus on NISQ computing---a paradigm analogous to early vacuum tube-based computing.  While NISQ lacks theoretical guarantees of achieving quantum advantage, it has nonetheless spurred dedicated efforts across all the aforementioned design domains to maximize the overall performance of NISQ systems.

Before the quantum analog of the transistor emerges---and whether such a breakthrough will ever occur remains highly uncertain---does this mean accurate quantum computation is unattainable in the interim? Currently, the mainstream solution is QEC: encoding a single logical qubit via the collaboration of thousands of physical qubits, essentially trading quantity for quality. With a well-designed quantum error correction code and a sufficiently large number of physical qubits, we can achieve arbitrary accuracy for the logical qubit. While this ``quantum vacuum tube'' stacking strategy will drive hardware costs up by a factor of 1,000 to 10,000, the exponential speedup that quantum computing offers for specific problems is more than sufficient to offset this overhead. Just as classical computers once transcended the limitations of the vacuum tube era, quantum computing is similarly striving to achieve a practical breakthrough---before the advent of a ``quantum transistor''---by leveraging error correction codes as a transitional solution.

In this section, we will not delve into the underlying theory of quantum error correction codes themselves---curious readers may refer to~\cite{Nielsen2010Quantum, gottesman2024surviving}---but instead focus on their design-related aspects. More specifically, we will explore four core design-centric topics: (1) the key considerations guiding QEC code design, (2) the design of syndrome extraction circuits for these codes, (3) the practical implementation of decoders, and (4) the formulation of schemes for logical quantum operations.

\subsection{Quantum error correction code}

As quantum computing systems are susceptible to noise, the quantum error correction code (QECC) is essential for quantum applications beyond the NISQ era. Choosing a suitable QECC is crucial for building a fault-tolerant quantum computing system.

A QECC is often denoted by $\llbracket n,k,d\rrbracket$ where $n,k$ and $d$ are parameters of the code: the QECC encodes $k$ logical qubits into $n$ physical qubits, and $d$ is the code distance, meaning that transforming one logical state to another requires operations on at least $d$ qubits. A high code rate $k/n$ is preferable because it corresponds to a small overhead: When encoding the same number of logical qubits, a code with a high code rate requires fewer physical qubits. The code distance $d$ reflects robustness against noise. In addition to these parameters, the hardware requirement is also an essential factor to take into account when choosing a QECC, as we will discuss below.

A QECC should be frequently checked to determine if any physical error has occurred by running some quantum circuit, so that a correction could be applied or subsequent operations could be modified accordingly. Such a quantum circuit is called a syndrome extract circuit (SEC). A large family of QECCs is the stabilizer code~\cite{gottesman1997stabilizer}, which is defined as the mutual eigenspace of a set of Pauli operators. The standard syndrome extraction circuit consists of initializing an ancilla qubit, applying controlled Pauli gates between the ancilla qubit and the physical qubits in the support of the stabilizer, and finally measuring the ancilla. To implement such circuits efficiently, the ancilla qubit should be connected to the physical qubits associated with it, which poses requirements for the hardware. In addition, a qubit with a higher degree of connection will likely suffer from a higher level of noise due to frequency collision (see~\Cref{{sec:control_optimization}}), so it is preferable to consider QECCs with a low degree on the superconducting platform.

\begin{figure}[!ht]
    \centering
    \begin{subfigure}[b]{0.3\textwidth}
        \includegraphics[width=\textwidth]{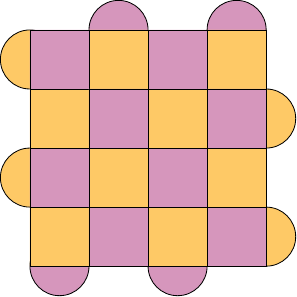}
        \caption{}
        \label{fig:surface_code}
    \end{subfigure}
    \hfill
    \begin{subfigure}[b]{0.6\textwidth}
        \includegraphics[width=\textwidth]{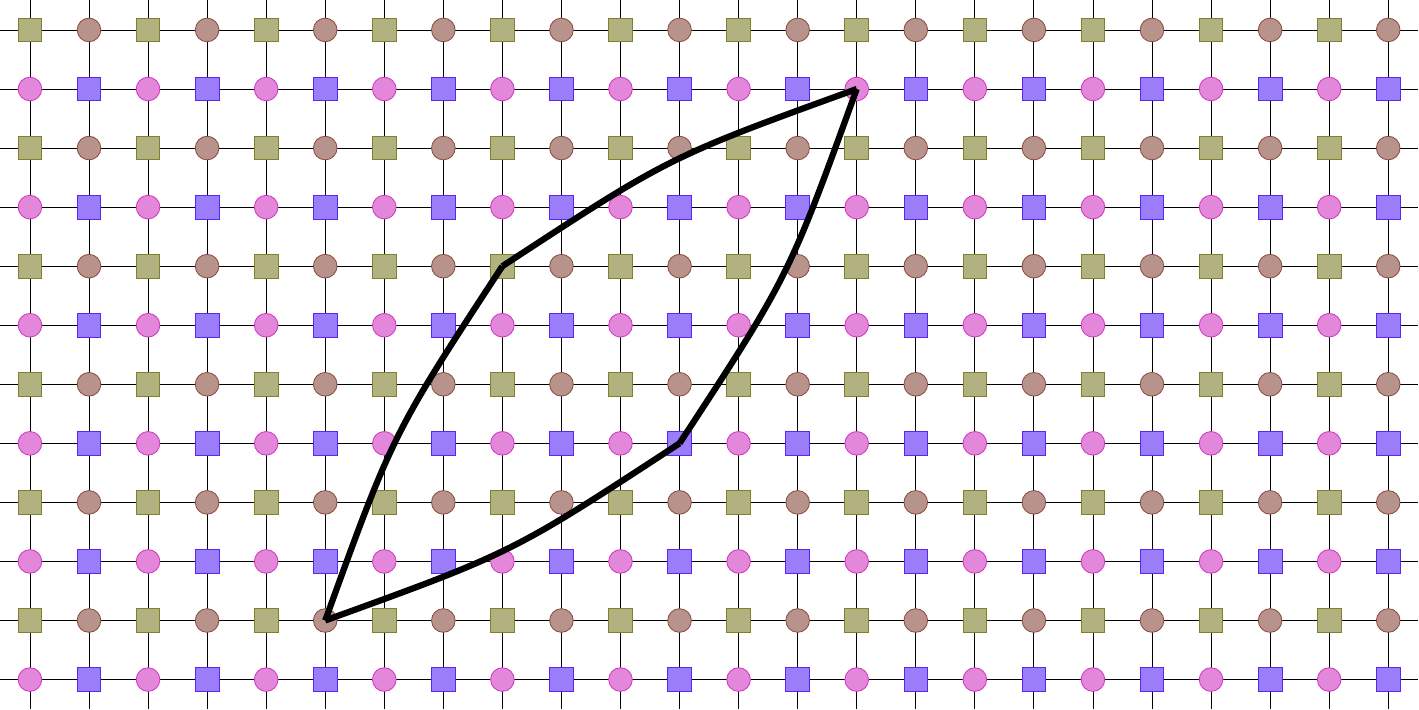}
        \caption{}
        \label{fig:bb_code}
    \end{subfigure}
    
    \caption{(a) The surface code of distance 5. The vertices represent data qubits, and the pink and yellow patches represent X and Z stabilizer checks, respectively. (b) BB code $\llbracket 144, 12, 12 \rrbracket$. The circles represent data qubits, and the squares represent ancilla qubits. In additional to the 2D grid with periodic boundary condition, there are long-range connections shown with thick lines. We only plotted some of the long-range edges for clarity.}
    \label{fig:codes}
\end{figure}

The most well-studied quantum error correction code for superconducting qubits is the surface code~\cite{bravyi1998quantum,freedman2001projective}, which only requires the connectivity of a 2D grid. The surface code with a code distance of up to 7 has been experimentally demonstrated on Google's Willow superconducting processor~\cite{Acharya2024Quantum}. The surface code of distance 5 is illustrated in~\Cref{fig:surface_code}. 2D color code~\cite{bombin2006topological} is another family of QECCs that can be implemented on a 2D surface. While they exhibit advantages in logical operations, their high connectivity requirements and lack of efficient decoders have limited research focus relative to the surface code.

The main drawback of the surface code and color codes is the substantial encoding overhead,  requiring a large number of physical qubits for each logical qubit. The surface code has parameter $\llbracket d^2, 1, d \rrbracket$, and it is estimated that hundreds or even thousands of physical qubits are needed to encode a single logical qubit~\cite{fowler2012surface}. The encoding overhead for the color codes is of the same order.

With the development of bump-bonding~\cite{rosenberg20173d,field2024modular,kosen2024signal,norris2025performancecharacterizationmultimodulequantum} and TSV~\cite{yost2020solid,mallek2021fabricationsuperconductingthroughsiliconvias,hazard2023characterization}, superconducting qubits and their connections are no longer restricted to a 2D layout. The bivariate bicycle code (BB code)~\cite{bravyi_high-threshold_2024} is a family of QECCs that can be implemented on a two-layer superconducting device with a much higher code rate. More specifically, the qubit connections of BB codes could be divided into two groups, and each could be placed on a 2D plane without crossing. Some of the BB codes with good parameters include $\llbracket 72, 12, 6 \rrbracket$, $\llbracket 90, 8, 10 \rrbracket$, $\llbracket 144, 12, 12 \rrbracket$, and $\llbracket 288, 12, 18 \rrbracket$, which have an order of magnitude higher code rate compared to the surface codes of the same code distance. An experimental demonstration of the $\llbracket 18, 4, 4 \rrbracket$ BB code has been reported in~\cite{wang2025demonstrationlowoverheadquantumerror}. The BB code $\llbracket 144, 12, 12 \rrbracket$ is demonstrated in~\Cref{fig:bb_code}.

\subsection{Syndrome extraction circuit}
After choosing a QECC for the system, there is additional freedom in the choice of SEC. The standard SEC initializes an ancilla for each stabilizer and then applies 2-qubit gates between each ancilla qubit and the data qubits associated with its stabilizer, followed by measurement of the ancilla qubits. In addition to this, there might be improved SEC designs that can relax hardware requirements. For example, the surface code could be implemented on a hexagonal lattice instead of a 2D square lattice~\cite{McEwen2023relaxinghardware}, reducing the qubit degree from 4 to 3. 

Due to fabrication problems, there might be defects in the 2D lattice of superconducting qubits, and adaptations should be made for surface codes to run on such devices. In~\cite{Siegel2023adaptivesurfacecode,wei_low-overhead_2025,Strikis2023quantumcomputing}, a method called superstabilizer was proposed, which allows the implementation of the surface code at the cost of reduced code distance. In~\cite{leroux_snakes_2024, debroy_luci_2024, zhou2024halmaroutingbasedtechniquedefect}, the accuracy loss has been significantly mitigated. Take~\cite{zhou2024halmaroutingbasedtechniquedefect} as an example. By leveraging an expanded 2-qubit gate set consisting of CNOT and iSWAP gates, the qubits could be routed around so that a functional ancilla qubit could be repurposed to measure a stabilizer associated with a defective qubit. By interleaving this type of special rounds and regular rounds, the defects could be mitigated without compromising the code distance.

For BB codes, it is possible to reduce the qubit degree from 6 to 5 by removing a short-range connection~\cite{Shaw2025lowering}. In~\cite{zhou2025louvrerelaxinghardwarerequirements}, the average degree could be further reduced to 4.5 and 4 with a limited increase in the logical error rate. The main observation behind this construction is that the structure of BB code allows reusing some of the connections. With an expanded gate set, the qubits could be routed so that a coupler could be reused in multiple stabilizer checks.

\subsection{Decoder}
After the syndrome data is obtained by running the SECs, a classical postprocessing process called ``decoding'' is required to infer the physical errors that occurred. The combined effect of the real physical errors and the inferred results should be a logical operator on the encoded qubits, and a logical error occurs if this combined operator is non-identity. There is often a trade-off between the accuracy and the efficiency of the decoder: The decoder with the lowest logical error rate often runs slower than the less accurate ones.

For any QECC, the decoder with the highest accuracy is the maximum likelihood decoder (MLD). Technically, one can enumerate all possible error configurations and sum up the probabilities corresponding to each logical operator. This is often too slow for practical purposes. Discussions on the MLD for the surface code could be found in~\cite{dennis2002topological}.

For a general QECC, a more efficient family of decoders is belief propagation (BP)- based decoders. BP~\cite{mackay1997near} has been widely used for classical error correction codes, but when applied to quantum codes, it often fails to generate a valid output due to the presence of degeneracies. Further post-processing should be applied to the output of BP, and some examples of such algorithms include ordered statistics decoding (OSD)~\cite{Panteleev2021degeneratequantum,roffe2020decoding}, localized statistics decoding (LSD)~\cite{hillmann2025localized}, etc. A recently proposed algorithm, Relay-BP~\cite{muller2025improvedbeliefpropagationsufficient}, achieves high decoding accuracy and exhibits inherent parallelism that facilitates efficient hardware implementation. Specifically, it has been successfully deployed on modern FPGA platforms~\cite{maurer2025realtimedecodinggrosscode}, enabling real-time decoding performance.

For the surface code, significantly more efficient decoders could be found based on the following observation: By decomposing the $Y$ errors into an $X$ error and a $Z$ error, the $X$ and $Z$ errors could be handled separately. Furthermore, each Pauli could flip at most 2 syndromes, so the decoding of the surface code could be reduced to a min-weight perfect matching (MWPM) problem on a graph. This essentially finds a single error pattern with max likelihood, instead of the set of error patterns that lead to the logical operator with max likelihood, and therefore there is a slight loss in accuracy. The blossom algorithm for the MWPM problem was first proposed in~\cite{Edmonds1973MatchingET}, and there were implementations of the algorithm with improved performance for surface code decoding~\cite{higgott2022pymatching,wu2023fusion}. An improved accuracy could be achieved by utilizing the correlation between $X$ and $Z$ errors and combining the BP and MWPM algorithms, as proposed in~\cite{higgott2023improved}. A decoder closely related to MWPM is the union-find (UF) decoder~\cite{Delfosse2021almostlineartime}, which finds a perfect matching on the graph that is close to the minimum weight. With the slight cost of accuracy, the union-find decoder has the advantage of being implementable on FPGA~\cite{liyanage2023scalable}.

On real quantum hardware, there might be complicated error models with sources not fully understood, and the performance of these traditional algorithms might be suboptimal compared to machine learning decoders. AlphaQubit~\cite{bausch2024learning} is one such example with the highest known accuracy for the surface code.

To avoid exponential backlog~\cite{Terhal2015QECreview}, the decoder should keep up with the speed of syndrome generation. Syndromes are generated with 1 $\mu s$ per round~\cite{ryananderson2021realization}, which poses a significant challenge to the decoder. One possible solution is to design parallel algorithms that utilize multiple processors to speed up the decoding. In addition to intrinsically parallel decoders such as Fusion Blossom~\cite{wu2023fusion}, most decoders could be parallelized with the sliding-window approach~\cite{tan2023scalable,Skoric2022ParallelWD,bombin2023modular,zhang2025learningneuraldecodingparallelism}. The multiple rounds of syndromes are divided into windows and each window could be decoded in parallel before being merged.

\subsection{Fault-tolerant logical operations}
To run a quantum algorithm on the encoded qubits, logical operations should be applied in addition to the syndrome extraction circuits. The mainstream approach is to compile the target circuit into Clifford gates plus a non-Clifford gate (often the $T$ gate or the Toffoli gate). For most QECCs, Clifford gates are relatively easy to implement, while non-Clifford gates require the generation of magic states and gate teleportation~\cite{bravyi2005universal}. 
logic-level
For surface code, a well-known approach to implement Clifford gates is lattice surgery~\cite{Horsman_2012,Litinski2018latticesurgery,fowler2019lowoverheadquantumcomputation,Litinski2019gameofsurfacecodes}, which measures the product of logical Pauli operators between multiple logical qubits. For joint measurement on logical qubits encoded in distant surface code patches, we either move them closer or use some of the physical qubits between them as routing space. A natural optimization problem is to compile the logical circuit so that lattice surgery operations could run in parallel without occupying the same routing space. Some early designs could be found in~\cite{Litinski2019gameofsurfacecodes}, and more efficient compilers include \url{latticesurgery.com}~\cite{Watkins2024highperformance}, LaSsynth~\cite{tan2024satscalpel} and TQEC~\cite{TQEC}.

Lattice surgery could be generalized to other QECCs~\cite{cross2024improvedqldpcsurgerylogical,swaroop2025universaladaptersquantumldpc,zhang2025time}. In~\cite{yoder2025tourgrossmodularquantum}, a fault-tolerant quantum computing architecture based on this generalized surgery construction with BB codes has been proposed.

The generation of high-fidelity magic states is generally considered costly. The traditional approach is magic state distillation~\cite{bravyi2005universal, Gidney2019efficientmagicstate, Litinski2019magicstate, Heussen2025efficientfault}, which produces a magic state with high fidelity by consuming multiple copies of low-fidelity magic states. Low-fidelity magic states could be obtained by state injection~\cite{landahl2014quantumcomputingcolorcodelattice, Li2015magicstate}. A different approach is based on the observation that color codes have transversal Clifford gates, i.e., logical Clifford gates could be implemented as gates applied individually on each physical qubit, so that magic states encoded with color codes can be prepared with high fidelity. This line of approaches started from~\cite{chamberland2020very, itogawa2025efficient} and culminated with the magic state cultivation scheme~\cite{gidney2024magicstatecultivationgrowing}, which greatly saves resources and can generate reasonably high-fidelity $\ket{T}$ states as cheap as CNOT gates in lattice surgery. The states generated are not encoded with the surface code, but they could still be used for applying non-Clifford gates in lattice surgery or fed into the magic state distillation protocol for even higher fidelity.

\subsection{What's next}
Besides theoretical explorations, the design of QECC and FTQC ultimately aims to facilitate the deployment of quantum hardware. While recent years have seen a range of experimental demonstrations of QECC---encompassing not only code implementation but also fast and accurate decoding---progress toward demonstrating logical operations is imminent. It is important to recognize, however, that these achievements remain confined to the realm of demonstrations; scalability persists as the critical barrier to unlocking FTQC's full potential. In this section, we have illustrated how cross-abstraction-layer approaches can enhance FTQC performance across various stages, such as defect mitigation and hardware deployment for BB codes. Significant opportunities thus remain for designing scalable, robust building blocks for FTQC systems.

\section{Logic-level Verification and Test}
In the realm of classical electronic design automation, verification and testing are critical processes to ensure system functionality and stability while identifying potential errors~\cite{wangEDAsvt2009}. Similarly, in the logic-level design phase of QDA, verification and testing play an essential role in guaranteeing that high-level components of quantum systems conform to design specifications. These components include logical quantum circuits, quantum programs, and other types of abstraction such as transition systems, which are all primary concerns of logic-level verification. This section provides a comprehensive overview of methodologies, tools, and domain-specific approaches for conducting logic-level verification and testing. The discussion is organized as follows: We start from recent progress in classical simulation of quantum circuits, and then explore key verification techniques, encompassing formal verification and simulation-based approaches, including decision diagram-based methods~\cite{WilleDD2023}, ZX-calculus~\cite{Peham2022Equivalence}, and proof assistants~\cite{ZhouCoqq2023,KissingerQuantomatic2015,CharetonQbricks2020,LiuQHLprover2019}. Subsequently, we address three specific verification scenarios: equivalence checking, program verification, and model checking.

\subsection{Classical simulation}

We begin with the classical simulation of quantum circuits. While classical simulation traces its roots to the early days of quantum information science, it has garnered renewed and intense interest since the publication of the ``quantum supremacy'' paper~\cite{Arute2019Quantum}. This renewed focus stems from a critical insight: understanding the limits of classical simulatability not only defines the boundaries of what quantum computers can achieve but also illuminates where their unique advantages---over classical counterparts---are likely to emerge.

The most basic quantum circuit simulation is \textbf{state-vector simulation} (Schrödinger simulator), tracking an $n$-qubit state via a $2^n$-amplitude vector $\ket{\psi}$ stored as a length-$2^n$ array, with $\ket{\psi}_{i_{n-1}\cdots i_0}$ at index $i_{n-1}2^{n-1}+\cdots+i_0$.

In general, when simulating a gate, all amplitudes need to be updated. For example, to perform a single qubit gate $U = \left(\begin{array}{cc}
    u_{00} & u_{01} \\
    u_{10} & u_{11}
\end{array}\right)$ on the $i$-th qubit of $\ket{\psi}$ to obtain $\ket{\psi'}$:

\begin{equation}
            \left\{\begin{aligned}
                \ket{\psi'}_{c_{n-1}\cdots c_{i+1}\textbf{\color{red}0}c_{i-1}\cdots c_0}=&u_{00}\ket{\psi}_{c_{n-1}\cdots c_{i+1}\textbf{\color{red}0}c_{i-1}\cdots c_0}
                +u_{01}\ket{\psi}_{c_{n-1}\cdots c_{i+1}\textbf{\color{red}1}c_{i-1}\cdots c_0}\\
                \ket{\psi'}_{c_{n-1}\cdots c_{i+1}\textbf{\color{red}1}c_{i-1}\cdots c_0}=&u_{10}\ket{\psi}_{c_{n-1}\cdots c_{i+1}\textbf{\color{red}0}c_{i-1}\cdots c_0}
                +u_{11}\ket{\psi}_{c_{n-1}\cdots c_{i+1}\textbf{\color{red}1}c_{i-1}\cdots c_0}
            \end{aligned}\right.
        \end{equation}

Simulating $m$ gates on an $n$-qubit state vector carries a computational complexity of $O(m2^n)$---an exponential scaling with $n$. However, through strategic optimizations, simulations for intermediate value of $n( \approx 50)$ become feasible. These optimizations leverage cache locality and parallelism: 
\begin{itemize}
    \item Performing even a single gate on $\ket{\psi}$ requires traversing and modifying all amplitudes on $\ket{\psi}$. Therefore, if we can fuse multiple smaller gates into a larger gate, we can reduce the number of times traversing the entire state vector.
    \item All amplitudes are grouped into pairs of $2$ amplitudes. These computations are all independent of each other and can be parallelized.
    \item The single qubit gate is performed on $2^{n-1}$ pairs of amplitude: $\ket{\psi}_{c_{n-1}\cdots c_{i+1}\textbf{\color{red}0}c_{i-1}\cdots c_0}$ and $\ket{\psi}_{c_{n-1}\cdots c_{i+1}\textbf{\color{red}1}c_{i-1}\cdots c_0}$. According to our previous memory layout assumption, the distance between the two amplitudes is exactly $2^i$. If we chunk the state vector into $2^L$-amplitude chunks where $L\geq i$, every amplitude will find its ``partner'' in the same chunk.
\end{itemize}

We will briefly explain several techniques for speeding-up the state-vector simulator:

\textbf{Local/Global Qubit Partitioning} exploits qubit ordering: lower-index (local) qubits yield cache-friendly gates, as $2^L$-amplitude cache blocks contain all needed amplitudes. Higher-index (global) qubits require cross-block access. Circuit segmentation by local gates, with matrix transposes between segments, optimizes locality. Controlled gates and diagonal gates (e.g., QMUX~\cite{Shende2006Synthesis}) remain local-friendly, as they avoid cross-block updates.

\textbf{Gate Fusion} reduces state traversals by combining consecutive gates. For $L$-qubit local segments:
\begin{enumerate}
\item Gate-by-gate: Processes gates sequentially in cache (saves bandwidth)~\cite{smelyanskiyQHiPSTERQuantumHigh2016}.
\item Full fusion: Combines gates into a $2^L{\times}2^L$ matrix, with one cache traversal~\cite{bayraktarCuQuantumSDKHighPerformance2023}.
\end{enumerate}

Full fusion benefits from fixed $O(2^{2L})$ complexity and optimized BLAS implementations but suffers overhead for few gates and incompatibility with controlled gate relaxations. Modern simulators (e.g., HyQuas~\cite{zhangHyQuasHybridPartitioner2021}, Atlas~\cite{xuAtlasHierarchicalPartitioning2024b}) use hybrid strategies to balance these trade-offs through greedy grouping or dynamic programming.

The \textbf{tensor network} formalism generalizes the matrix multiplication operations needed in state-vector simulations. 
A tensor is a generalization of the two-dimensional matrix with an arbitrary number of dimensions, and a tensor network is a mathematical object constructed by connecting multiple tensors. The value of a tensor network is another tensor, which can be computed using the natural algorithm of contraction.

An important aspect of the contraction procedure is the contraction order, which determines the time complexity and space complexity of the contraction. The ability to change the contraction order gives the tensor network approach a lot of flexibility. For example, the gate fusion strategy described in the previous section is merely a special case of contraction ordering. For another example, when simulating shallow circuits with limited connectivity, the optimal contraction order often uses a spatial-major order instead of the temporal-major order used in state-vector simulations. This takes advantage of the fact that different spatial regions in such a circuit are only loosely entangled. This approach works especially well when we need to know only a few amplitudes of the final state vector instead of all of them.

\textbf{Tensor network slicing.} Even with the optimal contraction order, the simulation cost for a sufficiently complex circuit will eventually grow exponentially with the circuit size. To push the boundaries of circuit simulation, it is a natural idea to utilize parallelism on a larger scale. A simple and often effective method of parallelizing tensor network contraction is \textbf{slicing}, where one tensor network $G$ is split into two simpler tensor networks $G_1$ and $G_2$ with the same shape, such that the value of $G$ is equal to the sum of the values of $G_1$ and $G_2$. The contraction cost of one part ($G_1$ or $G_2$) is usually somewhat higher than half of the contraction cost of $G$ due to some duplicate calculations between $G_1$ and $G_2$, but this overhead is usually quite small as long as the slicing is done at the optimal location. By iterating this procedure $k$ times, one can split a tensor network into $2^k$ parts, each more manageable than the original tensor network in terms of both time complexity and space complexity.

\textbf{Low-rank approximation.} By default, tensor network contraction is an exact algorithm; however, in practical use cases of quantum simulation, an approximate result is often sufficient. A method to further reduce the computational cost of tensor network contraction at the expense of precision is low-rank approximation, typically achieved by decomposing large tensors into matrix product states (MPS), which are a special kind of tensor network. Such a decomposition increases the number of vertices in the overall tensor network but decreases its local complexity, making it easier to transform and ultimately contract.

In general, an exact decomposition to MPS will not decrease the contraction cost due to the high bond dimension $\chi$ necessary, which is equal to the number of non-zero singular values of the tensor being decomposed. However, in practice, one can often keep only a few of the largest singular values and discard the rest without losing too much precision. Thus, the bond dimension $\chi$ can be a tunable parameter that allows a tradeoff between computational cost and precision.
%\subsubsection{Circuit cutting}

An extensive body of quantum circuit simulators has been developed~\cite{JavadiAbhari2024Quantum, steigerProjectQOpenSource2018, smelyanskiyQHiPSTERQuantumHigh2016, jonesQuESTHighPerformance2019, zhangHyQuasHybridPartitioner2021, xuAtlasHierarchicalPartitioning2024b, bayraktarCuQuantumSDKHighPerformance2023, suzukiQulacsFastVersatile2021, huang2021efficient, pan2022simulation, pan2022solving}, each distinguished by unique design objectives and optimizations tailored to specific hardware platforms. Examples of such hardware-specific enhancements include acceleration via SIMD and multithreading~\cite{suzukiQulacsFastVersatile2021, smelyanskiyQHiPSTERQuantumHigh2016}, compatibility with MPI clusters~\cite{haner05PetabyteSimulation2017}, optimization for GPUs~\cite{zhangHyQuasHybridPartitioner2021, xuAtlasHierarchicalPartitioning2024b}, and adaptation to many-core supercomputers~\cite{liQuantumSupremacyCircuit2018}. For a comprehensive overview of state-vector-based simulators, we direct readers to~\cite{jamadagniBenchmarkingQuantumComputer2024}. Notable advances in efficient tensor network-based simulations include the work by Huang et al.~\cite{huang2021efficient} and Pan et al.~\cite{pan2022simulation, pan2022solving}. 

\subsection{Verification techniques}
\textbf{Decision diagrams} fundamentally rely on data compression. By exploiting structural regularities and node sharing, along with practical techniques such as the unique table and computed table, decision diagrams (DD) can achieve up to exponential compression in the storage required for representing function mappings~\cite{BryantROBDD1986}. As the algebraic domains underlying decision diagrams have been generalized from Boolean values to more complex structures, various forms of decision diagrams~\cite{NiemannQMDD2016, HongTDD2022, SistlaCFLOBDD2024, WilleDD2023} with reduced ordered normal forms over the complex domain have been employed in a wide range of tasks for quantum computing verification. To establish a foundational representation, Quantum Multiple-Valued Decision Diagrams(QMDD) were formally refined for general quantum cases~\cite{NiemannQMDD2016}, enabling features such as variable order changes. DDSIM~\cite{wille2019advanced} subsequently proposed a new graph-based simulation approach by revisiting how quantum states and operations can be represented. Numerous variants of DDs have been developed by integrating concepts from physical and mathematical formalisms. Notable examples include Tensor Decision Diagrams (TDD)~\cite{HongTDD2022}, Local Invertible Map Decision Diagrams (LIMDD)~\cite{Vinkhuijzen2023limdddecision}, Feynman Decision Diagrams (FeynmanDD)~\cite{wang2025FeynmanDD}, and Context-Free-Language Ordered Decision Diagrams(CFLOBDD)~\cite{SistlaCFLOBDD2024}.

The variety of decision diagrams employed in quantum computing extends well beyond the aforementioned types. A framework for evaluating real-valued decision diagrams was established in~\cite{FargierKnowledge2014}, where the criteria of succinctness and traceability were introduced. Building on these dimensions, one can construct a knowledge compilation map that accommodates the diverse decision diagram structures designed for quantum computing.

The \textbf{ZX-calculus} is a graphical language employed in quantum computing, with its core relying on the use of ZX-diagrams to represent quantum operations~\cite{Coecke_2011}. In contrast to traditional quantum circuit models, the ZX-calculus expresses quantum operations as interconnected networks of fundamental graphical elements, such as Z-spiders and X-spiders, and implements rigorous reasoning and simplification through a complete set of rewriting rules~\cite{Backens_2014}. This feature enables the ZX-calculus to demonstrate distinct advantages in the field of quantum verification. In recent years, the ZX-calculus has been integrated into automated verification frameworks and software tools (e.g., PyZX~\cite{kissinger2020Pyzx}, MQT-QCEC~\cite{burgholzerQCECJKQTool2021}), demonstrating promising prospects in verifying the correctness of large-scale quantum circuit optimizations and the security of quantum protocols. Therefore, the ZX-calculus is increasingly recognized as one of the core methods in quantum verification research, possessing both theoretical completeness and practical engineering value.

\textbf{Proof assistants} have found widespread applications in the formalization of mathematics and in reasoning about program logics. Several proof assistants adapted for quantum contexts---such as Coq~\cite{ZhouCoqq2023}, Isabelle~\cite{LiuQHLprover2019}, Agda~\cite{Bian_2023}, and Why3~\cite{CharetonQbricks2020}---have been employed for the formal reasoning of quantum program logics and quantum circuit properties, as well as for tasks such as certified compilation and correctness proofs of quantum algorithms. From the perspective of their mathematical foundations, a wider range of classical logic–based tools has also been applied to logic-level quantum verification. For instance, SMT solvers~\cite{CharetonQbricks2020, BauersymQV2023} are commonly utilized in symbolic execution, while rewriting–based languages such as Maude have been adopted in symbolic quantum model checking~\cite{do2024symbolic}.

\medskip

In electronic design automation, simulation-based verification focuses on analyzing a circuit's input–output behavior under given testbenches, whereas formal verification relies on mathematical reasoning to establish and prove model properties~\cite{wangEDAsvt2009}. The aforementioned techniques can serve as a foundation for formal verification, but they are also widely employed as tools for the classical simulation of quantum circuits. Within logic-level quantum verification and testing, the scalability of matrix–vector–based simulation is inherently limited by the principles of quantum mechanics, necessitating the exploration of computational potential at the mathematical and formal levels. Another strategy for classical simulation from the perspective of tensor networks will be discussed in the next section.

\subsection{Equivalence checking}

Equivalence checking is a fundamental component in the compilation of quantum circuits, ensuring that the transformed quantum circuit preserves the original functionality. As previously discussed, this has been proven to be a QMA-complete problem~\cite{Janzing2005}. 

Direct matrix comparisons are computationally prohibitive, so practical methods rely on compact representations, such as decision diagrams or ZX-diagrams. Decision-diagram-based approaches enable efficient equivalence checking by exploiting redundancy~\cite{Hong2022Equivalence,Wille2022Tools,Burgholzer2020Advanced}, while the ZX-calculus allows graph-based reasoning via rewriting rules~\cite{Peham2022Equivalence}.
Tensor networks, such as those employing Matrix Product Operators, have also been demonstrated to be a scalable method for equivalence checking~\cite{Sander2025Equivalence}. Another approach relies on simulation, where equivalence is tested by evaluating circuit output for a selected set of input states~\cite{Burgholzer2020The,Burgholzer2021Random}, although at the cost of a trade-off between testing efficiency and accuracy. Other specialized approaches, such as partial equivalence checking~\cite{Chen2022Partial} and model counting~\cite{Mei2024Equivalence}, can also be employed for the equivalence checking of quantum circuits.

The advent of dynamic quantum circuits, which allow mid-circuit measurement and classically controlled operations, poses additional challenges due to their non-unitary behavior and data-dependent control flow. Novel decision-diagram-based algorithms have been proposed to characterize the functionality of these circuits using ensembles of linear operators~\cite{Hong2022Equivalence}.  
Another approach is to transform circuits that contain dynamic circuit primitives into unitary-only circuits to utilize existing verification techniques~\cite{Wille2022Handling}.

Approximate equivalence checking has emerged as an important direction for near-term quantum computing, where hardware noise and approximation-based compilation render exact equivalence overly restrictive. The relevant methods evaluate circuit equivalence up to a given fidelity threshold or diamond-norm bound~\cite{Hong2022Approximate,huang_approximation_2025}, providing practical assurance of functional correctness while tolerating small deviations.
    
\subsection{Program verification}
Formal verification of quantum programs, rooted in classical formal methods like Hoare logic~\cite{Hoare1969axiomatic,Reynolds1998PL}, aims to provide mathematically rigorous proofs of correctness, safety, and reliability for quantum algorithms and programs.
Over the past two decades, this field has evolved from early conceptual explorations to the development of logically complete frameworks and, more recently, to mechanized and automated tools.

The earliest attempts to generalize Hoare logic to the quantum domain emerged in the mid-2000s. Kakutani~\cite{Kakutani2009logic} proposed a Hoare-style logic for Selinger's Quantum Programming Language~\cite{Selinger04Towards}, extending probabilistic Hoare logic to density-matrix semantics. Chadha et al.~\cite{Chadha2006Reasoning} developed an ensemble semantics that combines classical and quantum information states, along with a corresponding logic system named Ensemble Exogenous Quantum Propositional Logic (EEQPL).
Building on the notion of quantum predicates proposed by D'Hondt and Panangaden~\cite{DHondt2006wlp}, Feng et al.~\cite{Feng2007proofrules} defined the weakest precondition (wp) and weakest liberal precondition (wlp) semantics for a simple quantum while-language. While these early frameworks were all sound for reasoning about quantum programs, they lacked the crucial property of completeness.

A major breakthrough came with the development of the first full-fledged Floyd–Hoare logic for quantum programs, which includes a novel proof of relative completeness~\cite{YingQHL2012}. This paper also introduces ranking functions to ensure the (almost-sure) termination of quantum programs. Subsequent extensions incorporated classical variables~\cite{FengQHL2021}, parallel composition~\cite{ying2022proof}, nondeterministic choices~\cite{feng2023verification}, and distributed quantum programs involving classical communication~\cite{feng2022verification}. To enhance automation, later studies formulated invariant~\cite{ying2017invariants} and ranking function generation~\cite{li2018algorithmic} as semidefinite programming problems. Alternative approaches include projector-based Hoare logic~\cite{Zhouapplied2019} and relational Hoare logic~\cite{barthe2019relational, Dominique2019qrhl}.
On the practical side, quantum Hoare logics have been implemented in interactive proof assistants~\cite{ZhouCoqq2023, Rand_2018, CharetonQbricks2020, LiuQHLprover2019}, and specialized variants based on stabilizer states~\cite{huang2025} have been successfully applied to verify real-world systems such as quantum error-correction codes. 

For a comprehensive overview of quantum program verification techniques, readers are referred to~\cite{ying2024foundations, Lewis2023formal}.

\subsection{Model-checking}

Quantum model checking is primarily applied during the design phase of quantum systems, such as quantum algorithms, programs, and circuits, to verify their properties automatically and exhaustively. While the expressive power of temporal logics enables the verification of complex properties in sophisticated systems~\cite{BaierPrinciples2008,clarke1997model}, the application of this technique to quantum systems faces amplified challenges, such as state-space explosion and complex probabilistic dynamics.

Early research in quantum model checking introduced exogenous quantum propositional logic~\cite{MateusTemporal2009}, including quantum computation tree logic and quantum linear-time logic, which were used to verify certain quantum communication protocols~\cite{Gay_Nagarajan_Papanikolaou_2009}. Subsequent work enriched the theoretical foundations for modeling and analyzing quantum systems. 
In particular, a notion of quantum Markov chains (QMCs) was introduced in~\cite{Ying2013ReachProb}, which is a pair $\langle \mathcal{H}, \mathcal{E}\rangle$, where $\mathcal{H}$ is a finite-dimensional Hilbert space and $\mathcal{E}$ is a trace-preserving quantum super-operator on it. Based on this model, a range of model-checking techniques have been developed~\cite{Ying2014modelLT,Guan2024Measurement,Li2015QMC}. 

An alternative but expressively equivalent model, called super-operator valued Markov chain (SVMC) was proposed in~\cite{FengModel2013}. An SVMC is a pair $\langle S, Q\rangle$, where $S$ is a countable set of classical states, and $Q$ 
assigns a super-operator to each pair of states such that for each $s \in S$, $\sum_{t \in S} Q(s, t)$ is trace-preserving. While expressively equivalent to the QMC model~\cite{Li2015QMC}, the SVMC formalism is particularly well suited for describing quantum programs and has been effectively applied to their verification and termination analysis~\cite{FengModel2013,feng2017model,Feng2013reachRecur}.
 
 Recent years have seen growing interest in developing practical quantum model checkers~\cite{Fengqpmc2015, Daiqreach2024, do2024symbolic, Guan2024Measurement}. These efforts seek to leverage optimization strategies from classical model checking, such as symbolic~\cite{McMillan_1993_sym} and bounded model checking~\cite{biere2021bounded}, and integrate them with existing classical verification frameworks. Techniques such as decision diagrams are also being explored to enhance the scalability of quantum model checking~\cite{Lukas2020Improved,Daiqreach2024}.

 For readers seeking a deeper technical understanding, the monograph~\cite{Ying_Feng_2021} offers a comprehensive treatment of quantum model checking, covering topics such as quantum graph theory, fixed-point algorithms, and various temporal logics tailored for quantum systems.
 
 \subsection{What's next}

Despite the significant progress in logic-level verification and testing for quantum systems, scalability and efficiency remain the foremost challenges. Existing formal verification and model-checking techniques often suffer from exponential resource demands, making them impractical for large-scale or noisy quantum programs. A promising future direction is to exploit structural properties and domain-specific symmetries in quantum circuits and programs---such as stabilizer subspaces, tensor network decompositions, and commutativity patterns---to develop more scalable symbolic and hybrid verification methods.

Another important line of research lies in distributed and communication-centric quantum systems, including quantum networks and cryptographic protocols. These systems involve intricate interactions between classical and quantum information, requiring compositional and probabilistic reasoning frameworks that go beyond current Hoare-style and model-checking approaches. The integration of formal verification with quantum cryptographic analysis, for example, verifying the security properties of protocols like QKD or teleportation-based schemes, represents an essential and challenging goal.

Finally, bridging the gap between logic-level verification and practical testing will be crucial for end-to-end reliability in quantum software engineering. This involves coupling static formal methods with dynamic testing, runtime verification, and certified compilation. Advances in tool automation, decision-diagram optimization, and proof assistant integration will further enhance the applicability of formal methods, enabling robust verification workflows across the entire quantum software stack.

\section{Summary and Outlook}

In this paper, we have presented a comprehensive review of the emerging field of QDA, detailing its foundations, challenges, methodologies, and tool development. We have explored the critical aspects of the QDA, ranging from compiling quantum algorithms into quantum circuits to the physical implementations of quantum computing hardware, with a particular focus on superconducting quantum computing. We have discussed the key stages in the QDA workflow, including quantum instruction set design, quantum circuit synthesis, verification and simulation, quantum error correction, chip design and simulation, control design, decoherence modeling, test data analysis, package and cryogenic setup design, and TCAD, enabling a holistic understanding of designing quantum computing systems. We emphasized the necessity of bridging the gap between computer science-driven logic-level design and physics-oriented physical-level design, highlighting the importance of co-design methodologies that integrate both perspectives to advance scalable and practical quantum computing. 

Looking toward the future, the development of QDA is becoming a cornerstone in realizing scalable quantum computation. As quantum hardware continues to advance, the demand for QDA tools and methodologies will grow increasingly urgent. Beyond the forward-looking insights outlined in the ``What's Next'' of each section, several grand challenges stand out, each intertwined with transformative opportunities that will shape the future not only of QDA but also of quantum computing in general.
\begin{itemize}
    \item \textbf{Simulation: }As the number of qubits increases, the computational cost of exactly simulating quantum computing processes grows exponentially. Therefore, it is necessary to investigate multi-scale modeling strategies for quantum computing systems, which involve abstractions and approximations, from the chip's geometric patterns and environmental noise to the logical operations and error models in quantum circuits, while balancing simulation efficiency and accuracy. Furthermore, empirical compensation models built using measurement data are crucial for quickly and accurately predicting the properties and operations of quantum processors. 
    
    \item \textbf{Co-design: }We have showcased several examples of bridging physical-level and logic-level design, including quantum instruction set design, quantum error correction code design, and their execution. However, most of these system improvements remain human-designed. A key challenge is establishing hardware-software design frameworks that account for factors such as instruction set extensions, chip layout, parameter constraints, noise characteristics (types and amplitudes), and manufacturing process variations—all of which impact quantum algorithm performance. Developing automated, integrated tools and workflows capable of co-optimizing quantum circuits and hardware is, therefore, crucial.
    
    \item \textbf{Fault-tolerance: }QECC and FTQC represent the mainstream approaches to realizing the transformative potential of quantum computing. However, QECC itself demands careful hardware-software co-design. Ultimately, to streamline such design workflows in the FTQC era, future QDA tools must integrate architectural frameworks and algorithmic capabilities. These should support three key objectives: (1) the design and optimization of error-correcting codes, (2) compilation passes tailored for FTQC, and (3) the development of quantum hardware compatible with these codes. Notably, many of these critical directions remain underexplored.
    
    \item \textbf{Multiplexed control architecture: }Multiplexed control is a promising architecture for scaling up superconducting quantum computers~\cite{Shi2023multiplexedControl, zhao2024MultiplexedControl}. This technique significantly reduces the number of control lines needed to manipulate qubits, which simplifies the cryogenic system and reduces the need for complex electronics~\cite{Acharya2023cryo-CMOSmultiplexer}. 
    By using shared control lines, the number of wires can be drastically reduced. 
    However, this approach presents challenges, notably the risk of crosstalk between qubits and control lines. 
    Another trade-off is the increased complexity of the control system. While the number of wires is reduced, the control electronics must be able to generate and route the multiplexed signals with high precision. 
    When compiling quantum circuits, it is imperative to consider the limitations that multiplexing imposes, particularly the limitation on the number of parallel gate operations, which is restricted by the number of control lines~\cite{Richter2025multiplexedControl}.
    Achieving an optimal design that balances the number of control lines and the execution time of a quantum algorithm necessitates that the QDA tools incorporate the workflow, from routing shared control lines and placing components for parasitic coupling suppression to compiling the quantum circuit compatible with a specific implementation of the hardware's multiplexing.
    
    \item \textbf{Integration with classical EDA: }Leveraging the expertise and tools developed in classical EDA will be helpful for advancing the QDA. For example, place and route tools from classical VLSI design can be adapted for qubit placement and routing on a quantum chip, minimizing crosstalk and ensuring signal integrity. Simulation tools, such as those used for parameter extraction and signal integrity analysis, can be employed to model and mitigate the impact of noise and interference on control errors. Formal verification techniques can be applied to verify the correctness of quantum circuit designs and ensure that they meet specifications. Optical proximity correction (OPC), a technique used to compensate for diffraction and other optical effects during lithography, can also be adapted to improve the yield of quantum device fabrication. Leveraging these established methodologies and tools allows for the acceleration of QDA development.
        
    \item \textbf{Artificial intelligence: }The AI techniques can be applied to various aspects of QDA, such as quantum circuit optimization, decoding in quantum error correction, quantum optimal control~\cite{Sivak2022ModelFree, Nguyen2024ReinforcementLearningPulse, Genois2024quantumoptimalcontrolsuperconducting}, measurement data analysis~\cite{Castelano2023physicsinformedneuralnetworks, Kung2025Automatic}, and quantum device design~\cite{Ai2025SQCwithGNN}. However, applying AI in QDA also presents challenges. These include the need for large, high-quality datasets for training AI models; however, quantum computing experiments are typically costly, resulting in limited data. Another challenge is whether AI-generated designs can be realized experimentally. Since the performance of quantum computing is very sensitive to certain design parameters, such as the geometry of qubit components and the shape of control pulses, AI models should be able to generate designs that precisely meet the requirements. Addressing these challenges will be crucial in the application of AI for QDA.
\end{itemize}

In conclusion, QDA stands as a rapidly evolving field that is critical to unlocking the transformative potential of quantum computing. Research into QDA methodologies, coupled with the advancement of associated tools, serves three pivotal purposes: it streamlines the automation of quantum design processes, accelerates progress in quantum hardware development, and elevates the overall performance of quantum systems through hardware-software co-design.

As quantum hardware technologies continue to advance, particularly in the dimension of scalability, and quantum computing tasks grow increasingly intricate, QDA will assume an ever more indispensable role. By addressing these evolving challenges, it will pave the way for the development of scalable, robust quantum computers---a key milestone in realizing the full promise of quantum computing.

\section*{Acknowledgment}

This work is supported by Zhongguancun Laboratory. 
J.~C. and Z.~J. would like to thank Prof. Mingsheng Ying for pointing out the significance of quantum design automation nearly a decade ago. J.~C. and H.-H.~Z. would like to thank their former colleagues at the Alibaba Quantum Laboratory---with special thanks to Yaoyun Shi---for building a small yet full-stack team, as well as for the extensive discussions and exchanges this team conducted at the time. These efforts were, in essence, the initial impetus behind this work. L.~K. thanks Runshi Zhou for helping with the generation of~\Cref{fig:codes}. Z.~J. is partially supported by National Key Research and Development Program of China (Grant No.\ 2023YFA1009403), National Natural Science Foundation of China (Grant No.\ 12347104), Beijing Natural Science Foundation (Grant No.\ Z220002). Y.~F. was partially supported by the National Natural Science Foundation of China under Grant 92465202.

\bibliographystyle{quantum}
\bibliography{QDA}

\end{document}